\documentclass[12pt, a4paper]{article}
\usepackage{algpseudocode}
\usepackage{algorithm}
\usepackage{amsfonts}
\usepackage{amsmath}
\usepackage{amssymb}
\usepackage{amsthm}
\usepackage{array}
\usepackage{bm}
\usepackage{booktabs}
\usepackage{caption}
\usepackage{color}
\usepackage{enumerate}
\usepackage{float}
\usepackage[stable, bottom, marginal]{footmisc}
\usepackage{geometry}
\usepackage{graphicx}
\usepackage{lineno}
\usepackage{mathrsfs}
\usepackage{multirow}
\usepackage[authoryear,round]{natbib}
\usepackage[normalem]{ulem}
\usepackage{subfigure}
\usepackage[figuresright]{rotating}
    \setcitestyle{authoryear, open = {(}, close = {)}}
\usepackage{changes}
\usepackage{hyperref}
\usepackage{placeins}
\newtheoremstyle{highlightremark} % 样式名
  {3pt}    % 上方间距
  {3pt}    % 下方间距
  {\itshape} % 正文样式（斜体）
  {}       % 缩进
  {\bfseries} % 标题样式（加粗 = highlight）
  {.}      % 标题后标点
  { }      % 标题后空格
  {}       % 标题格式
\theoremstyle{highlightremark}
\newtheorem{remark}{Remark}

\begin{document}

\begin{flushleft}
    {\bfseries \Large Robust optimal reconciliation for hierarchical time series forecasting with M-estimation}
\end{flushleft}

\vskip 1cm
\noindent
\textbf{Authors:}
\vskip 0.3cm
\noindent
Zhichao Wang$^{1, 2,}$, \\
Shanshan Wang$^{1, 3,*}$, \\
Wei Cao$^{1,*}$, \\
Fei Yang$^4$.

\vskip 1cm
\noindent
\textbf{Affiliations:}
\begin{enumerate}[{$^1$}]
    \setlength{\itemsep}{0pt}
    \setlength{\parsep}{0pt}
    \setlength{\parskip}{0pt}
    \item School of Economics and Management, Beihang University, Beijing 100191, China;
    \item Data Management Department, Industrial and Commercial Bank of China Limited,
    Beijing, China;
   \item MOE Key Laboratory of Complex System Analysis and Management Decision, Beihang University, Beijing, China;
   \item Department of Mathematics, The University of Manchester, Manchester, GB.
\end{enumerate}

\vskip 0.5cm
\noindent
\textbf{Correspondence:}
\vskip 0.3cm
\noindent
* Corresponding author. \\
Correspondence to: School of Economics and Management, Beihang University, Beijing 100191, China. \\
E-mail address: sswang@buaa.edu.cn (S.S.~Wang), caowei08@buaa.edu.cn (W. Cao).

\newpage

\vskip 2cm
\noindent
{\bfseries \large Abstract}
\vskip 1cm
\noindent

Aggregation constraints, arising from geographical or sectoral division, frequently emerge in a large set of time series. Coherent forecasts of these constrained series are anticipated to conform to their hierarchical structure organized by the aggregation rules. To enhance its resilience against potential irregular series, we explore the robust reconciliation process for hierarchical time series (HTS) forecasting. We incorporate M-estimation to obtain the reconciled forecasts by minimizing a robust loss function of transforming a group of base forecasts subject to the aggregation constraints. The related minimization procedure is developed and implemented through a modified Newton-Raphson algorithm via local quadratic approximation. Extensive numerical experiments are carried out to evaluate the performance of the proposed method, and the results suggest its feasibility in handling numerous abnormal cases (for instance, series with non-normal errors). The proposed robust reconciliation also demonstrates excellent efficiency when no outliers exist in HTS. Finally, we showcase the practical application of the proposed method in a real-data study on Australian domestic tourism.


\newpage

\section{Introduction} \label{Section_Introduction}
 Hierarchical time series (HTS) often refer to a collection of associated time series that form a hierarchy \citep{Hyndman2016Fast}, in which some series are aggregates of others.
In numerous applications, forecasts for all series within such a hierarchical system are of practical importance. For example, a country’s airline passenger volume can be geographically decomposed into provincial-, city-, and airport-level series. The most disaggregated series, such as passenger volumes at the airport or city level, constitute the bottom level of the hierarchy, while their aggregation forms higher-level series, with the national total representing the top level. Ignoring these hierarchical relationships may lead to incoherent or suboptimal forecasts, whereas explicitly accounting for the hierarchical structure enables coherent estimation across levels and supports more accurate forecasting and effective policy formulation. More comprehensive applications of HTS forecasting can be found in studies on tourism \citep{Athanasopoulos2009Hierarchical, Karmy2019Hierarchical, Ashouri2021Fast},
stock price index \citep{Sohnab2007Hierarchical,mattera2024improving}, indicator trends \citep{Kremer2015The, Shang2017Reconciling} and various types of energy demand \citep{jeon2019probabilistic,spiliotis2020cross,nystrup2020temporal,taieb2021hierarchical}.

Challenges in forecasting these time series stem from the intrinsic aggregation constraints within the hierarchy.
Specifically, a coherent forecast of an aggregated series ought to be the sum of the forecasts of all its direct components. One common approach, known as the ``bottom-up (BU)" method, is to obtain the forecasts of all series at the bottom level and sum them up along the hierarchy as the results of the related aggregated series \citep{Orcutt1968Data}.
This method has found diverse applications \citep[e.g.,][]{Duarte2007Forecasting, Ghedamsi2016Modeling, Lee2017Forecasting}.
Another method called ``top-down", adopting the opposite concept, forecasts the most aggregated series at the top level and allocates the result to each series below in a historical or estimated proportion \citep{Gross1990Disaggregation, Luna2011Top}. For reviews of these approaches, including a discussion of their advantages and disadvantages, see \cite{Dangerfield1992Top, Widiarta2009Forecasting, Williams2011Top}, among others.

The aforementioned methods conduct the forecasting of HTS within the traditional framework of multivariate time series. That is, they treat the hierarchical structure primarily as an intermediate device for obtaining the final forecasts and fail to explicitly incorporate the inherent hierarchical relationships into the forecasting process. To address this limitation, the forecast reconciliation method was proposed. It involves two steps: first, separate forecasts are generated for all series; then, these forecasts are adjusted to ensure coherence with aggregation constraints. 

For example, \cite{Athanasopoulos2009Hierarchical} first applied forecast reconciliation to Australian domestic tourism data and showed that it outperforms the ``bottom-up" and ``top-down" approaches.  Subsequently, \cite{Hyndman2011Optimal} introduced the optimal combination approach based on generalized least squares, which reconciles base forecasts of HTS using ordinary least squares (OLS) estimation. \cite{Hyndman2016Fast} extended this framework to account for heteroscedasticity by developing a weighted least squares (WLS) estimator. Later, \cite{Athanasopoulos2017Forecasting} proposed structure-based weights associated with aggregation levels in a temporal hierarchy, and \cite{kourentzesCrosstemporalCoherentForecasts2019} proposed cross-temporal forecast reconciliation.  
\cite{Wickramasuriya2019Optimal} proposed the forecast reconciliation known as ``MinT" and theoretically derived an explicit reconciled solution by minimizing the trace of the covariance matrix (i.e., the sum of the variances) of all reconciled forecast errors. Numerous empirical studies demonstrate that forecast reconciliation methods and their subsequent extensions have achieved remarkable improvement in HTS forecasting \citep[e.g.,][]{Yang2017Reconciling, Villegas2018Supply, Bai2019Distributed, Li2019A, Karmy2019Hierarchical, Oliveira2019Assessing}. In addition, some theoretical insights into the performance of forecast reconciliations methods has been provided by \cite{vanervenGameTheoreticallyOptimalReconciliation2015} and  \cite{wickramasuriyaPropertiesPointForecast2021}. For a more comprehensive demonstration of the superior performance of reconciliation approaches, see \cite{athanasopoulos2024forecast}.

As mentioned earlier, the reconciliation method generates forecasts that respect the inherent hierarchical structure, or, more generally, it is a post-forecasting process intended to improve the quality of forecasts for systems of linearly constrained multiple time series \citep{girolimetto2024cross}. This may improve the forecast accuracy of the hierarchy overall. However, conversely, the improvement in forecast accuracy in individual series at some particular forecast level may not be significant or even worse \citep{Wickramasuriya2019Optimal, pritulargaStochasticCoherencyForecast2021}. To address this issue, \cite{hollyman2021understanding} developed a level conditional coherent (LCC) method which keeps particular concern levels unchanged. \cite{difonzoForecastCombinationBased2021} subsequently studied the theoretical properties of LCC. A limitation of LCC is that the preserved series must comprise all series from a single hierarchical level, which may be impractical in real applications. To relax this restriction, \cite{zhang_optimal_2023} extended LCC by allowing a subset of series from different hierarchical levels to remain unchanged.

Although numerous methods have been proposed for forecasting HTS, most assume that base forecasts are unbiased. In practice, some series may exhibit irregular behavior, leading to inaccurate base forecasts. For example, less regular series may produce forecasts that deviate systematically from their expectations, with non-normal or large residuals--hereafter referred to as ``bad" forecasts. Such forecasts can disproportionately affect MinT reconciliation, as the least-squares (LS) loss is sensitive to outliers. Consequently, irregularity in a subset of series may propagate through the hierarchy via the reconciliation process. To address this issue, \citet{ben2019regularized, wickramasuriyaPropertiesPointForecast2021} relaxed the unbiasedness constraint on the reconciliation matrix and introduced a regularized regression framework for hierarchical forecasting, demonstrating empirical improvements over MinT in terms of total mean squared error. More recently, \citet{wang2025optimal} proposed a time-series selection procedure that improves reconciliation accuracy by filtering out biased base forecasts. These findings highlight the importance of robustness in HTS forecasting, particularly for large-scale hierarchies, and motivate the development of reconciliation procedures that mitigate the influence of potentially poor base forecasts. 

Furthermore, several robust forecast reconciliation approaches also have been proposed recently. \citet{lila2022forecasting} and \citet{meira2023novel} extend the MinT framework by incorporating the Huber and least absolute deviation (LAD) losses, respectively, while \citet{aikawa2025hierarchical} develops a robust optimization framework that accounts for uncertainty in the estimated covariance matrix and formulates the reconciliation problem as a semidefinite program. Although effective, these methods are tied to specific loss functions or rely on computationally intensive optimization schemes. 

In this paper, we propose a robust forecast reconciliation procedure based on M-estimation. As an alternative to least squares estimation, M-estimation is well suited for data contaminated with outliers. We first obtain base forecasts for the hierarchical time series using standard forecasting approaches, either independently or jointly. Reconciled forecasts are then obtained by minimizing an M-estimation objective that measures reconciliation cost through a user-specified loss function, such as the LAD or Huber loss \citep{Huber1973Robust}. Compared with existing robust reconciliation methods \citep{lila2022forecasting, meira2023novel}, the proposed approach offers two key advantages. 
First, it generalizes LS-based MinT estimator to accommodate any nonnegative convex loss function, thereby unifying a broad class of robust reconciliation procedures within a single framework. Second, the method is implemented using a modified Newton–Raphson algorithm based on the local quadratic approximation of \citet{Fan2001Variable}. Relative to the iterative reweighted least squares (IRLS) and simplex-based algorithms commonly used for Huber and LAD reconciliation, the Newton-based approach provides superior computational efficiency and scalability, making it particularly attractive for large and complex hierarchical systems. Moreover, when the least squares loss is specified, the iterative solution reduces to the explicit MinT estimator, extending the theoretical results of \citet{Wickramasuriya2019Optimal} to a robust reconciliation setting. We refer to the proposed method as RoME throughout the remainder of the paper.

To demonstrate the outstanding performance of proposed method, we conduct numerical experiments on both simulated and real-world data. A remarkable finding from simulation results is that, by employing the LAD loss function, the simple OLS-based reconciliation \citep{Hyndman2011Optimal} would achieve desirable (even the best) improvement in mid- and long-term forecasting. This property suggests a simplified robust reconciliation procedure without complex designs for the covariance matrix of base forecast errors, which may offer a feasible alternative to prevent the misspecification of the covariance structure of forecast errors over an ahead-forecasting period. Our methods and data have
been incorporated into the R package \textit{RoME}, freely available on \href{https://github.com/Weiccao/RoME}{GitHub}.

The remainder of this paper is organized as follows. Section~\ref{Section_Reconciliation} investigates the robust forecast reconciliation procedure with M-estimation and related computational algorithm.
Some preliminaries for the idea of reconciliation are also provided in this section.
Section~\ref{Section_Simulation} conducts extensive numerical experiments to verify the performance of the proposed robust procedure. Section~\ref{Section_Application} performs a real-data study on Australian domestic tourism to illustrate the effectiveness of the proposed method.
Some conclusions and discussions are finally given in Section~\ref{Section_Discussion}.

\section{Robust forecast reconciliation for HTS} \label{Section_Reconciliation}

Before stating the main procedure, we first introduce some notations and briefly review the MinT forecast reconciliation in Section~\ref{Section_Preliminaries}.
For simplification, we use commas and semicolons in matrix expressions to indicate that the adjacent blocks in a matrix are organized by column and by row, respectively.
All series in a hierarchy are ordered in a column vector from the most aggregated to disaggregated levels.

\subsection{Preliminaries} \label{Section_Preliminaries}

\begin{figure}[htbp]
	\centering
	\includegraphics[width = 10cm]{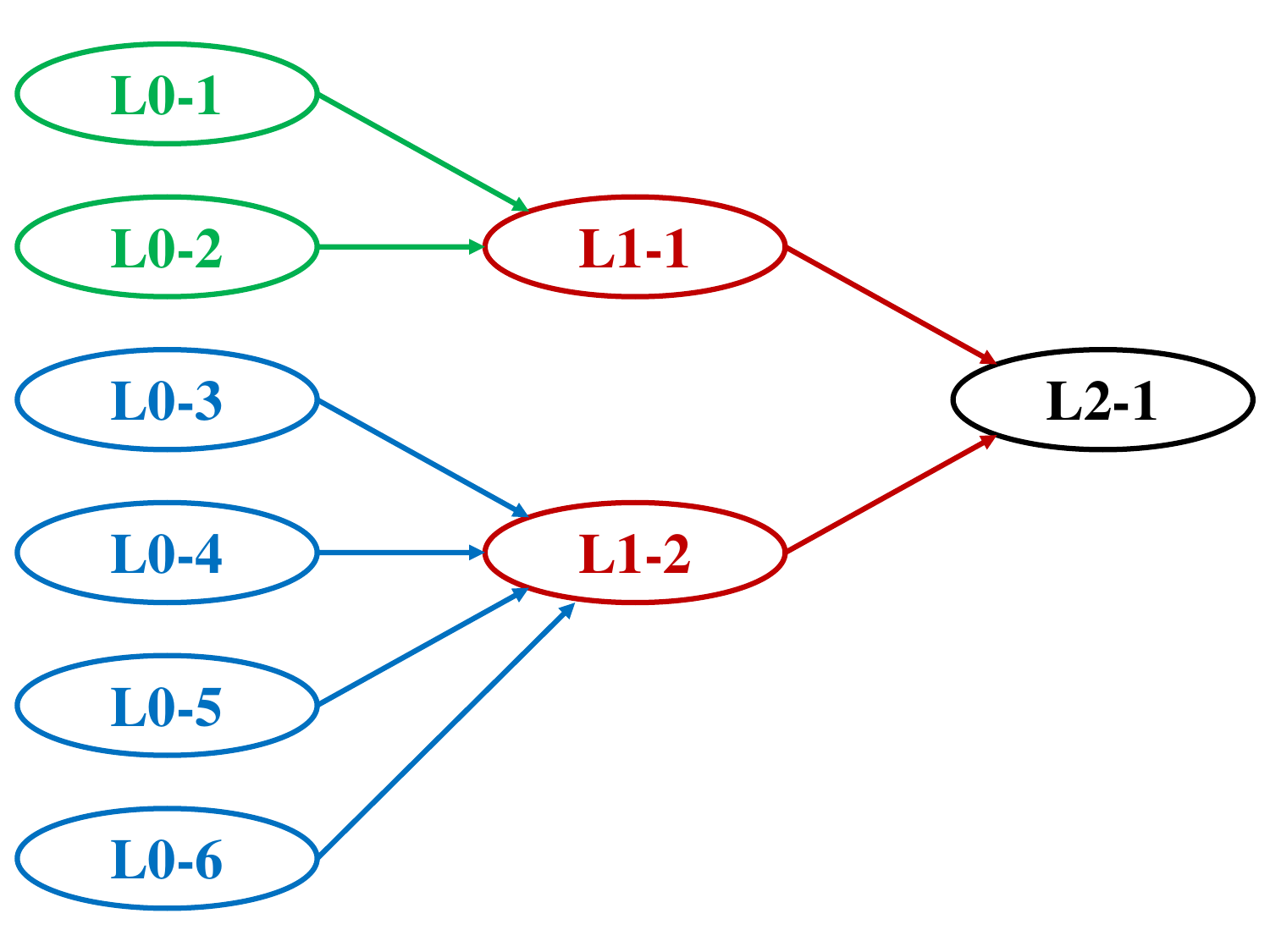} \\
	\caption{\rm Paradigm of a two-level HTS depicted by the hierarchical tree. This hierarchy contains $n = 6$ series at the bottom level, where the first 2 (in green) and the other 4 (in blue) of them separately aggregate two series (in red) at the middle level, and all series finally constitute the most aggregated one (in black) at the top level. The summing matrix is $\bm{S} = (\mathbf{1}_6^{\top}; \mathbf{D}_{2, 4}; \mathbf{I}_6)$, where $\mathbf{1}_6$, $\mathbf{D}_{2, 4}$ and $\mathbf{I}_6$ correspond to the 1, 2 and 6 series at the top (``L2"), middle (``L1") and bottom (``L0") levels, respectively.}
	\label{Figure_Paradigm}
\end{figure}

Consider an HTS, say $\mathcal{Y}_{\mathcal{I}} = \{ \bm{y}_t \in \mathbb{R}^n: t \in \mathcal{I} \}$ with a total of $n$ series in the hierarchy, the aggregation constraints can be formulated in a bottom-up form, i.e.,
    \begin{equation} \label{Equation_BU}
        \bm{y}_t = \bm{G} \bm{b}_t
        \quad
        (t \in \mathcal{I}),
    \end{equation}
where $\mathcal{I}$ denotes the time set, $\bm{b}_t$ contains the series, say $n_b$ ($n_b < n$), at the bottom level of $\bm{y}_t$ at time $t$, and $\bm{G}$ of dimension $n \times n_b$ denotes the summing matrix corresponding to the related aggregation rules.
For example, the summing matrix of the two-level HTS 
$ \bm{G} = (\bm{1}_6^{\top}; \bm{C}_{2, 4}; \mathbf{I}_6)$ is illustrated in Figure~\ref{Figure_Paradigm}, where 
$\bm{C}_{2, 4} =  \text{diag} (\mathbf{1}_2^{\top}, \mathbf{1}_4^{\top})$
is blocked diagonal of dimension $2 \times 6$, and $\mathbf{1}_\cdot$ and $\mathbf{I}_\cdot$ denote respectively the unit column vector and unit matrix of related dimensions indicated by the subscript.

\begin{remark} \label{re:hts}
We define HTS as an $n$-dimensional multivariate time series arranged in a hierarchical structure. However, \cite{panagiotelisForecastReconciliationGeometric2021} pointed out that data considered within the HTS framework need not strictly follow a hierarchy. For example, temporal hierarchies may involve grouped structures \citep{Athanasopoulos2017Forecasting} or cross-temporal structures that incorporate both cross-sectional and temporal dimensions \citep{kourentzesCrosstemporalCoherentForecasts2019}, for which forecast reconciliation is also applicable.
For simplicity, in this article we use the term HTS according to the definition given in \eqref{Equation_BU}, and focus on its forecast reconciliation.
\end{remark}

Given a group of $h$-step-ahead base forecasts at time $t$, denoted by $\hat{\bm{y}}_t (h)$, the expected reconciled forecasts, denoted by $\tilde{\bm{y}}_t (h)$, from general linear reconciliation procedure can be expressed as:

    \begin{equation} \label{Equation_Reconciliation}
        \tilde{\bm{y}}_t (h) = \bm{S} \bm{G}_h \hat{\bm{y}}_t (h)
    \end{equation}
with reconciliation matrix $\bm{G}_h$ of dimension $n_b \times n$ to be determined.
From (\ref{Equation_Reconciliation}), 
$\bm{G}_h$ distinguishes various forecasting methods for HTS (not only the family of forecast reconciliation) as it specifies the way of linearly combining the base forecasts as the reconciled ones for the bottom level.
For example, $\bm{G}_h = (\mathbf{0}_{n_b \times m^\ast} , \mathbf{I}_{n_b})$ coincides with the BU method, where $\mathbf{0}_{n_v \times m^\ast}$ denotes the zero matrix of dimension $n \times m^\ast$, and $m^\ast = n - n_b$. For more values of $\bm{G}_h$ and corresponding methods, see \cite{Athanasopoulos2009Hierarchical}.

In HTS forecasting, the base forecasts are usually assumed to be unbiased. That is, given the historical information, say $\mathcal{Y}_{\mathcal{I}_t}$ with $\mathcal{I}_t$ denoting the time set up to $t$, we have $\mathbb{E} [\hat{\bm{y}}_t (h) | \mathcal{Y}_{\mathcal{I}_t}] = \mathbb{E} [\bm{y}_{t + h} | \mathcal{Y}_{\mathcal{I}_t}]$.
If we further require the reconciliation procedure to result unbiased forecasts, then $\bm{G}_h$ is subject to (s.t.)
    \begin{equation} \label{Equation_Constraint}
        \bm{S} \bm{G}_h \bm{S} = \bm{S}
        \quad \text{or} \quad
        \bm{G}_h \bm{S} = \mathbf{I}_n.
    \end{equation}
Thus, the objective of the reconciliation procedure is to find an appropriate $\bm{G}_h$ under the assumption of unbiasedness (\ref{Equation_Constraint}).

Let $\hat{\bm{\varepsilon}}_t (h)$ and $\tilde{\bm{\varepsilon}}_t (h)$ denote the error vectors of related base and reconciled forecasts, i.e., $\hat{\bm{\varepsilon}}_t (h) = \bm{y}_{t + h} - \hat{\bm{y}}_t (h)$ and $\tilde{\bm{\varepsilon}}_t (h) = \bm{y}_{t + h} - \tilde{\bm{y}}_t (h)$, respectively. \cite{Wickramasuriya2019Optimal} developed the covariance structure of $\tilde{\bm{\varepsilon}}_t (h)$ as:
    \begin{equation} \nonumber
        \text{Var} {\big (} \tilde{\bm{\varepsilon}}_t (h) {\big |} \mathcal{Y}_{\mathcal{I}_t} {\big )} = \bm{S} \bm{G}_h \bm{W}_h \bm{G}_h^{\top} \bm{S}^{\top},
    \end{equation}
where $\bm{W}_h$ is the covariance matrix of mean-zero $\hat{\bm{\varepsilon}}_t (h)$, i.e., $\bm{W}_h = \mathbb{E} [ \hat{\bm{\varepsilon}}_t (h) \hat{\bm{\varepsilon}}_t^{\top} (h) | \mathcal{Y}_{\mathcal{I}_t} ]$.
By minimizing the trace of the aforementioned conditional covariance matrix, their MinT reconciliation derived the optimal value of
$\bm{G}_h$, denoted as
    \begin{equation} \nonumber
       \bm{G}_{\rm MinT} (h) = (\bm{S}^{\top} \bm{W}_h^{-1} \bm{S})^{-1} \bm{S}^{\top} \bm{W}_h^{-1},
    \end{equation}
with minimum forecast error from the perspective of the whole HTS. For computation efficiency, \cite{Wickramasuriya2019Optimal} reformulated the trace minimization problem as the following constrained WLS-based estimation with respect to (w.r.t.)~$\check{\bm{y}}_t (h)$, given the base forecasts $\hat{\bm{y}}_t (h)$:
    \begin{equation} \label{Equation_Optimization-MinT}
        \begin{array}{cl}
            \mathop{\text{minimize}}\limits_{\check{\bm{y}}_t (h) \in \mathbb{R}^m} & {{\big (} \check{\bm{y}}_t (h) - \hat{\bm{y}}_t (h) {\big )}^{\top} \bm{W}_h^{-1} {\big (} \check{\bm{y}}_t (h) - \hat{\bm{y}}_t (h) {\big )}} \\
            \text{s.t.} & \bm{U}^{\top} \check{\bm{y}}_t (h) = \mathbf{0}_{m^\ast},
        \end{array}
    \end{equation}
where $\bm{U} = (\mathbf{I}_{m^\ast}, -\bm{S}_0)$ of dimension $m^\ast \times n$ is given with $\bm{S}_0$ of dimension $m^\ast \times n_b$ piled up by the first $m^\ast$ row(s) of $\bm{S}$.
The optimal value of $\check{\bm{y}}_t (h)$ in (\ref{Equation_Optimization-MinT}) refers to the reconciled forecasts by MinT, and the optimization problem (\ref{Equation_Optimization-MinT}) leads to the equivalent solution to 
$\bm{G}_{\rm MinT} (h)$ as:
    \begin{equation} \label{Equation_P-MinT}
         \bm{G}_{\rm MinT} (h) = \bm{J} - \bm{J} \bm{W}_h \bm{U} (\bm{U}^{\top} \bm{W}_h \bm{U})^{-1} \bm{U}^{\top}
    \end{equation}
with constant matrix $\bm{J} = (\mathbf{0}_{n_b \times m^\ast}, \textbf{I}_{n_b})$ of dimension $n_b \times n$.

\begin{remark}
    \label{re:mint}
    Note that, from \eqref{Equation_Optimization-MinT}, the reconciliation procedure can be viewed as finding forecasts that minimize the cost of transforming the base forecasts to satisfy the aggregation constraints; in MinT, this cost is measured using a least-squares criterion.
\end{remark}

The specific value of 
$\bm{G}_h$ relies on $\bm{W}_h$.
Since the identification of $\bm{W}_h$, though possible, is of great challenge, \cite{Wickramasuriya2019Optimal} presented five alternative designs for $\bm{W}_h$, including OLS, WLSv (WLS with variance scaling), WLSs (WLS with structural scaling), Sample (full-sample estimation), and Shrink (Sample with off-diagonal shrinkage).
Table~\ref{Table_Design-W} summaries the expressions and assumptions of these covariance structure designs.
We note that $\bm{W}_h$ for a larger $h$ in these designs is either independent of the estimated covariance matrix or associated only with the estimate for $h = 1$, i.e.,
    \begin{equation} \nonumber
        \widehat{\bm{W}}_1 = T^{-1} \sum\limits_{t = 1}^T {\big (} \bm{y}_t - \hat{\bm{y}}_{t - 1} (1) {\big )} {\big (} \bm{y}_t - \hat{\bm{y}}_{t - 1} (1) {\big )}^{\top}
    \end{equation}
    with sample size $T + 1$.

\begin{table}[htbp]
	\caption {\rm Alternative designs for the covariance matrix of base forecast errors.}
	\label{Table_Design-W}
	\setlength\tabcolsep{8pt}
	\renewcommand{\arraystretch}{1.2}
	\begin{center}
		\footnotesize
		\begin{tabular}{p{1.3cm} p{4.3cm} p{8.6cm}}
			\toprule
			Method & Expression for $\bm{W}_h$ & Assumption \\
			\midrule
			OLS & $k_h \cdot \mathbf{I}_n$$^a$ & Series are uncorrelated with equal forecast variances. \\
			WLSv & $k_h \cdot \text{diag} (\widehat{\bm{W}}_1)$
            & Series are uncorrelated with different forecast variances estimated by the covariance matrix for $h = 1$. \\
			WLSs & $k_h \cdot \text{diag} (\bm{S} \bm{1}_n)$ & Forecast variances are equal for series at the bottom level and combined as those in the aggregated levels along the hierarchical structure. \\
			Sample & $k_h \cdot \widehat{\bm{W}}_1$ & Series are correlated and the correlation is measured by the estimated covariance matrix for $h = 1$. \\
			Shrink & $k_h [ \lambda \cdot \text{diag} (\widehat{\bm{W}}_1) + (1 - \lambda) \widehat{\bm{W}}_1 ]$ & The correlation of series is measured by $\widehat{\bm{W}}_1$ after an off-diagonal shrinkage with constant $\lambda \in [0, 1]$. \\
			\bottomrule
			\multicolumn{3}{l}{$^a$ $k_h$ is a given positive constant.}
		\end{tabular}
	\end{center}
\end{table}
As indicated by the MinT procedure described above, the reconciliation framework conceptually consists of two steps that are carried out separately. In this setting, the base forecasts can be viewed as given inputs obtained from historical data. Consequently, different choices of the forecast origin and horizon, denoted by $t$ and $h$ in $\hat{\bm{y}}_t(h)$, have little impact on the subsequent reconciliation step. In some cases, these indices can therefore be omitted without causing ambiguity.

From the perspective of the equivalent optimization formulation as stated in Remark \ref{re:mint}, forecast reconciliation starts with a collection of base forecasts and enforces the aggregation constraints at an associated cost, with the reconciled forecasts obtained by minimizing this cost. In the MinT framework, the cost is measured by a quadratic form with respect to the deviations between the base and reconciled forecasts.
However, LS loss function in MinT is sensitive to outliers, i.e., potential bad forecasts of HTS. As the linear combinations of all base forecasts, the reconciled results for the other well-behaving series with relatively accurate base forecasts may be perturbed by those with bad forecasts. This motivates the need of developing a robust reconciliation procedure to enhance the robustness of reconciled forecasts throughout the hierarchy.

\subsection{Robust procedure and implementation} \label{Section_RoME}

Under the assumption that $\bm{W}_h$ is positive-definite, we standardize the base forecasts and denote the resulting quantities as follows:
    \begin{equation} \nonumber
        \hat{\bm{y}}_t^\ast (h) = \bm{W}_h^{-1 / 2} \hat{\bm{y}}_t (h) = {\big (} \hat{y}_{t,1}^\ast (h), \hat{y}_{t,2}^\ast (h), \cdots, \hat{y}_{t,n}^\ast (h) {\big )}^\top.
    \end{equation}
Then, we propose the constrained convex minimization w.r.t. $\check{\bm{y}}_t^\ast (h)$:
    \begin{equation} \label{Equation_Optimization-RoME}
       \begin{array}{cl}  
        \mathop{\text{minimize}}\limits_{\check{\bm{y}}_t^\ast (h) \in \mathbb{R}^n} & {\sum\limits_{i = 1}^n \rho {\big (} | \check{y}_{t,i}^\ast (h) - \hat{y}_{t,i}^\ast (h) | {\big )}} \\
        \text{s.t.} & \bm{U}^{\top} \bm{W}_h^{1 / 2} \check{\bm{y}}_t^\ast (h) = \mathbf{0}_{m^\ast},
        \end{array}
    \end{equation}
where
$\check{\bm{y}}_t^\ast (h) = {\big (} \check{y}_{t,1}^\ast (h), \check{y}_{t,2}^\ast (h), \cdots, \check{y}_{t,n}^\ast (h) {\big )}^{\top}$ to be determined serves as the standardized result of $\check{\bm{y}}_t (h)$ in (\ref{Equation_Optimization-MinT}), and $\rho (| \cdot |)$ indicates a suitable choice of loss function. The optimization problem (\ref{Equation_Optimization-RoME}) clarifies the strategy of the proposed robust reconciliation procedure with M-estimation, as we refer to RoME, and the solution to $\check{\bm{y}}_t^\ast (h)$, denoted as 
$\check{\bm{y}}_t^\ast (h) = {\big (} \check{y}_{t,1}^\ast (h), \check{y}_{t,2}^\ast (h), \cdots, \check{y}_{t,n}^\ast (h) {\big )}^{\top}$ implies the expected reconciled forecasts, denoted by $\tilde{\bm{y}}_t (h)$, i.e.,
    \begin{equation} \nonumber
        \tilde{\bm{y}}_t (h) = \bm{W}_h^{1 / 2} \tilde{\bm{y}}_t^\ast (h).
    \end{equation}
Then, the Lagrangian function for (\ref{Equation_Optimization-RoME}) is expressed as
    \begin{equation} \label{Equation_Lagrangian}
        \mathcal{L} {\big (} \check{\bm{y}}_t^\ast (h); \bm{\lambda} {\big )} = \sum\limits_{i = 1}^n \rho {\big (} | \check{y}_{t,i}^\ast (h) - \hat{y}_{t,i}^\ast (h) | {\big )} + \bm{\lambda}^{\top} \bm{U}^{\top} \bm{W}_h^{1 / 2} \check{\bm{y}}_t^\ast (h),
    \end{equation}
where $\bm{\lambda}$ of dimension $m^\ast$ denotes the Lagrangian multipliers.

Finding the optimal value of $\check{\bm{y}}_t^\ast (h)$ that minimizes (\ref{Equation_Lagrangian}) faces great challenge due to the unspecific form of loss function.
Generally, the loss function can be any nonnegative convex function.
Some popular examples, for any $x \in \mathbb{R}$, include the $l_p$-regression estimate as $\rho (| x |) = |x|^p$ for some $p \in [1, 2]$; the Huber's estimate as
    \begin{equation} \nonumber
        \rho (|x|) = \frac{1}{2} |x|^2 \cdot \xi (|x| \leq k) + (k |x| - \frac{1}{2} k^2) \cdot \xi (|x| > k)
    \end{equation}
    for some positive constant $k$, where $\xi (\cdot)$ denotes the indicator function taking 1 (true) or 0 (false); and the quantile regression as $\rho (x) = q x_+ + (1 - q) (-x)_+$ for $0 < q < 1$, where the subscript ``$+$" denotes the positive operator, e.g., $x_+ = x \cdot \xi (x \geq 0)$.
Specifically, $\rho (| \cdot |)$ corresponds to the ordinary LS estimate when $p = 2$ or $k \rightarrow + \infty$, and the LAD estimate when $p = 1$ or $q = 1 / 2$.

\begin{figure}[ht]
	\begin{center}
		\subfigure[Loss function.]{
			\resizebox{7.2cm}{5.4cm}{\includegraphics{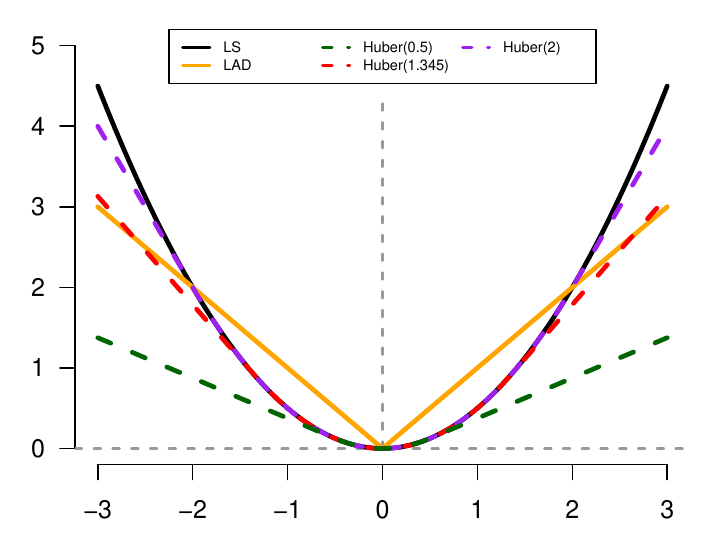}}
		}
		\qquad
		\subfigure[Derivative loss function.]{
			\resizebox{7.2cm}{5.4cm}{\includegraphics{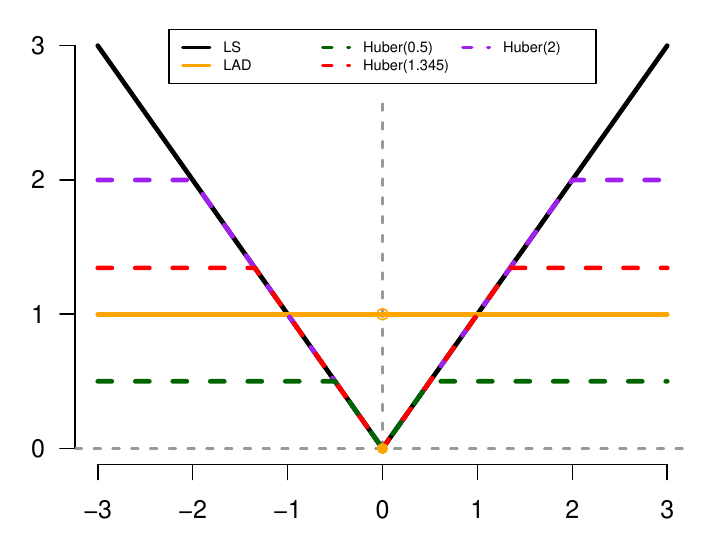}}
		}
	\end{center}
	\caption{\rm Curves of LS, LAD, Huber's loss functions and their derivatives, where ``Huber(0.5)", ``Huber(1.345)" and ``Huber(2)" correspond to the Huber's loss function with $k = 0.5$, 1.345 and 2, respectively.}
	\label{Figure_LossFunction}
\end{figure}

We mainly focus on three typical loss functions, i.e., LS, LAD and Huber's estimates, in this paper.
Figure~\ref{Figure_LossFunction} illustrates the difference of them and their derivatives in measuring the loss of the deviation from the origin. Specifically, LS loss function shows the familiar parabolic shape that rises sharply with the increase of the distance from the origin; by contrast, LAD loss function, showing a fold line, reacts relatively slowly to that increase in LS; and Huber's loss function falls in between, which coincides with LS for the points close to the origin and changes into the straight line like LAD for those far away.
The derivatives of three loss functions exhibits different properties.
The derivative of LS loss function reduces to the identical function, i.e., $\rho^\prime (| x |) = |x|$; that of LAD loss function is the constant function except for the origin, i.e., $ \rho^\prime ( |x| ) = \xi (|x| > 0) \quad (x \neq 0)$; and for Huber's loss function, the derivative combines that of the LS's and LAD's as
	\begin{equation} \nonumber
        \rho^\prime (|x|) = |x| \cdot \xi (|x| \leq k) + k \cdot \xi (|x| > k).
    \end{equation}

\begin{remark}
    \label{re:linearconst}
    Inspired by a referee, forecast reconciliation is a post-forecasting process intended, in a more general sense, to improve the quality of forecasts for systems of linearly constrained multiple time series \citep{Hyndman2011Optimal, panagiotelisForecastReconciliationGeometric2021}, of which HTS form a proper subset. Moreover, forecast reconciliation may be applied to temporal \citep{Athanasopoulos2017Forecasting} and cross-temporal \citep{kourentzesCrosstemporalCoherentForecasts2019} problems as well, not only cross-sectional ones. The proposed RoME method is formulated within the HTS framework, and it is also worth investigating within a more general forecast reconciliation framework. For example, in the empirical analysis, we found that the proposed method is also applicable to HTS with ``unbalanced" hierarchies. We will outline several potential analytical extensions in Section \ref{Section_Discussion}.
\end{remark}

\subsection{Implementation}\label{Section_implement}

The aforementioned examples demonstrate that the general loss function may not be differentiable at the origin (e.g., LAD), or correspond to a complicate derivative (e.g., the Huber's), which increases the difficulty of deriving an explicit solution to the minimization of (\ref{Equation_Lagrangian}) in a consistent way.
To address these problems, we conduct a local approximation of general loss function in a quadratic form, which was first suggested by \cite{Fan2001Variable} and referred to as local quadratic approximation (LQA).
Within the framework of LQA, the optimization of the Lagrangian function (\ref{Equation_Lagrangian}) can be carried out numerically through a modified Newton-Raphson algorithm.
In this paper, we adapt the perturbed LQA algorithm presented in \cite{Hunter2005Variable} for the robust constrained optimization and derive the iterative algorithm of the reconciled forecasts.

Specifically, let 
$\check{\bm{y}}_t^{\ast^{(\omega)}} (h) = {\big (} \check{y}_{t,1}^{\ast^{(\omega)}} (h), \check{y}_{t,2}^{\ast^{(\omega)}} (h), \cdots, \check{y}_{t,n}^{\ast^{(\omega)}} (h) {\big )}^{\top}$
denote the present value of $\check{\bm{y}}_t^\ast (h)$ at iteration $\omega$, where the superscript $\omega$ indexes the iteration, and $\omega = 0$ indicates particularly the initialization.
Suppose that the obtained $\check{\bm{y}}_t^{\ast^{(\omega)}} (h)$ is close enough to the expected value for (\ref{Equation_Optimization-RoME}), denoted as $\tilde{\bm{y}}_t^\ast (h) = \mathbf{W}_h^{-1 / 2} \tilde{\bm{y}}_t (h)$, the related cost of reconciliation will synchronously approach the optimal one, denoted as $\check{e}_{t,i}^{\ast^{(\omega)}} (h) = \check{y}_{t,i}^{\ast^{(\omega)}} (h) - \check{y}_{t,i}^{\ast^{(\omega)}} (h)$ and $\tilde{e}_{t,i}^\ast (h) = \tilde{y}_{t,i}^\ast (h) - \tilde{y}_{t,i}^\ast (h)$, respectively, for
 $i = 1, 2, \cdots, n$.
Therefore, the derivative of $\rho (| \cdot |)$ at $\tilde{e}_{t,i}^{\ast^{(\omega)}} (h)$, denoted by $[ \rho (| \tilde{e}_{t,i}^{\ast^{(\omega)}} (h) |) ]^\prime$, can be approximated as:
    \begin{equation} \label{Equation_derivative}
        {\big [} \rho {\big (} | \tilde{e}_{t,i}^{\ast^{(\omega)}} (h) | {\big )} {\big ]}^\prime = \rho^\prime {\big (} | \tilde{e}_{t,i}^{\ast^{(\omega)}} (h) | {\big )} \cdot \text{sign} (\tilde{e}_{t,i}^{\ast^{(\omega)}} (h)) \approx \frac{\rho^\prime (| \tilde{e}_{t,i}^{\ast^{(\omega)}} (h) |)} {| \tilde{e}_{t,i}^{\ast^{(\omega)}} (h) |} \tilde{e}_{t,i}^\ast (h).
    \end{equation}
Substituting $\tilde{e}_{t,i}^\ast (h)$ by $(\tilde{e}_{t,i}^\ast (h) + \check{e}_{t,i}^{\ast^{(\omega)}} (h)) / 2$, we can further extend (\ref{Equation_derivative}) to be:
    \begin{equation} \nonumber
        {\big [} \rho {\big (} | \tilde{e}_{t,i}^{\ast^{(\omega)}} (h) | {\big )} {\big ]}^\prime \approx \frac{1}{2} \frac{\rho^\prime (|\check{e}_{t,i}^{\ast^{(\omega)}} (h)|)} {|\check{e}_{t,i}^{\ast^{(\omega)}} (h)|} {\big (} \tilde{e}_{t,i}^\ast (h) + \check{e}_{t,i}^{\ast^{(\omega)}} (h) {\big )},
    \end{equation}
which leads to the quadratic approximation of $\rho (| \tilde{e}_{t,i}^\ast (h) |)$, i.e.,
    \begin{equation} \label{Equation_Approximation}
        \rho {\big (} | \tilde{e}_{t,i}^\ast (h) | {\big )} \approx \rho {\big (} | \check{e}_{t,i}^{\ast^{(\omega)}} (h) | {\big )} + \frac{1}{2} \frac{\rho^\prime (|\check{e}_{t,i}^{\ast^{(\omega)}} (h)|)} {|\check{e}_{t,i}^{\ast^{(\omega)}} (h)|} {\big (} (\tilde{e}_{t,i}^\ast (h))^2 - (\check{e}_{t,i}^{\ast^{(\omega)}} (h))^2 {\big )}.
    \end{equation}
Thus, after omitting the terms irrelevant to $\tilde{e}_{t,i}^\ast (h)$ in (\ref{Equation_Approximation}), the proposed perturbed LQA algorithm updates the reconciled forecasts from $\check{\bm{y}}_t^{\ast^{(\omega)}} (h)$ as
    \begin{equation} \label{Equation_Update}
        \check{\bm{y}}_t^{\ast^{(\omega + 1)}} (h) = \mathop{\arg\min}\limits_{\check{\bm{y}}_t^\ast (h)} {\Big \{} \sum_{i = 1}^m \frac{1}{2} \frac{\rho^\prime (| \check{e}_{t,i}^{\ast^{(\omega)}} (h) |)} {| \check{e}_{t,i}^{\ast^{(\omega)}} (h) | + \varsigma} {\big (} \check{y}_{t,i}^\ast (h) - \hat{y}_{t,i}^\ast (h) {\big )}^2 + \bm{\lambda}^{\top} \bm{U}^{\top} \mathbf{W}_h^{1 / 2} \check{\bm{y}}_t^\ast (h) {\Big \}}
    \end{equation}
    with a slight positive perturbation, say $\varsigma = 10^{-8}$.
The first-order necessary conditions of (\ref{Equation_Update}) indicate that $\check{\bm{y}}_t^{\ast^{(\omega + 1)}} (h)$ satisfies   
    \begin{equation} \label{Equation_Condition}
        (\bm{D}_h^{(\omega)})^{-1} {\big (} \check{\bm{y}}_t^{\ast^{(\omega + 1)}} (h) - \hat{\bm{y}}_t^\ast (h) {\big )} + \mathbf{W}_h^{1 / 2} \bm{U} \bm{\lambda} = \mathbf{0}_n,
    \end{equation}
where $\bm{D}_h^{(\omega)}$ of dimension $n \times n$ is diagonal matrix with the $i$-th diagonal element being $\frac{ | \check{e}_{t,i}^{\ast^{(\omega)}} (h) | + \varsigma} {\rho^\prime (| \check{e}_{t,i}^{\ast^{(\omega)}} (h) |)}$ ($i = 1, 2, \cdots, n$).
Finally, the iterative solution can be easily obtained as
    \begin{equation} \label{Equation_y-RoME}
        \check{\bm{y}}_t^{(\omega + 1)} (h) = {\big (} \mathbf{I}_m - \bm{W}_h^{1 / 2} \mathbf{D}_h^{(\omega)} \bm{W}_h^{1 / 2} \bm{U} (\bm{U}^{\top} \bm{W}_h^{1 / 2} \mathbf{D}_h^{(\omega)} \bm{W}_h^{1 / 2} \bm{U})^{-1} \bm{U}^{\top} {\big )} \hat{\bm{y}}_t (h),
    \end{equation}
$\bm{G}_{\rm RoME}^{(\omega + 1)} (h)$, as
    \begin{equation} \label{Equation_P-RoME}
        \bm{G}_{\rm RoME}^{(\omega + 1)} (h) = \bm{J} - \bm{J} \bm{W}_h^{1 / 2} \mathbf{D}_h^{(\omega)} \bm{W}_h^{1 / 2} \bm{U} (\bm{U}^{\top} \bm{W}_h^{1 / 2} \mathbf{D}_h^{(\omega)} \bm{W}_h^{1 / 2} \bm{U})^{-1} \bm{U}^{\top}.
    \end{equation}
See details in Appendix~\ref{App_A1}.

The aforementioned iterative procedure for RoME reconciliation is convergent due to the convex objective function.
Therefore, the initial value of $\check{\bm{y}}_t^\ast (h)$ can be any satisfying the aggregation constraints, where the simplest one is suggested to be 
$\check{\bm{y}}_t^{\ast^{(0)}} (h) = \mathbf{0}_n$.
In this paper, the convergence criterion is that the maximum difference between the updated and previous reconciled forecasts, say $\check{\bm{y}}_t^{(\omega)} (h)$ and $\check{\bm{y}}_t^{(\omega - 1)} (h)$, is small enough; specifically, the algorithm converges if they first realize,
    \begin{equation} \nonumber
        {\big \|} \check{\bm{y}}_t^{(\omega)} (h) - \check{\bm{y}}_t^{(\omega - 1)} (h) {\big \|}_\infty < \epsilon,
    \end{equation}
    for a given convergence threshold, say $\epsilon = 10^{-4}$, where $\| \cdot \|_\infty$ denotes the maximum norm of a vector.
Meanwhile, the algorithm will also stop if it exceeds the iteration limit, say $\omega_{\rm max} = 1000$.
As to be verified in the numerical experiments, the proposed algorithm with specific loss functions under the aforementioned convergence criteria will realize consistent and efficient results.
We report the computation time of the proposed robust reconciliation procedure and MinT in Appendix~\ref{App_B1}, and the computation burden caused by the iterative procedure would be acceptable and worthy in view of the benefits of RoME on the reconciled forecasts from MinT.

In summary, we describe the computational algorithm of the proposed RoME reconciliation procedure for HTS forecasting in Algorithm~\ref{alg}.
To feature the core process, we have removed the indicators of the latest observation time and ahead-forecasting step, as discussed in Section~\ref{Section_Preliminaries}.

\begin{algorithm}[ht]
	\caption{\textbf{RoME reconciliation procedure for HTS forecasting.}}
	\vskip 0.2cm
	\textbf{Input:}
	The base forecast $\hat{\bm{y}}$ and its estimated covariance matrix $\widehat{\bm{W}}$; the summing matrix $\bm{S}$ and related matrix $\bm{U}$; the initial value $\check{\bm{y}}^{(0)}$; the positive perturbation $\varsigma$; the convergence threshold $\epsilon$; the iteration limit $\omega_{\rm max}$.
	\vskip 0.2cm
	\textbf{Output:}
	The reconciled forecasts $\tilde{\bm{y}}$.
	\begin{algorithmic}[1]
		\State Set $\omega = 0$;
		\Repeat
		\State compute $\check{\bm{e}}^{\ast^{(\omega)}}$ and $\mathbf{D}^{(\omega)}$:
        \begin{eqnarray*}
			& & \check{\bm{e}}^{\ast^{(\omega)}} = \widehat{\bm{W}}^{-1 / 2} (\check{\bm{y}}^{(\omega)} - \hat{\bm{y}}) := (\check{e}_1^{\ast^{(\omega)}}, \check{e}_2^{\ast^{(\omega)}}, \cdots, \check{e}_n^{\ast^{(\omega)}})^{\top}, \\
			& & \mathbf{D}^{(\omega)} = \text{diag} {\Big (} \frac{|\check{e}_1^{\ast^{(\omega)}}| + \varsigma} {\rho^\prime (| \check{e}_1^{\ast^{(\omega)}} |)}, \frac{|\check{e}_2^{\ast^{(\omega)}}| + \varsigma} {\rho^\prime (| \check{e}_2^{\ast^{(\omega)}} |)}, \cdots, \frac{|\check{e}_n^{\ast^{(\omega)}}| + \varsigma} {\rho^\prime (| \check{e}_n^{\ast^{(\omega)}} |)} {\Big )},
		\end{eqnarray*}
		\State compute $\check{\bm{y}}^{(\omega + 1)}$:
		\begin{equation*}
			\check{\bm{y}}^{(\omega + 1)} = {\big (} \mathbf{I}_m - \widehat{\bm{W}}^{1 / 2} \mathbf{D}^{(\omega)} \widehat{\bm{W}}^{1 / 2} \bm{U} (\bm{U}^{\top} \widehat{\bm{W}}^{1 / 2} \mathbf{D}^{(\omega)} \widehat{\bm{W}}^{1 / 2} \bm{U})^{-1} \bm{U}^{\top} {\big )} \hat{\bm{y}},
		\end{equation*}
		\State let $\omega := \omega + 1$;
		\Until
		\begin{equation*}
			\| \check{\bm{y}}^{(\omega)} - \check{\bm{y}}^{(\omega - 1)} \|_\infty < \epsilon
			\quad {\rm or} \quad
			\omega > \omega_{\rm max};
		\end{equation*}
		\State \Return
		\begin{equation*}
			\tilde{\bm{y}} := \check{\bm{y}}^{(\omega)}.
		\end{equation*}
	\end{algorithmic}
	\label{alg}
\end{algorithm}

\subsection{Some issues} \label{Section_Issue}

In this section, we first clarify the relationship between the proposed RoME and latest MinT reconciliations.
As the generalization of forecast reconciliation within the robust framework, the constrained optimization problem (\ref{Equation_Optimization-RoME}) extends the LS-based minimization in MinT to the general loss function.
The related iterative computational algorithm for RoME can be verified to keep consistent with the explicit solution in MinT.
Actually, when LS loss function is specified, the objective function reduces to
    \begin{equation} \nonumber
       \frac{1}{2} \sum_{i = 1}^n {\big |} \check{y}_{t,i}^\ast (h) - \hat{y}_{t,i}^\ast (h) {\big |}^2,
    \end{equation}
which is proportional to $(\check{\bm{y}}_t (h) - \hat{\bm{y}}_t (h))^{\top} \bm{W}_h^{-1} (\check{\bm{y}}_t (h) - \hat{\bm{y}}_t (h) )$ in \eqref{Equation_Optimization-MinT}, since $\check{\bm{y}}_t^\ast (h) = \bm{W}_h^{-1 / 2} \check{\bm{y}}_t (h)$.
If we ignore the slight perturbation in $\mathbf{D}_h^{(\omega)}$, it will equivalently equal the unit matrix for any $\omega$.
Therefore, the iterative solution to
$\bm{G}_h$ is simplified into (\ref{Equation_P-MinT}), indicating that the algorithm will immediately finish at the reconciled forecasts by MinT.
Similar conclusions can also be drawn for (\ref{Equation_P-RoME}). These show that the proposed constrained optimization problem in RoME is in line with the equivalent LS-based estimation in MinT when LS loss function is employed.

Then, we analogously reformulate the proposed constrained optimization problem for RoME from the perspective of the idea of trace minimization in MinT.
Specifically, let $\tilde{\bm{\varepsilon}}_t (h) = {\big (} \tilde{\varepsilon}_{t,1} (h), \tilde{\varepsilon}_{t,2} (h), \cdots, \tilde{\varepsilon}_{t,n} (h) {\big )}^{\top}$ denote the reconciled forecast errors of $\tilde{\bm{y}}_t (h)$, the expectation of the general loss function~w.r.t.~$\tilde{\bm{\varepsilon}}_t (h)$, given the historical information, say $\mathcal{Y}_{\mathcal{I}_t}$, can be defined in component-wise manner, i.e., $\mathbb{E} [ \rho (| \tilde{\varepsilon}_{t,i} (h) |) | \mathcal{Y}_{\mathcal{I}_t} ]$ ($i = 1, 2, \cdots, n$).
As a generalization of the variance of $\tilde{\varepsilon}_{t,i}(h)$, the expectation above measures the average deviation of the reconciled forecast errors from the origin through the general
loss function.
By minimizing the sum of $\mathbb{E} [ \rho (| \tilde{\varepsilon}_{t,i} (h) |) | \mathcal{Y}_{\mathcal{I}_t} ]$ for $i = 1, 2, \cdots, n$, we can formulate the constrained optimization problem for RoME as:
    \begin{equation} \label{Equation_Model-RoME}
        \begin{array}{cl}
            \mathop{\text{minimize}}\limits_{\bm{G}_h \in \mathbb{R}^{n_b \times n}} & \sum\limits_{i = 1}^n \mathbb{E} {\big [} \rho (| \tilde{\varepsilon}_{t,i} (h) |) {\big |} \mathcal{Y}_{\mathcal{I}_t} {\big ]} \\
            \text{s.t.} & \bm{G}_h \bm{S} = \mathbf{I}_{n_b}.
        \end{array}
    \end{equation}
When LS loss function is specified, the objective function in (\ref{Equation_Model-RoME}) will reduce to the sum of the variances of all components in $\tilde{\bm{\varepsilon}}_t (h)$, i.e., the trace of the covariance matrix of $\tilde{\bm{\varepsilon}}_t (h)$, as considered in MinT.
The reconciliation matrix can be identified theoretically from (\ref{Equation_Model-RoME}), however, with no explicit expression either.
By adapting the quadratic approximation in (\ref{Equation_Approximation}), the iterative solution from the reconciled forecasts $\bm{G}_{\rm RoME}^{(\omega)} (h)$
can be derived as
    \begin{equation} \label{Equation_P-RoME-equivalent}
        \bm{G}_{\rm RoME}^{(\omega + 1)} (h) = {\big (} \bm{S}^{\top} (\mathbb{W}_h^{(\omega)})^{-1} \bm{S} {\big )}^{-1} \bm{S}^{\top} (\mathbb{W}_h^{(\omega)})^{-1},
    \end{equation}
where $\mathbb{W}_h^{(\omega)} = (\bm{\Omega}_h^{(\omega)})^{1 / 2} \bm{W}_h (\bm{\Omega}_h^{(\omega)})^{1 / 2}$, $\bm{\Omega}_h^{(\omega)}$ of dimension $n$ consists of the diagonal elements $\varpi_{hi}^{(\omega)}$ for $i = 1, 2, \cdots, n$ as
    \begin{equation} \nonumber
         \varpi_{hi}^{(\omega)} = \frac{\rho^\prime (| \tilde{\varepsilon}_{t,i}^{(\omega)} (h) |)} {| \tilde{\varepsilon}_{t,i}^{(\omega)} (h) |} = \frac{\rho^\prime (| \bm{S}_i \bm{G}_{\rm RoME}^{(\omega)} (h) \hat{\bm{\varepsilon}}_t (h) |)} {| \bm{S}_i \bm{G}_{\rm RoME}^{(\omega)} (h) \hat{\bm{\varepsilon}}_t (h) |},
    \end{equation}
    and the row vector $\bm{S}_i$ denotes the $i$-th row of $\bm{S}$.
See details in Appendix~\ref{App_A2}.

The iterative solution (\ref{Equation_P-RoME-equivalent}) 
also satisfies the property of global convergence.
Compared with the explicit expression in MinT, it substitutes the covariance structure $\bm{W}_h$ with a weighted quadratic approximation, i.e., $\mathbb{W}_h^{(\omega)}$, which reduces to (\ref{Equation_P-MinT}) when $\rho (|\cdot|)$ coincides with LS loss function.
The expectation minimization above provides limited insight into the proposed robust reconciliation procedure.
Thus, we prefer to formulate the RoME reconciliation in terms of the constrained optimization problem by (\ref{Equation_Optimization-RoME}) as it embodies the essential idea and process of the proposed method.

Next, we discuss the specific value of the dynamic weight, i.e, $\bm{D}_h^{(\omega)}$, for LAD loss function since it does not reduce to the unit matrix, and will be irreversible when any reconciliation cost is zero. Appropriate weights help to reconcile the base forecasts to be with the aggregation constraints, however, we need to prevent the weights from being extremely large for some series.
Two implementation solutions would be available.
One is \emph{Perturbation}; that is, introduce a positive perturbation to $\bm{D}_h^{(\omega)}$, say $\varrho = 10^{-8}$, such that $\bm{D}_h^{(\omega)}$ is reversible.
Actually, the introduction of $\varrho$ adds an equal basic weight for every series in the hierarchy, which guarantees the upper limit of the dynamic weights in implementation.
The other is \emph{Huber approximation}.
Recalling the relationship between LAD and Huber's loss functions, we may conduct LAD loss function through the Huber's one with a small $k$.
In that case, the narrow band of the LS part centered on the origin ensures the weights corresponding to extremely small reconciliation cost to be 1, with the others' less then 1.
As verified in Appendix~\ref{App_B2}, the reconciled results for two implementation solutions are extremely the same.
Therefore, we adopt the Huber approximation throughout the paper since it avoids introducing the additional perturbation to $\mathbf{D}_h^{(\omega)}$, and maintains the expressions presented in Section~\ref{Section_RoME}.

Moreover, the appropriate value of the constant $k$ in Huber's loss function, suggested by \cite{Huber1981Robust}, lies in the range between 1 and 2.
In most R functions for robust analysis (e.g., \emph{rlm} in package \emph{MASS}), it defaults to be 1.345, which will achieve 95\% efficiency in theory when errors of all series are standard normal.
In this paper, we adopt this value and introduce the estimated standard deviation of the base forecast residuals, denoted by $\hat{\sigma}$, for accommodating the scale of HTS.
Thus, the parameter is set to be $k = 1.345 \cdot \hat{\sigma}$ in both numerical experiments and real-data study.

Finally, within the traditional framework of univariate or multivariate time series forecasting, numerous methods can be conducted for obtaining a group of unbiased base forecasts.
Regarding as a separate step before the reconciliation procedure, we shall not put much technical discussion on the forecasting in this paper, and simply implement it through some automatic algorithms built up by \cite{Hyndman2008Automatic}.
Specifically, two popular approaches, i.e., ARIMA (autoregressive integrated moving average) and ETS (exponential smoothing), are performed in R project by package \emph{forecast} with functions \emph{auto.arima} and \emph{ets}, respectively.

\section{Numerical experiments} \label{Section_Simulation}

In this section, we report some simulation results to illustrate the performance of the proposed robust reconciliation procedure.
Five issues are considered, i.e., the types of non-Gaussian series, possible efficiency loss for normal cases, proportion of bad forecasts, correlation among HTS, and complexity of hierarchy in Sections~\ref{Section_NonGaussian}--\ref{Section_Complexity}, respectively.

The behavior of a series is described through the distribution of its random error.
Specifically, we generate the data of an HTS with $n_b$ series in the bottom level from the autoregressive model:
    \begin{equation} \label{Equation_Generation}
                b_{t} = \alpha_i b_{t - 1, i} + \sigma_i \varepsilon_{t,i}
        \quad
        (t = 1, 2, \cdots, T; \ i = 1,2, \cdots, n_b),
    \end{equation}
where $T$ indicates the sample size, $b_{t,i}$ and $\varepsilon_{t,i}$ denote the observations of the $i$-th most disaggregated series and its related error at time $t$, and $\alpha_i$ and $\sigma_i$ are respectively. The autoregressive coefficient and variation parameter that may vary with series.
The data of the aggregated series are then obtained based on those for the bottom level by (\ref{Equation_BU}), which finally finishes the data generation of the whole HTS, i.e., $\bm{y}_t$, containing $n$ series.
The sample size is fixed as $T = 192$, where the first 180 and latest 12 observations are used for conducting the forecast reconciliation and evaluating the forecasting accuracy, respectively. For every case, we replicate the simulation 1000 times independently.

We evaluate the forecasting accuracy of competing models using the root mean square error (RMSE), which provides a natural measure of the magnitude of average squared forecast errors. Specifically, consider the $h$-step-ahead reconciled forecasts from $\bm{y}_t$, say $\tilde{\bm{y}}_t (h) = {\big (} \tilde{y}_{t,1} (h), \tilde{y}_{t,2} (h), \cdots, \tilde{y}_{t,n} (h) {\big )}^{\top}$ ($h = 1, 2, \cdots, H$), $\bm{y}_{t + h} = (y_{t + h, 1}, y_{t + h, 2}, \cdots, y_{t + h, n})^{\top}$ denotes the corresponding observation of $\tilde{\bm{y}}_t (h)$, then the RMSE measure for these reconciled forecasts over a period of out-of-sample forecasting, say $H$ terms, is 
    \begin{equation} \nonumber
        \text{RMSE} {\big (} \tilde{\bm{y}}_t (h), \bm{y}_{t + h}; h = 1, 2, \cdots, H {\big )} = {\Big (} \frac{1}{nH} \sum_{h = 1}^H \sum_{i = 1}^n {\big (} y_{t + h, i} - \tilde{y}_{t,i} (h) {\big )}^2 {\Big )}^{1 / 2},
    \end{equation}
where both the ahead-forecasting steps (i.e., $h$) and series at different aggregation levels are treated comparably, with equal weights. To eliminate the impact of data magnitude, we use the percentage decrease (or increase) in average RMSE relative to the incoherent base forecasts, as in \cite{Wickramasuriya2019Optimal, wickramasuriya2020optimal}, defined as:
    \begin{equation} \nonumber
        \frac{\Big(\text{RMSE}_{\text{Compete}} - \text{RMSE}_{\text{Base}} \Big)}{\text{RMSE}_{\text{Base}}}\times 100, 
    \end{equation}
where $\text{RMSE}_{\text{Compete}}$ denotes the RMSE of the reconciled forecasting method (e.g., MinT or RoME), and $\text{RMSE}_{\text{Base}}$ denotes the RMSE of the incoherent base forecasts. A negative value indicates a reduction in average RMSE relative to the base forecasts.

For comparison, the MinT reconciliation procedure is implemented as a special case of RoME with LS loss function, and the results show no differences to those obtained through package \emph{hts}. For unbalanced hierarchies, we also found that RoME with the LS loss function is equivalent to point forecast reconciliation using the \emph{FoReco} package \citep{FoReco2025}, with the covariance matrix of base forecast errors specified as OLS, WLSv, WLSs, or Shrink, as shown in Table \ref{Table_Design-W}. For more details, see Appendix \ref{App_B5}.
The shrinkage parameter $\lambda$ is determined as the scale and location invariant estimator presented in \cite{Schafer2005A}, and the estimation is also realized through package \emph{hts}, as illustrated by \cite{Hyndman2024hts}.
We note that the covariance matrix $\bm{W}_h$ in this package is estimated based on the in-sample base forecast residuals, which is also employed in this paper.

We identify a specific RoME model by merging the names of the loss function and covariance design used.
For example, LAD-OLS, as to be discussed later, indicates the reconciliation procedure with LAD loss function under OLS covariance design.

\subsection{Non-Gaussian series} \label{Section_NonGaussian}

Many forecasting approaches are investigated on the assumption of Gaussian white noise in theory. However, this assumption does not always hold, and the misspecification of model may probably lead to bad forecasts with non-Gaussian residuals.

\begin{figure}[ht]
	\centering
	\includegraphics[width = 12cm]{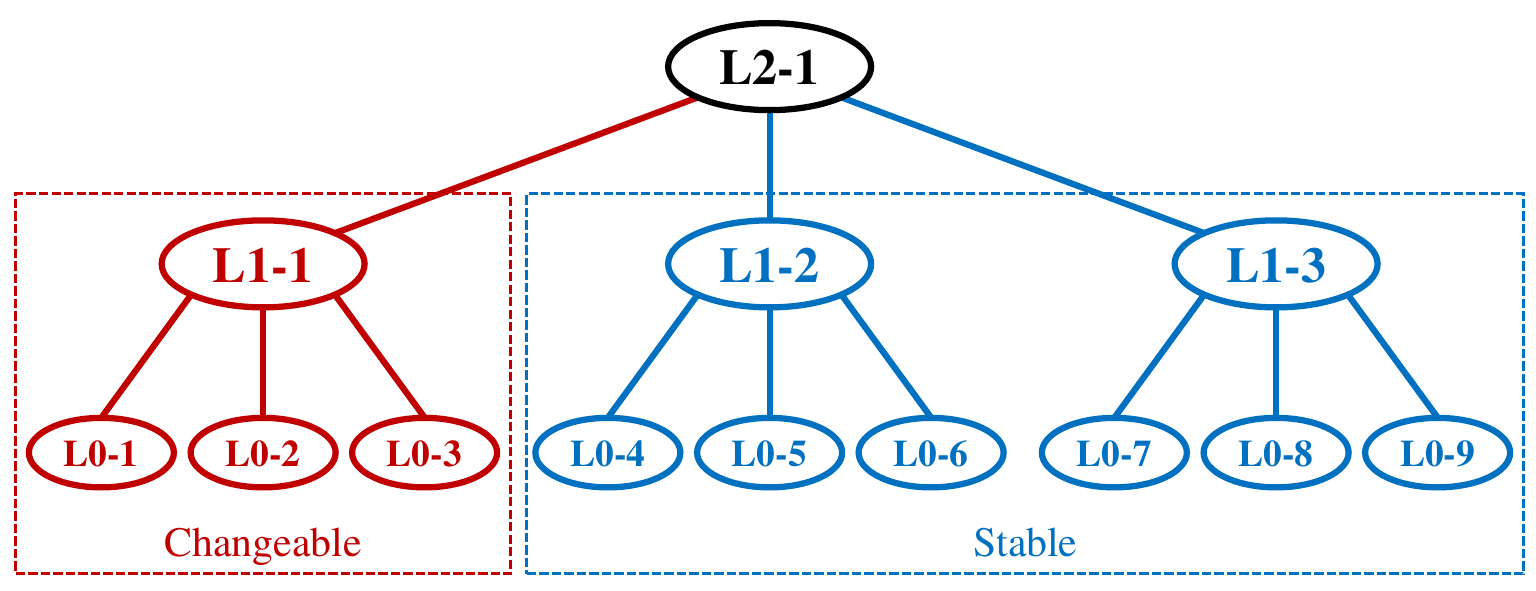} \\
	\caption{\rm Hierarchical tree for the numerical experiments in Section~\ref{Section_NonGaussian}. This hierarchy contain $n=13$ series, with 9 series (``L0-$i$", $i = 1, 2, \cdots, 9$) at the bottom level, 3 series (``L1-1", ``L1-2" and ``L1-3") at the middle level, and the most aggregated one (``L2-1") at the top level. The ``Changeable" group (in red) indicates the series of which errors have non-normal distributions, the ``Stable" one (in blue) indicates the other Gaussian series, and the ``All" group indicates the whole HTS.}
	\label{Figure_NonGaussian}
\end{figure}

We first consider the hierarchical structure with $n_b = 9$ as illustrated in Figure~\ref{Figure_NonGaussian}, and assume that the first three series at the bottom level (i.e., the ``Changeable" group) have irregular errors.
Here we consider three types of non-normal distributions as follows:
    \begin{enumerate}[{\ (a)}]
        \setlength{\itemsep}{0pt}
        \setlength{\parsep}{0pt}
        \setlength{\parskip}{0pt}
        \item the mixture normal distribution, denoted as $0.9 \mathcal{N} (0, 1) + 0.1 \mathcal{N} (0, 3^2)$;
        \item the standard $t$ distribution with 3 degrees of freedom, denoted as $t (3)$;
        \item the standard Cauchy distribution, denoted as ${\rm Cauchy} (0, 1)$.
    \end{enumerate}

The errors of  remaining  six series at the bottom level are assumed to be of standard normality, and the correlation coefficient between any two of them, say the $i$-th and $i^\prime$-th ones ($i = 4, 5, \cdots, n_b$), is $\varrho^{|i - i^\prime|}$ with $\varrho = 0.4$.
In Model~(\ref{Equation_Generation}), the autoregressive coefficient and variation parameter are $\alpha_j = 0.8$ and $\sigma_j = 0.5$, respectively.
Tables~ \ref{Table_NonGaussian-mixture}--\ref{Table_NonGaussian-cauchy} compare the reconciled forecasts by RoME with three types of loss functions, including MinT, under three types of non-normal distributions, where the base forecasts are obtained through the automatic ARIMA algorithm.

\begin{table}[htbp]
	\caption {\rm Out-of-sample forecasting performances of RoME with the base forecasts obtained by ARIMA under the mixture normal distribution in Section~\ref{Section_NonGaussian}. The row ``Base" reports the averages of RMSE for the related base forecasts, and the others report percentage increases of average RMSE between the alternative models and the benchmark. The columns ``Changeable", ``Stable" and ``All" correspond to the related groups of series, and the sub-columns ``$h = 1$", ``$1 \sim 6$" and ``$1 \sim 12$" indicate the out-of-sample forecasting periods of $H = 1$, 6 and 12, respectively. The best results are highlighted in bold.}
	\label{Table_NonGaussian-mixture}
	\setlength\tabcolsep{4.5pt}
	\renewcommand{\arraystretch}{1.2}
	\begin{center}
		\footnotesize
		\begin{tabular}{cccccccccccccc}
			\toprule
			\multirow{2}{*}{$\rho$} & \multirow{2}{*}{$\bm{W}_h$} & & \multicolumn{3}{c}{Changeable} & & \multicolumn{3}{c}{Stable} & & \multicolumn{3}{c}{All} \\
			\cline{4-6} \cline{8-10} \cline{12-14}
			& & & $h = 1$ & $1 \sim 6$ & $1 \sim 12$ & & $h = 1$ & $1 \sim 6$ & $1 \sim 12$ & & $h = 1$ & $1 \sim 6$ & $1 \sim 12$ \\
			\midrule
			\multirow{5}[1]{*}{LS$^a$} & OLS & & -0.506 & -1.174 & -1.234 & & -0.322 & -0.367 & -0.531 & & -0.559 & -1.017 & -1.242 \\
			& WLSv & & -0.655 & -1.415 & -1.644 & & -0.437 & -1.092 & -1.422 & & -0.744 & -1.530 & -1.987 \\
			& WLSs & & -0.595 & -1.312 & -1.513 & & -0.442 & -0.954 & -1.250 & & \textbf{-0.751} & -1.448 & -1.861 \\
			& Sample & & -0.166 & -0.119 & -0.105 & & 0.910 & -0.112 & -0.520 & & 0.374 & -0.297 & -0.719 \\
			& Shrink & & \textbf{-0.802} & -1.553 & -1.882 & & -0.357 & -1.380 & -1.885 & & -0.743 & \textbf{-1.750} & -2.369 \\
            \midrule
			\multirow{5}[1]{*}{LAD} & OLS & & -0.604 & \textbf{-1.653} & \textbf{-2.170} & & \textbf{-0.564} & \textbf{-1.399} & \textbf{-2.057} & & -0.748 & -1.692 & \textbf{-2.407} \\
			& WLSv & & 0.225 & -0.157 & -0.162 & & -0.035 & -0.645 & -0.828 & & -0.136 & -0.527 & -0.794 \\
			& WLSs & & 0.225 & -0.157 & -0.162 & & -0.036 & -0.645 & -0.828 & & -0.136 & -0.527 & -0.794 \\
			& Sample & & 28.720 & 40.840 & 46.890 & & 30.771 & 48.702 & 56.931 & & 39.310 & 60.244 & 71.473 \\
			& Shrink & & 8.051 & 13.182 & 14.276 & & 8.252 & 14.870 & 17.925 & & 11.668 & 20.859 & 24.447 \\
            \midrule
			\multirow{5}[1]{*}{Huber} & OLS & & -0.506 & -1.198 & -1.351 & & -0.322 & -0.388 & -0.625 & & -0.559 & -1.037 & -1.363 \\
			& WLSv & & -0.655 & -1.393 & -1.690 & & -0.437 & -1.131 & -1.512 & & -0.744 & -1.537 & -2.090 \\
			& WLSs & & -0.595 & -1.298 & -1.559 & & -0.442 & -1.006 & -1.382 & & \textbf{-0.751} & -1.467 & -1.996 \\
			& Sample & & 0.770 & 13.740 & 19.993 & & 1.722 & 15.346 & 22.612 & & 1.891 & 21.523 & 31.499 \\
			& Shrink & & -0.792 & 1.088 & 2.412 & & -0.358 & 0.161 & 1.517 & & -0.662 & 1.492 & 3.690 \\
			\midrule
			\multicolumn{2}{c}{Bottom-up} & & 0.226 & -0.156 & -0.161 & & -0.035 & -0.645 & -0.828 & & -0.136 & -0.527 & -0.793 \\
			\midrule
			\multicolumn{2}{c}{Base} & & 0.714 & 1.156 & 1.300 & & 0.678 & 1.027 & 1.146 & & 0.880 & 1.344 & 1.492 \\
			\bottomrule
			\multicolumn{14}{l}{$^a$ indicates the MinT reconciliation.}
		\end{tabular}
	\end{center}
\end{table}

\begin{table}[htbp]
	\caption {\rm Out-of-sample forecasting performances of RoME with the base forecasts obtained by ARIMA under the standard $t$ distribution in Section~\ref{Section_NonGaussian}. The rows, columns and sub-columns denote the same as Table~\ref{Table_NonGaussian-mixture}. The best results are highlighted in bold.}
	\label{Table_NonGaussian-t}
	\setlength\tabcolsep{4.5pt}
	\renewcommand{\arraystretch}{1.2}
	\begin{center}
		\footnotesize
		\begin{tabular}{cccccccccccccc}
			\toprule
			\multirow{2}{*}{$\rho$} & \multirow{2}{*}{$\bm{W}_h$} & & \multicolumn{3}{c}{Changeable} & & \multicolumn{3}{c}{Stable} & & \multicolumn{3}{c}{All} \\
			\cline{4-6} \cline{8-10} \cline{12-14}
			& & & $h = 1$ & $1 \sim 6$ & $1 \sim 12$ & & $h = 1$ & $1 \sim 6$ & $1 \sim 12$ & & $h = 1$ & $1 \sim 6$ & $1 \sim 12$ \\
			\midrule
			\multirow{5}[1]{*}{LS} & OLS & & -0.534 & -0.954 & -1.029 & & -0.381 & -0.725 & -0.924 & & -0.568 & -1.060 & -1.306 \\
			& WLSv & & -0.935 & -1.257 & -1.297 & & \textbf{-0.684} & -1.113 & -1.474 & & \textbf{-0.840} & -1.412 & -1.769 \\
			& WLSs & & -0.945 & -1.212 & -1.245 & & -0.635 & -1.094 & -1.419 & & -0.819 & -1.397 & -1.728 \\
			& Sample & & 0.799 & 0.648 & 0.433 & & 0.706 & 0.253 & 0.079 & & 0.644 & 0.324 & 0.163 \\
			& Shrink & & -0.821 & -1.433 & -1.660 & & -0.603 & -1.294 & -1.740 & & -0.706 & -1.550 & -2.020 \\
            \midrule
			\multirow{5}[1]{*}{LAD} & OLS & & \textbf{-1.062} & \textbf{-1.941} & \textbf{-2.368} & & -0.592 & \textbf{-1.531} & \textbf{-2.207} & & -0.825 & \textbf{-1.796} & \textbf{-2.367} \\
			& WLSv & & -0.831 & -0.135 & 0.218 & & -0.593 & -0.493 & -0.743 & & -0.507 & -0.347 & -0.417 \\
			& WLSs & & -0.835 & -0.039 & 0.382 & & -0.593 & -0.499 & -0.745 & & -0.546 & -0.300 & -0.305 \\
			& Sample & & 33.123 & 51.402 & 57.030 & & 28.182 & 51.841 & 60.816 & & 38.247 & 67.975 & 77.907 \\
			& Shrink & & 6.115 & 12.435 & 14.043 & & 6.433 & 14.147 & 17.114 & & 9.323 & 18.769 & 22.299 \\
            \midrule
			\multirow{5}[1]{*}{Huber} &  OLS & & -0.534 & -1.005 & -1.116 & & -0.381 & -0.751 & -0.998 & & -0.568 & -1.109 & -1.421 \\
			& WLSv & & -0.935 & -1.372 & -1.429 & & \textbf{-0.684} & -1.122 & -1.474 & & \textbf{-0.840} & -1.494 & -1.846 \\
			& WLSs & & -0.945 & -1.291 & -1.318 & & -0.635 & -1.130 & -1.457 & & -0.819 & -1.483 & -1.803 \\
			& Sample & & 1.465 & 16.838 & 23.837 & & 1.478 & 16.373 & 24.007 & & 1.801 & 23.776 & 33.475 \\
			& Shrink & & -0.865 & 0.585 & 2.219 & & -0.637 & 0.580 & 2.228 & & -0.768 & 1.310 & 3.953 \\
			\midrule
			\multicolumn{2}{c}{Bottom-up} & & -0.835 & -0.041 & 0.381 & & -0.592 & -0.493 & -0.743 & & -0.545 & -0.295 & -0.303 \\
			\midrule
			\multicolumn{2}{c}{Base} & & 0.821 & 1.400 & 1.611 & & 0.670 & 1.015 & 1.131 & & 0.929 & 1.451 & 1.632 \\
			\bottomrule
		\end{tabular}
	\end{center}
\end{table}

\begin{table}[htbp]
	\caption {\rm Out-of-sample forecasting performances of RoME with the base forecasts obtained by ARIMA under the standard Cauchy distribution in Section~\ref{Section_NonGaussian}. The columns and sub-columns denote the same as Table~\ref{Table_NonGaussian-mixture}. The best results are highlighted in bold.}
	\label{Table_NonGaussian-cauchy}
	\setlength\tabcolsep{3.5pt}
	\renewcommand{\arraystretch}{1.2}
	\begin{center}
		\footnotesize
		\begin{tabular}{cccccccccccccc}
			\toprule
			\multirow{2}{*}{$\rho$} & \multirow{2}{*}{$\bm{W}_h$} & & \multicolumn{3}{c}{Changeable} & & \multicolumn{3}{c}{Stable} & & \multicolumn{3}{c}{All} \\
			\cline{4-6} \cline{8-10} \cline{12-14}
			& & & $h = 1$ & $1 \sim 6$ & $1 \sim 12$ & & $h = 1$ & $1 \sim 6$ & $1 \sim 12$ & & $h = 1$ & $1 \sim 6$ & $1 \sim 12$ \\
			\midrule
			\multirow{5}[1]{*}{LS} & OLS & & -1.596 & -1.095 & -0.808 & & 6.048 & 15.100 & 19.003 & & -1.356 & -0.948 & -0.695 \\
			& WLSv & & -2.591 & -1.662 & -1.156 & & -0.581 & -0.831 & -1.109 & & -2.379 & -1.676 & -1.189 \\
			& WLSs & & -2.052 & -1.562 & -1.173 & & 3.769 & 7.899 & 10.144 & & -1.873 & -1.524 & -1.144 \\
			& Sample & & 138.047 & 112.120 & 90.017 & & 2.217 & 3.299 & 3.600 & & 134.588 & 111.580 & 89.854 \\
			& Shrink & & -1.878 & -1.122 & -0.799 & & -0.588 & -0.850 & -1.117 & & -1.701 & -1.139 & -0.830 \\
            \midrule
			\multirow{5}[1]{*}{LAD} & OLS & & \textbf{-3.106} & \textbf{-2.238} & \textbf{-1.593} & & \textbf{-0.654} & -0.821 & \textbf{-1.242} & & -2.621 & \textbf{-1.942} & \textbf{-1.392} \\
			& WLSv & & -2.131 & -0.956 & -0.606 & & -0.496 & -0.402 & -0.657 & & -2.010 & -1.086 & -0.753 \\
			& WLSs & & -0.483 & -0.836 & -0.764 & & -0.510 & -0.411 & -0.659 & & 0.078 & -0.619 & -0.618 \\
			& Sample & & --$^a$ & -- & -- & & 46.860 & 78.545 & 90.950 & & -- & -- & -- \\
			& Shrink & & -- & -- & -- & & 2.648 & 5.276 & 5.951 & & -- & -- & -- \\
            \midrule
			\multirow{5}[1]{*}{Huber} & OLS & & -1.567 & -1.282 & -0.980 & & 5.887 & 10.733 & 12.309 & & -1.334 & -1.131 & -0.863 \\
			& WLSv & & -3.052 & -1.563 & -1.046 & & -0.578 & -0.834 & -1.118 & & \textbf{-2.795} & -1.583 & -1.096 \\
			& WLSs & & -1.724 & -1.613 & -1.351 & & 2.989 & 4.841 & 5.219 & & -1.476 & -1.584 & -1.343 \\
			& Sample & & 383.842 & 554.349 & 537.220 & & 6.295 & 10.265 & 14.137 & & 376.151 & 553.403 & 537.482 \\
			& Shrink & & -0.265 & -0.368 & -0.526 & & -0.530 & \textbf{-0.852} & -1.091 & & -0.019 & -0.344 & -0.544 \\
			\midrule
			\multicolumn{2}{c}{Bottom-up} & & -0.480 & -0.835 & -0.764 & & -0.481 & -0.380 & -0.643 & &     0.080 & -0.619 & -0.618 \\
			\midrule
			\multicolumn{2}{c}{Base} & & 9.564 & 28.426 & 46.187 & & 0.667 & 0.998 & 1.123 & & 6.635 & 19.393 & 31.439 \\
			\bottomrule
			\multicolumn{14}{l}{$^a$ indicates the value larger than 1000.}
		\end{tabular}
	\end{center}
\end{table}

As shown in Tables~\ref{Table_NonGaussian-mixture}--\ref{Table_NonGaussian-cauchy}, LAD-OLS behaves well in three groups of series, realizing the maximum decreases of reconciled forecast errors, especially for mid- and long-term forecasting.
On the contrary, the results for MinT with LS loss function, though not the best, are generally comparable.
For short-term forecasting, the improvement of reconciliation procedure, including MinT, from the base forecasts is relatively small; it tends to increase with the ahead-forecasting step.
This result implies that, in a short ahead-forecasting period, the information on the hierarchical structure of HTS may contribute less than that on historical data of every series to the final reconciled result.
Due to the complicate covariance structure of non-normal distributions, using the full-sample covariance matrix (i.e., Sample covariance design) leads to a large bias in most cases, and the shrinkage process may greatly reduce the bias.

An interesting observation is that when introducing a robust loss function, we may not need to put a complicate covariance design of the covariance matrix of the out-of-sample forecasting errors, and employing the simple OLS one would achieve reasonable (or even the best) reconciled results.

\subsection{Efficiency loss for normal cases} \label{Section_EfficiencyLoss}

\begin{figure}[ht]
	\begin{center}
		\subfigure[For the variation of normality.]{
			\resizebox{7.2cm}{5.4cm}{\includegraphics{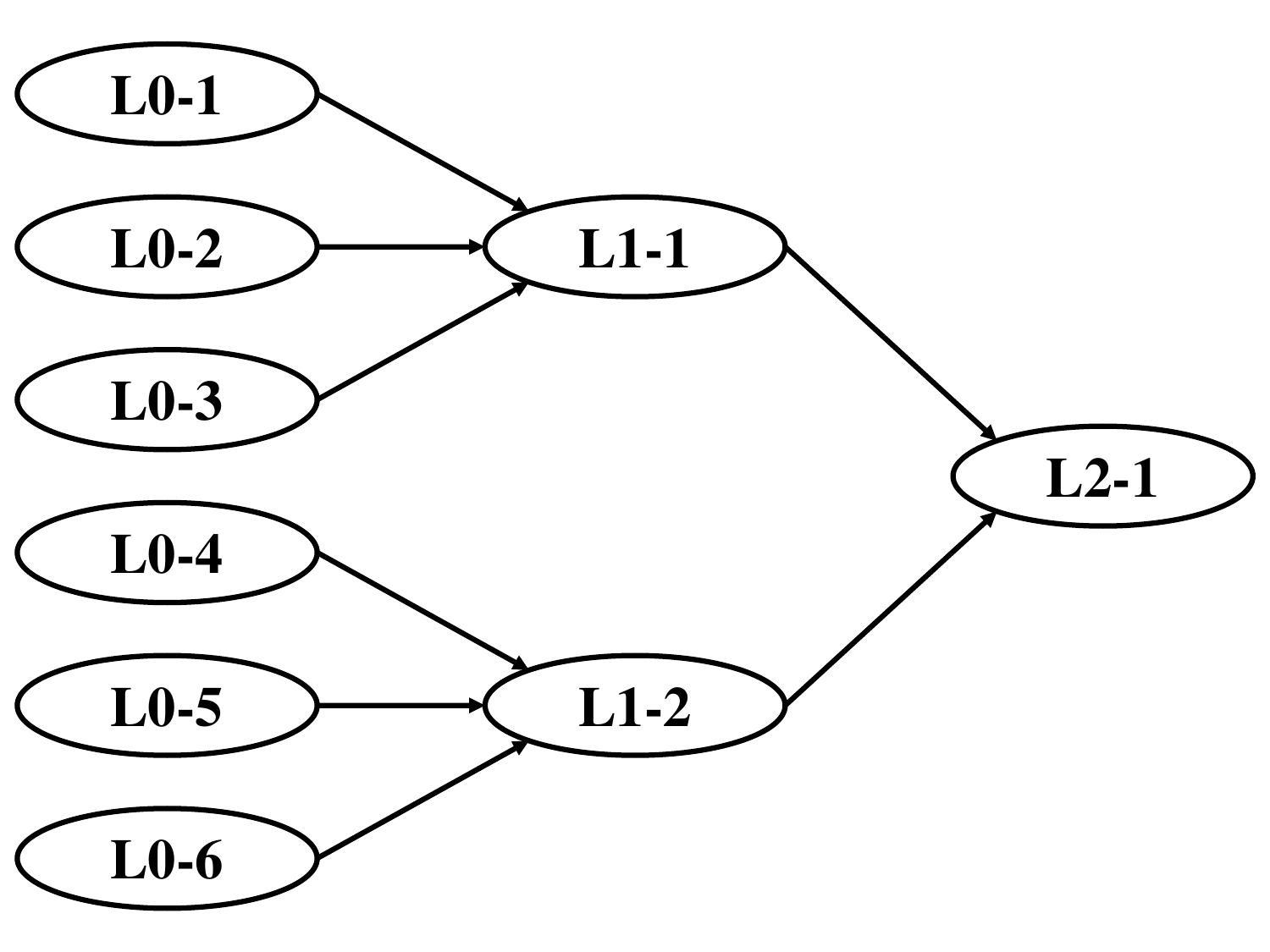}}
		}
		\qquad
		\subfigure[For the scale of hierarchy.]{
			\resizebox{7.2cm}{5.4cm}{\includegraphics{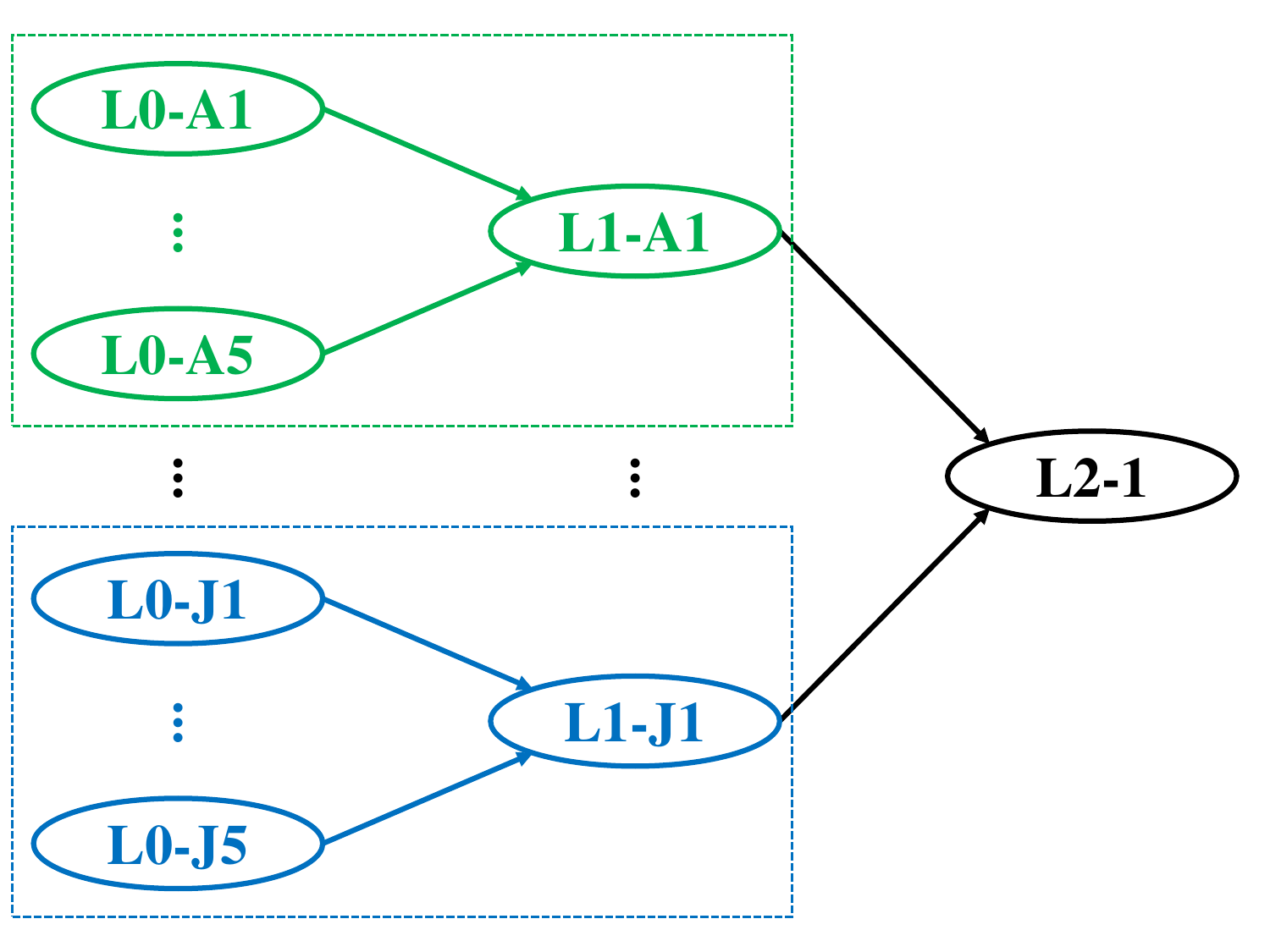}}
		}
	\end{center}
	\caption{\rm Hierarchical trees for the numerical experiments in Section~\ref{Section_EfficiencyLoss}. The summing matrix corresponding to the hierarchy of (a) in the left panel is $\bm{S} = (\mathbf{1}_6^{\top}; \text{diag} (\mathbf{1}_3^{\top}, \mathbf{1}_3^{\top}); \mathbf{I}_6)$, indicating the 1, 2 and 6 series (``L2-1"; ``L1-1" and ``L1-2"; and ``L0-$i$" for $i = 1, 2, \cdots, 6$) at the top, middle and bottom levels, respectively. For (b) in the right panel, the family of the two-level hierarchies follow the aggregation rules that series (e.g., ``L1-A1") at the middle level are aggregated in order by every 5 series (e.g., ``L0-A$i$", $i = 1, 2, \cdots, 5$) at the bottom level.}
	\label{Figure_EfficiencyLoss}
\end{figure}

Then, we evaluate the possible efficiency loss of RoME in normal cases for two issues, i.e., (a) the variation of normality, and (b) the scale of hierarchy. Figure~\ref{Figure_EfficiencyLoss} introduces the hierarchical structures used for both cases.

Specifically, in Model (\ref{Equation_Generation}), all series at the bottom level are assumed to be standard normal, where the correlation coefficients between any two of them remain $\varrho^{|i - i^\prime|}$ with $\varrho = 0.4$.
(a) For the variation of normality, $n_b = 6$, $\alpha_i = 0.6$, and $\sigma_i$ varies between 0.5 and 3; and (b) for the scale of hierarchy, $n_b$ increases from 10 to 50, $\alpha_i = 0.8$ and $\sigma_i = 1$.
Tables \ref{Table_EfficiencyLoss-sigma} and \ref{Table_EfficiencyLoss-n} report the results for the reconciled forecasts for two issues, where the base forecasts are obtained by ARIMA.

As shown in Tables~\ref{Table_EfficiencyLoss-sigma} and \ref{Table_EfficiencyLoss-n}, MinT is more suitable for short-term forecasting (i.e., $h = 1$), especially that with Shrink covariance design.
When all series in the hierarchy are Gaussian, the assumption of MinT with LS loss function coincides with the reality of HTS; therefore, the reconciled forecasts by MinT for a short ahead-forecasting step would be better.
For the mid- and long-term ones, LAD-OLS behaves remarkably.
Moreover, the reconciled forecasts by RoME with Huber's loss function are similar to those by MinT.
Specifically, (a) for the variation of normality, the improvement of the reconciliation procedure from the base forecasts is generally limited (below 1\%) in the normal case.
When the standard deviation of the error is small (i.e., $\sigma = 0.5$), indicating a group of nearly perfect base forecasts, MinT would behave better than RoME, even for mid- and long-term forecasting.
(b) For the scale of hierarchy, the robust reconciliation procedure achieves a significant improvement from the base forecasts, especially for relatively long-term forecasting.

\begin{table}[htbp]
	\caption {\rm Percentage increases of average RMSE for the whole HTS between RoME and the base forecasts obtained by ARIMA under different variations of normality in Section~\ref{Section_EfficiencyLoss}. The columns denote the same as the sub-columns of Table~\ref{Table_NonGaussian-mixture}, and the sub-columns ``LS", ``LAD" and ``Huber" correspond to the specific loss functions. The best results are highlighted in bold.}
	\label{Table_EfficiencyLoss-sigma}
	\setlength\tabcolsep{4pt}
	\renewcommand{\arraystretch}{1.1}
	\begin{center}
		\footnotesize
		\begin{tabular}{cccccccccccccc}
			\toprule
			\multirow{2}{*}{Variation} & \multirow{2}{*}{$\bm{W}_h$} & & \multicolumn{3}{c}{$h = 1$} & & \multicolumn{3}{c}{$1 \sim 6$} & & \multicolumn{3}{c}{$1 \sim 12$} \\
			\cline{4-6} \cline{8-10} \cline{12-14}
			& & & LS & LAD & Huber & & LS & LAD & Huber & & LS & LAD & Huber \\
			\midrule
			\multirow{5}[1]{*}{$\sigma = 0.5$} & OLS & & -0.410 & \textbf{-0.484} & -0.410 & & -0.372 & \textbf{-0.497} & -0.372 & & -0.339 & -0.424 & -0.339 \\
			& WLSv & & -0.450 & -0.198 & -0.450 & & -0.482 & -0.171 & -0.480 & & \textbf{-0.448} & -0.194 & -0.446 \\
			& WLSs & & -0.467 & -0.199 & -0.467 & & -0.482 & -0.178 & -0.479 & & -0.440 & -0.200 & -0.436 \\
			& Sample & & 0.652 & 34.590 & 1.034 & & 0.462 & 35.119 & 3.315 & & 0.275 & 32.126 & 3.882 \\
			& Shrink & & -0.204 & 10.745 & -0.140 & & -0.411 & 12.002 & 0.151 & & -0.419 & 11.119 &  0.373 \\
            \midrule
			\multirow{5}[1]{*}{$\sigma = 1$} & OLS & & -0.416 & -0.671 & -0.416 & & -0.369 & \textbf{-0.731} & -0.392 & & -0.362 & \textbf{-0.721} & -0.391 \\
			& WLSv & & -0.781 & \textbf{-0.915} & -0.785 & & -0.665 & -0.675 & -0.683 & & -0.650 & -0.622 & -0.656 \\
			& WLSs & & -0.680 & \textbf{-0.915} & -0.681 & & -0.592 & -0.674 & -0.621 & & -0.582 & -0.623 & -0.599 \\
			& Sample & & 0.207 & 30.267 & 0.330 & & 0.406 & 32.585 & 2.593 & & 0.330 & 30.943 & 2.599 \\
			& Shrink & & -0.638 & 7.525 & -0.685 & & -0.464 & 9.817 & 0.007 & & -0.485 & 9.797 & 0.179 \\
            \midrule
			\multirow{5}[1]{*}{$\sigma = 1.5$} & OLS & & -0.402 & -0.588 & -0.402 & & -0.353 & \textbf{-0.678} & -0.363 & & -0.330 & \textbf{-0.713} & -0.347 \\
			& WLSv & & -0.777 & \textbf{-0.963} & -0.777 & & -0.609 & -0.446 & -0.629 & & -0.591 & -0.474 & -0.614 \\
			& WLSs & & -0.670 & \textbf{-0.963} & -0.670 & & -0.554 & -0.444 & -0.577 & & -0.528 & -0.477 & -0.555 \\
			& Sample & & 0.350 & 28.073 & 1.024 & & 0.499 & 33.037 & 2.239 & & 0.281 & 31.189 & 2.300 \\
			& Shrink & & -0.656 & 7.437 & -0.677 & & -0.481 & 10.261 & -0.454 & & -0.506 & 9.824 & -0.386 \\
            \midrule
			\multirow{5}[1]{*}{$\sigma = 2$} & OLS & & -0.413 & \textbf{-0.699} & -0.413 & & -0.329 & \textbf{-0.628} & -0.329 & & -0.303 & \textbf{-0.543} & -0.303 \\
			& WLSv & & -0.662 & -0.630 & -0.662 & & -0.604 & -0.441 & -0.607 & & -0.485 & -0.285 & -0.489 \\
			& WLSs & & -0.620 & -0.631 & -0.620 & & -0.575 & -0.437 & -0.578 & & -0.465 & -0.283 & -0.469 \\
			& Sample & & 0.459 & 31.272 & 0.686 & & 0.328 & 32.931 & 2.388 & & 0.380 & 31.034 & 2.760 \\
			& Shrink & & -0.465 & 8.569 & -0.465 & & -0.468 & 10.080 & -0.344 & & -0.398 & 10.175 & -0.240 \\
            \midrule
			\multirow{5}[1]{*}{$\sigma = 3$} & OLS & & -0.371 & -0.578 & -0.371 & & -0.351 & \textbf{-0.740} & -0.352 & & -0.343 & \textbf{-0.709} & -0.347 \\
			& WLSv & & -0.721 & \textbf{-0.885} & -0.721 & & -0.619 & -0.516 & -0.631 & & -0.624 & -0.564 & -0.640 \\
			& WLSs & & -0.627 & \textbf{-0.885} & -0.627 & & -0.570 & -0.510 & -0.583 & & -0.567 & -0.561 & -0.585 \\
			& Sample & & 0.589 & 32.033 & 0.968 & & 0.613 & 32.393 & 2.297 & & 0.444 & 30.244 & 2.646 \\
			& Shrink & & -0.628 & 8.080 & -0.637 & & -0.474 & 9.868 & -0.453 & & -0.474 & 9.659 & -0.251 \\
			\bottomrule
		\end{tabular}
	\end{center}
\end{table}

\begin{table}[htbp]
	\caption {\rm Percentage increases of average RMSE for the whole HTS between RoME and the base forecasts obtained by ARIMA under different scales of hierarchy in Section~\ref{Section_EfficiencyLoss}. The rows, columns and sub-columns denote the same as Table~\ref{Table_EfficiencyLoss-sigma}. The best results are highlighted in bold.}
	\label{Table_EfficiencyLoss-n}
	\setlength\tabcolsep{4pt}
	\renewcommand{\arraystretch}{1.2}
	\begin{center}
		\footnotesize
		\begin{tabular}{cccccccccccccc}
			\toprule
			\multirow{2}{*}{Scale} & \multirow{2}{*}{$\bm{W}_h$} & & \multicolumn{3}{c}{$h = 1$} & & \multicolumn{3}{c}{$1 \sim 6$} & & \multicolumn{3}{c}{$1 \sim 12$} \\
			\cline{4-6} \cline{8-10} \cline{12-14}
			& & & LS & LAD & Huber & & LS & LAD & Huber & & LS & LAD & Huber \\
			\midrule
			\multirow{5}[1]{*}{$n = 10$} & OLS & & -0.582 & -0.966 & -0.582 & & -1.114 & -2.000 & -1.234 & & -1.344 &  -2.463 & -1.640 \\
			& WLSv & & \textbf{-1.100} & -0.981 & -1.097 & & -2.065 & -1.468 & -2.123 & & -2.374 & -1.494 & -2.479 \\
			& WLSs & & -1.019 & -0.981 & -1.018 & & -1.925 & -1.468 & -2.032 & & -2.216 & -1.494 & -2.426 \\
			& Sample & & 0.369 & 49.243 & 2.764 & & -0.341 & 86.897 & 32.354 & & -0.450 & 103.726 &  47.465 \\
			& Shrink & & -0.898 & 14.991 & -0.779 & & \textbf{-2.118} & 32.607 & 3.955 & & \textbf{-2.579} & 40.702 &   9.299 \\
            \midrule
			\multirow{5}[1]{*}{$n = 20$} & OLS & & -0.415 & -0.737 & -0.415 & & -0.856 & \textbf{-1.968} & -0.926 & & -1.016 & \textbf{-2.643} & -1.247 \\
			& WLSv & & -0.878 & -0.598 & -0.888 & & -1.864 & -1.254 & -1.909 & & -2.250 & -1.384 & -2.370 \\
			& WLSs & & -0.847 & -0.598 & -0.861 & & -1.778 & -1.254 & -1.878 & & -2.157 & -1.384 & -2.394 \\
			& Sample & & 0.477 & 67.725 & 7.339 & & 0.685 & 117.718 & 64.818 & & 0.839 & 137.135 & 90.184 \\
			& Shrink & & \textbf{-0.944} & 11.254 & -0.801 & & -1.951 & 24.939 & 3.652 & & -2.397 & 30.418 &  6.545 \\
            \midrule
			\multirow{5}[1]{*}{$n = 30$} & OLS & & -0.340 & -0.661 & -0.340 & & -0.670 & -1.800 & -0.705 & & -0.812 & \textbf{-2.411} & -0.905 \\
			& WLSv & & -0.890 & -0.812 & -0.894 & & -1.745 & -1.311 & -1.813 & & -2.071 & -1.413 & -2.219 \\
			& WLSs & & -0.816 & -0.812 & -0.831 & & -1.606 & -1.311 & -1.734 & & -1.923 & -1.414 &  -2.172 \\
			& Sample & & 0.800 & 80.027 & 15.139 & & 2.048 & 140.365 & 95.401 & & 2.462 & 163.702 & 124.544 \\
			& Shrink & & \textbf{-0.937} & 7.775 & -0.784 & & \textbf{-1.823} & 18.704 & 1.718 & & -2.195 & 23.346 &  4.094 \\
            \midrule
			\multirow{5}[1]{*}{$n = 40$} & OLS & & -0.293 & -0.704 & -0.293 & & -0.589 & \textbf{-1.827} & -0.601 & & -0.725 & \textbf{-2.536} & -0.793 \\
			& WLSv & & -0.834 & -0.667 & -0.835 & & -1.544 & -0.962 & -1.562 & & -2.001 & -1.158 &  -2.056 \\
			& WLSs & & -0.783 & -0.668 & -0.792 & & -1.474 & -0.962 & -1.516 & & -1.915 & -1.158 & -2.046 \\
			& Sample & & 1.161 & 105.806 & 23.133 & & 3.109 & 176.993 & 120.836 & & 3.881 & 202.798 & 155.168 \\
			& Shrink & & \textbf{-0.908} & 7.342 & -0.663 & & -1.667 & 16.802 & 1.786 & & -2.144 & 20.469 &  3.740 \\
            \midrule
			\multirow{5}[1]{*}{$n = 50$} & OLS & & -0.257 & -0.831 & -0.257 & & -0.521 & -1.660 & -0.535 & & -0.646 & \textbf{-2.282} & -0.703 \\
			& WLSv & & -1.266 & -1.233 & -1.290 & & -1.845 & -1.503 & \textbf{-1.886} & & -2.189 & -1.583 & -2.220 \\
			& WLSs & & -1.138 & -1.233 & -1.168 & & -1.683 & -1.503 & -1.753 & & -2.014 & -1.583 &  -2.105 \\
			& Sample & & 2.230 & 114.385 & 27.136 & & 4.770 & 198.350 & 144.367 & & 5.573 & 231.640 & 188.251 \\
			& Shrink & & -1.227 & 3.826 & \textbf{-1.308} & & -1.868 & 11.161 & 0.274 & & -2.267 & 15.331 &  1.975 \\
			\bottomrule
		\end{tabular}
	\end{center}
\end{table}

We may conclude from the simulation results above that, compared with MinT, RoME would not lead to significant efficiency loss when all series in the hierarchy are normally distributed.
To some extent, the proposed procedure shows its robustness for not only the potential non-Gaussian series, but also the possible accumulation of forecast errors, though of normality, with the increase of the ahead-forecasting step. Since RoME can provide competitive (or even the best) forecast results, it is suggested to carry out the proposed robust reconciliation procedure for HTS forecasting when no outliers exist.

\subsection{Proportion of bad forecasts} \label{Section_Proportion}

\begin{figure}[ht]
	\centering
	\includegraphics[width = 12cm]{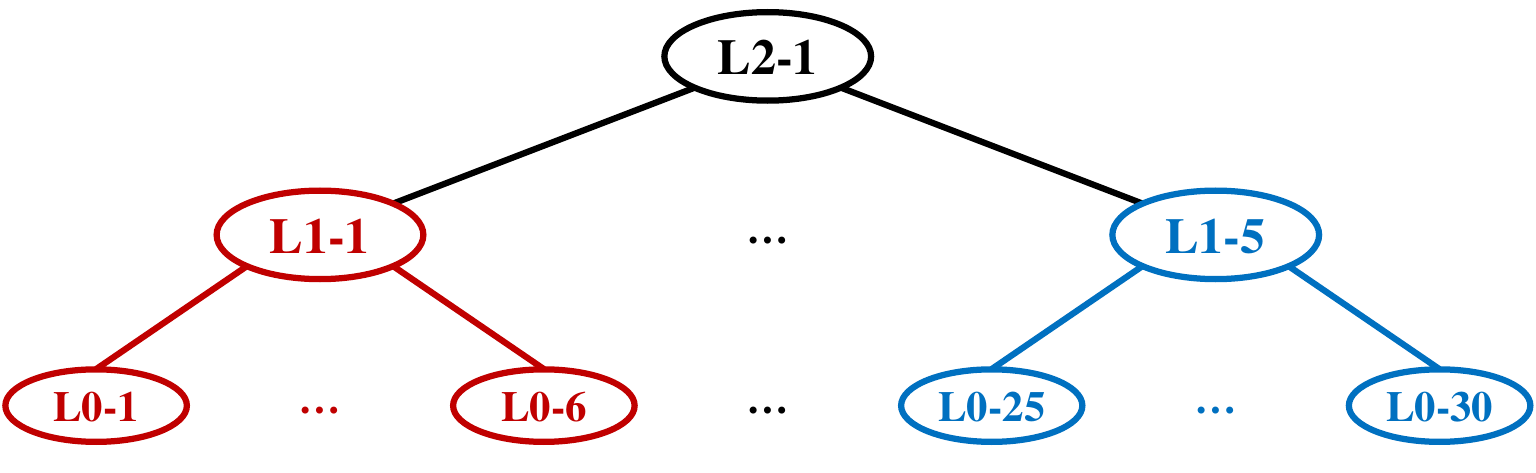} \\
	\caption{\rm Hierarchical tree for the numerical experiments in Section~\ref{Section_Proportion}. This hierarchy, following the similar aggregation rules of that in Figure~\ref{Figure_EfficiencyLoss}(b) that series at the middle level are aggregated in order by every 6 series at the bottom level, contains $m = 36$ series, with 30 series (``L0-$i$", $i = 1, 2, \cdots, 30$) at the bottom level, 5 series (``L1-$j$", $j = 1, 2, \cdots, 5$) in the middle level, and the most aggregated one (``L2-1") in the top level.}
	\label{Figure_Proportion}
\end{figure}

Next, we consider the influence on the reconciled results by RoME from different proportions of bad forecasts for HTS via a relatively larger hierarchy with $n_b = 30$, as illustrated in Figure~\ref{Figure_Proportion}, where the ``Irregular" group consists of the first several series at the bottom level in a proportion, and the others constitute the ``Normal" group.

\begin{figure}[htbp]
	\begin{center}
		\subfigure[Bottom level: $h = 1$, $1 \sim 6$ and $1 \sim 12$.]{
			\begin{minipage}[t]{15.6cm}
				\includegraphics[width = 5cm]{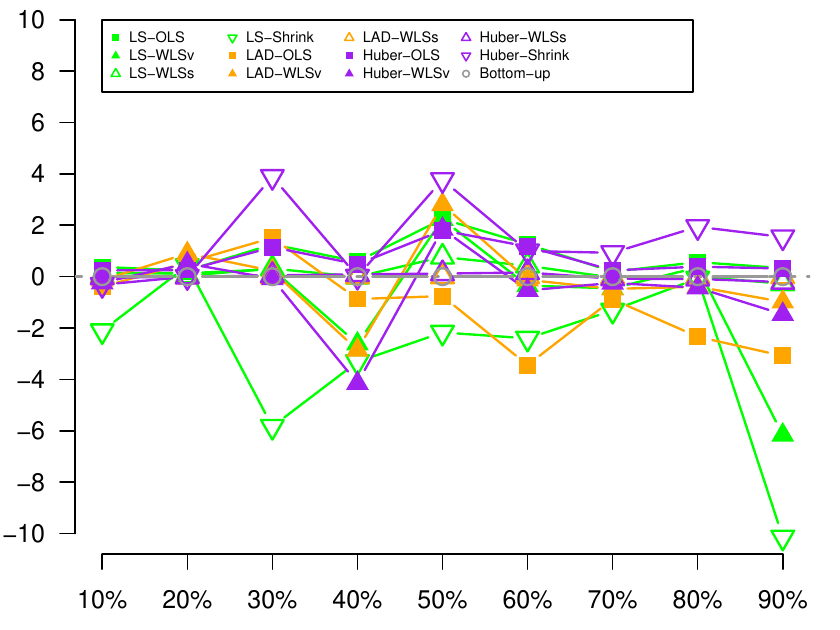}
				\
				\includegraphics[width = 5cm]{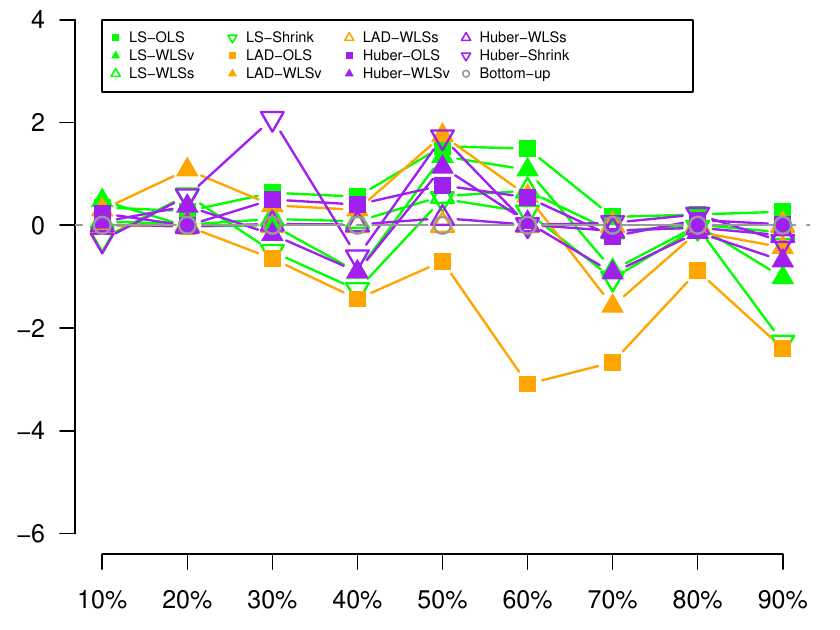}
				\
				\includegraphics[width = 5cm]{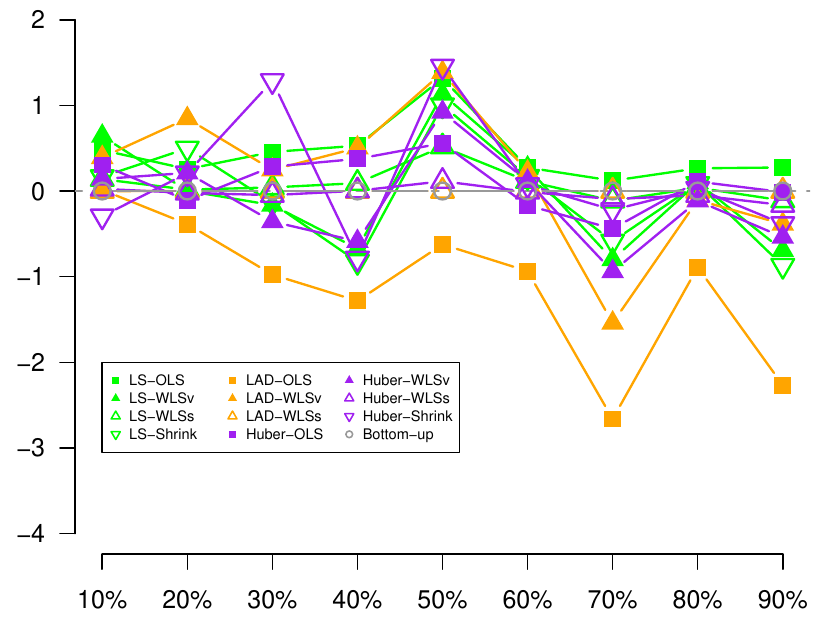}
			\end{minipage}
		}
	\end{center}
	\begin{center}
		\subfigure[Aggregated levels: $h = 1$, $1 \sim 6$ and $1 \sim 12$.]{
			\begin{minipage}[t]{15.6cm}
				\includegraphics[width = 5cm]{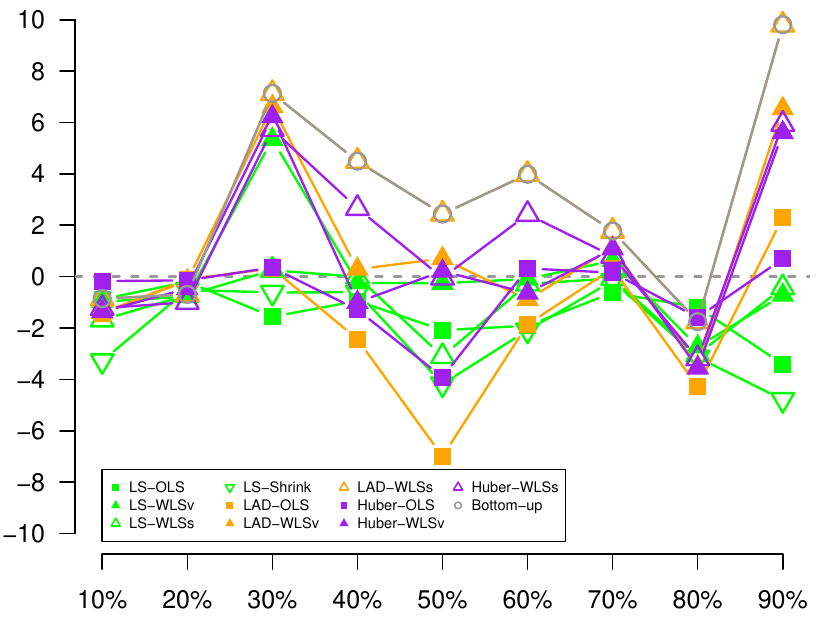}
				\
				\includegraphics[width = 5cm]{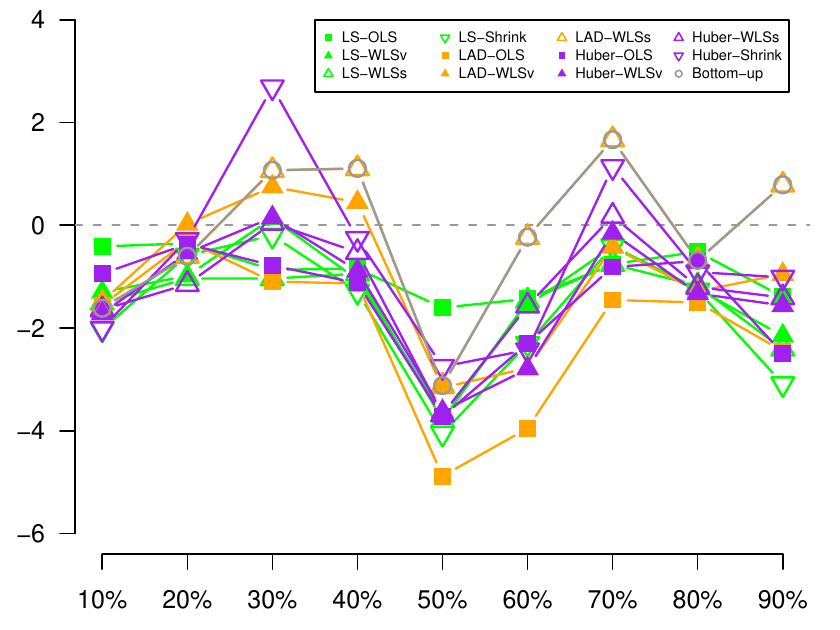}
				\
				\includegraphics[width = 5cm]{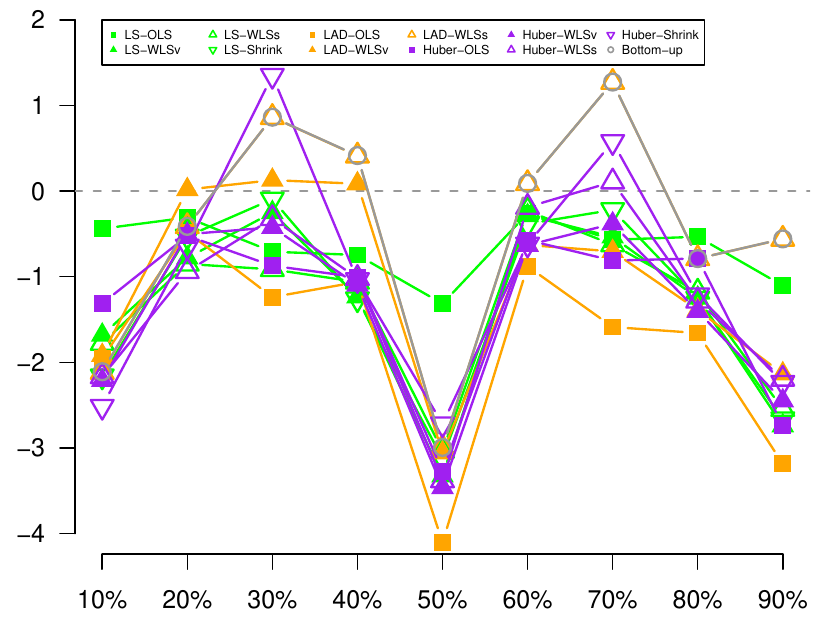}
			\end{minipage}
		}
	\end{center}
	\begin{center}
		\subfigure[Whole hierarchy: $h = 1$, $1 \sim 6$ and $1 \sim 12$.]{
			\begin{minipage}[t]{15.6cm}
				\includegraphics[width = 5cm]{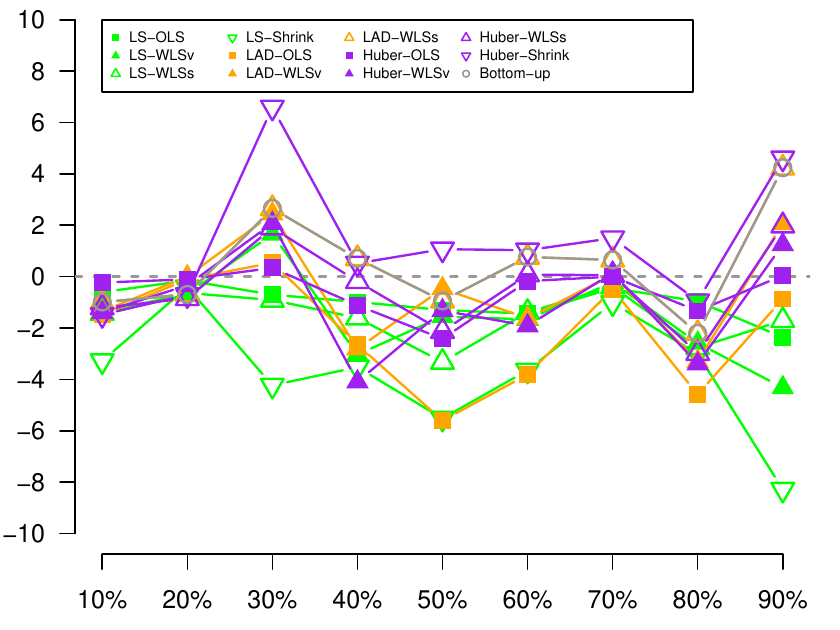}
				\
				\includegraphics[width = 5cm]{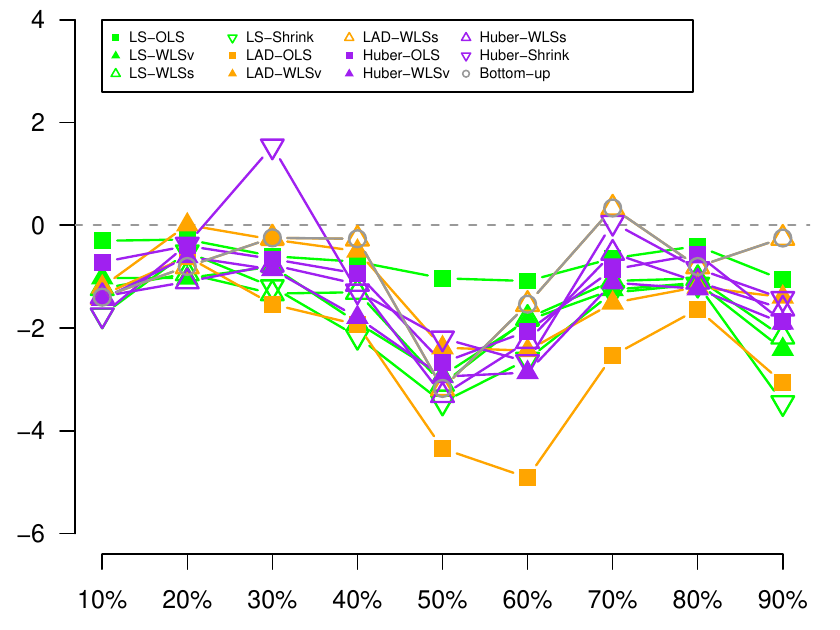}
				\
				\includegraphics[width = 5cm]{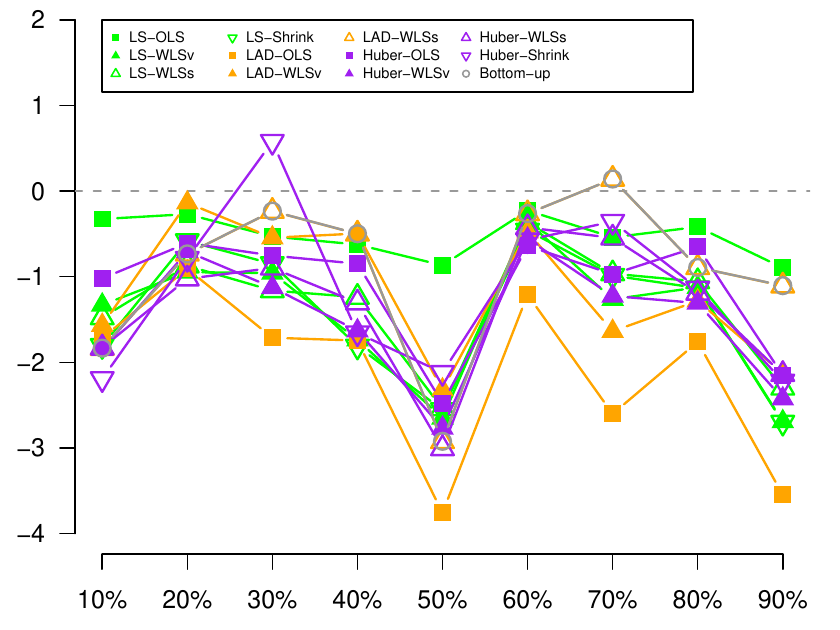}
			\end{minipage}
		}
	\end{center}
	\caption{\rm Line charts for the percentage increases by RoME from average RMSE of the base forecasts obtained by ARIMA against the proportion (in percentage) of bad forecasts in Section~\ref{Section_Proportion}. The panels in the sub-figures from left to right correspond to the out-of-sample forecasting periods of $H = 1$, 6 and 12, where the aggregated levels contain the middle and top levels. The horizontal axis in every single figure denotes the percentage of non-Gaussian series in HTS. The lines exceeding the upper limits of the vertical axis, i.e., 10\%, 4\% and 2\% for the left, middle and right panels, are not displayed for concentrating the valuable information on the others.}
	\label{Figure_Proportion-results}
\end{figure}

Specifically, in Model (\ref{Equation_Generation}), $\alpha_i$ is uniformly taken over $[0.6, 0.8]$, which extends the constant settings in Sections \ref{Section_NonGaussian} and \ref{Section_EfficiencyLoss}, and $\sigma_i = 0.5$.
The distributions of the errors for the ``Irregular" group are selected, with equal probabilities, from three types of non-normal distributions in Section~\ref{Section_NonGaussian}, and those for the ``Normal" group are standard normal and independent with each other.
We increase the proportion of non-Gaussian series at the bottom level from 10 to 90 percentages, i.e., for the first 3, ..., 27 ones.
Figure~\ref{Figure_Proportion-results} reports the train of the percentage increases of RMSE for alternative RoME models with increasing proportions of non-Gaussian series in HTS, where the base forecasts are obtained by ARIMA.

As shown in Figure~\ref{Figure_Proportion-results}, the lines of the average percentage increases of RMSE by OLS-LAD show a significant downward trend, indicating that the reconciled results are getting better, with the increase of the proportion of irregular series in HTS.
By contrast, the overall performances of the other alternative models exert few differences in the proportion of irregular series.
Compared with the benchmark, most of the reconciliation procedures improve the forecasting accuracy from the base forecasts, and LAD-OLS achieves the maximum percentage decrease, especially for the aggregation levels and the whole hierarchy.
However, those procedures using the full-sample covariance matrix (i.e., Sample covariance design) may behave extremely bad as most of their related lines exceed the upper limit (i.e., 10\%) and are not depicted in the figures.
Due to the diversified types of potential non-Gaussian series in HTS, it is difficult to identify the covariance structure of their forecast errors; therefore, using a complicate and insecure estimate in the reconciliation procedure may possibly lead to a worse result.

\subsection{Correlation among HTS} \label{Section_Correlation}

The complex covariance structure of non-normal distributions, such as the three considered in Section~\ref{Section_NonGaussian}, makes it difficult to characterize cross-series correlations. To isolate the effect of cross-series correlation, we therefore adopt the hierarchical structure illustrated in Figure~\ref{Figure_NonGaussian} while assuming that all series are Gaussian, excluding the ``Changeable" group.

\begin{figure}[htbp]
	\begin{center}
		\subfigure[Bottom level: $h = 1$, $1 \sim 6$ and $1 \sim 12$.]{
			\begin{minipage}[t]{15.6cm}
				\includegraphics[width = 5cm]{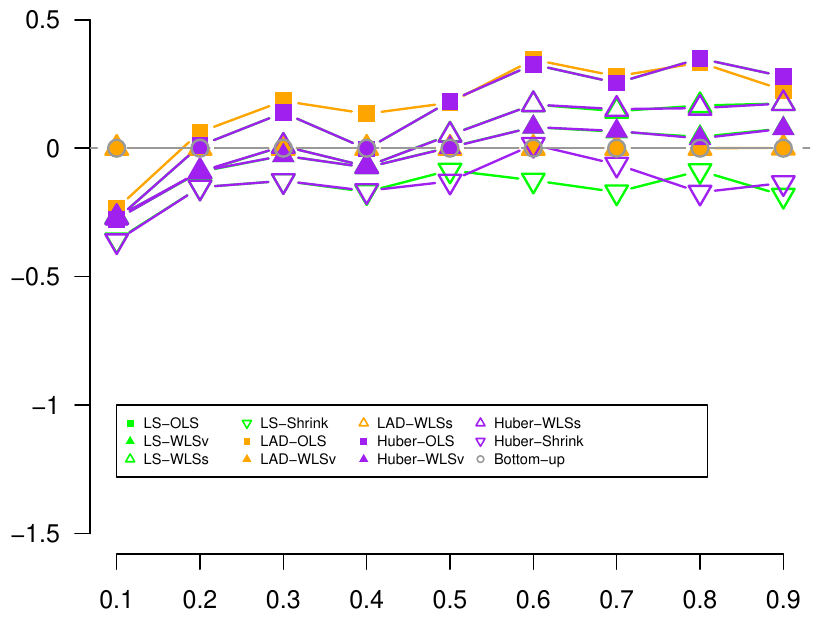}
				\
				\includegraphics[width = 5cm]{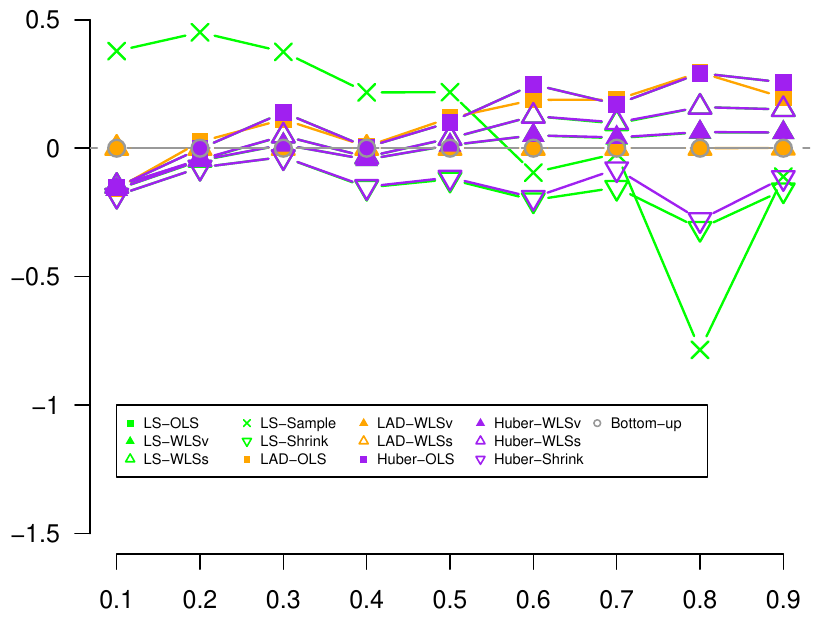}
				\
				\includegraphics[width = 5cm]{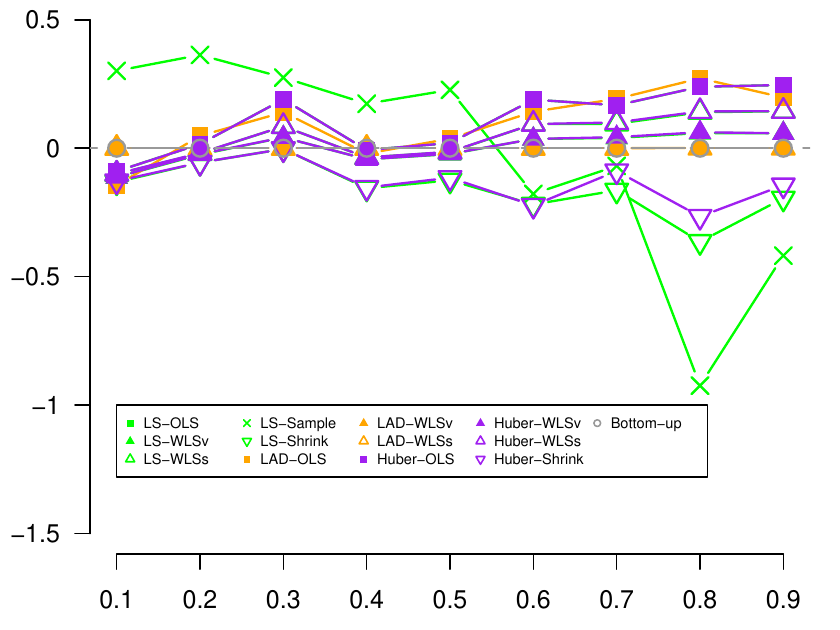}
			\end{minipage}
		}
	\end{center}
	\begin{center}
		\subfigure[Aggregated levels: $h = 1$, $1 \sim 6$ and $1 \sim 12$.]{
			\begin{minipage}[t]{15.6cm}
				\includegraphics[width = 5cm]{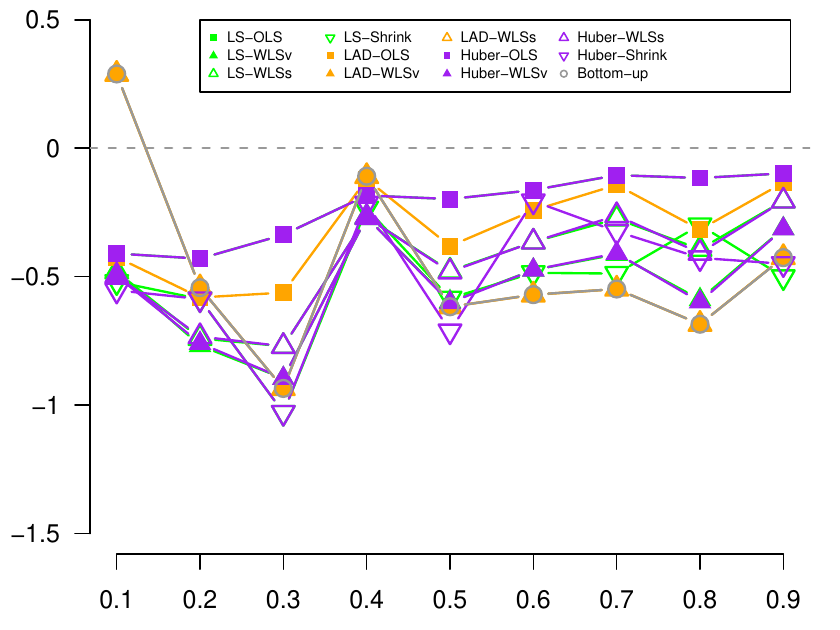}
				\
				\includegraphics[width = 5cm]{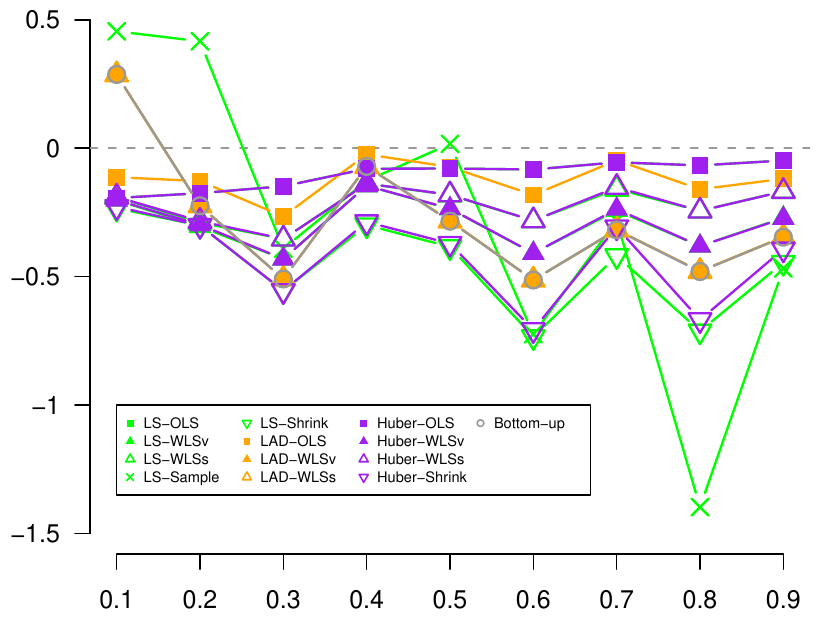}
				\
				\includegraphics[width = 5cm]{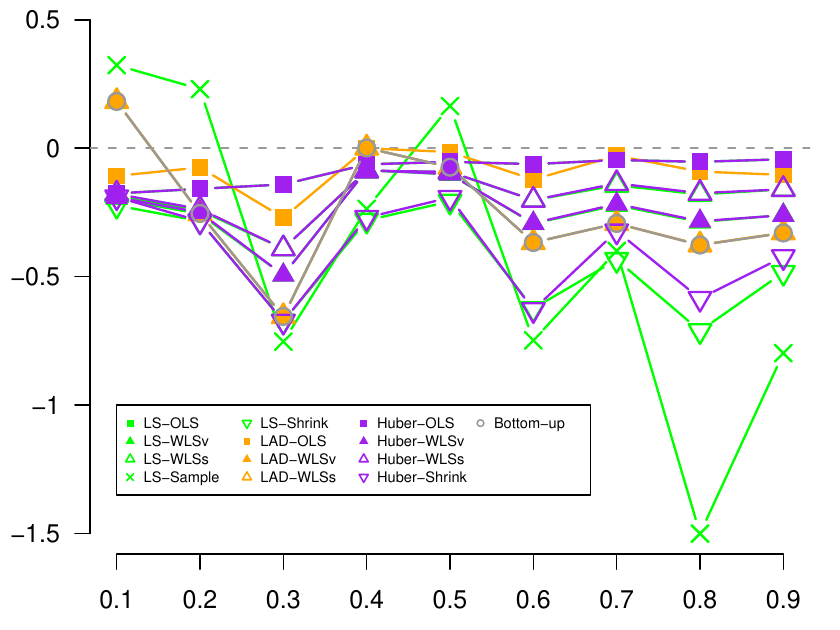}
			\end{minipage}
		}
	\end{center}
	\begin{center}
		\subfigure[Whole hierarchy: $h = 1$, $1 \sim 6$ and $1 \sim 12$.]{
			\begin{minipage}[t]{15.6cm}
				\includegraphics[width = 5cm]{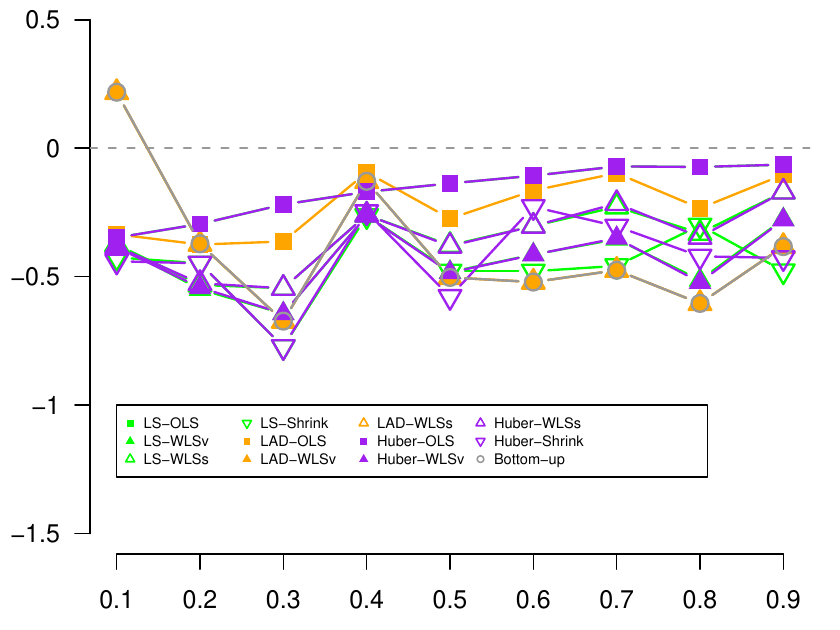}
				\
				\includegraphics[width = 5cm]{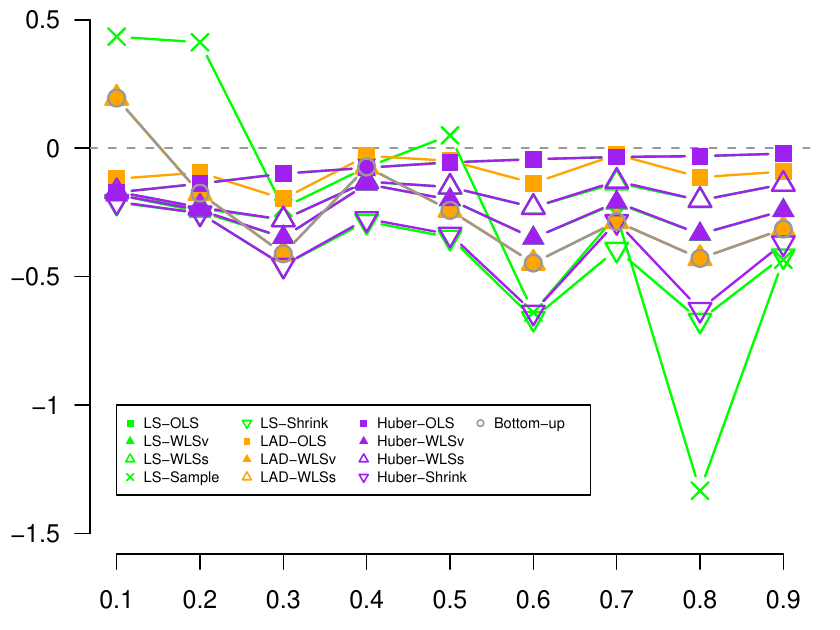}
				\
				\includegraphics[width = 5cm]{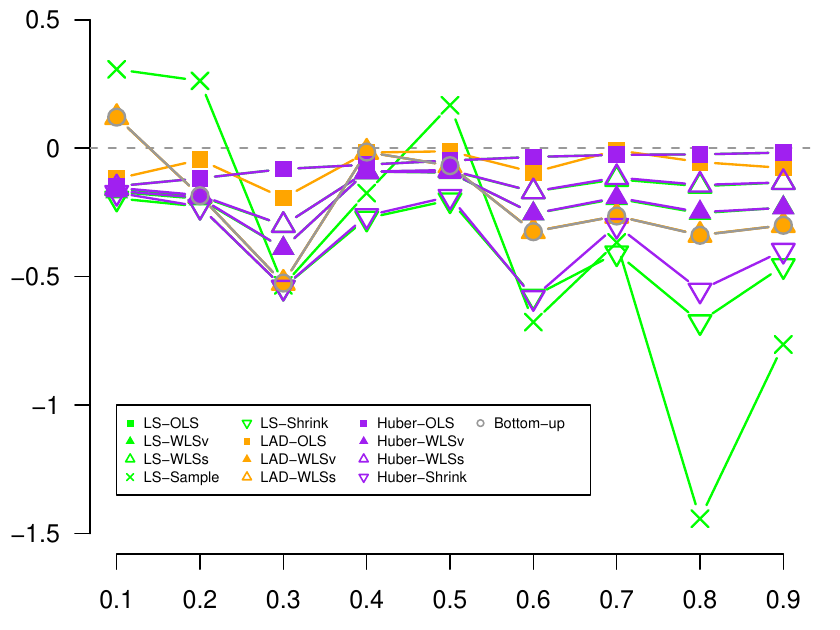}
			\end{minipage}
		}
	\end{center}
	\caption{\rm Line charts for the percentage increases by RoME from average RMSE of the base forecasts obtained by ETS against the correlation among HTS in Section~\ref{Section_Correlation}. The panels in the sub-figures, as well as the aggregated levels wherein, denote the same as Figure~\ref{Figure_Proportion-results}. The horizontal axis in every single figure denotes the correlation coefficient between any two most disaggregated series. The lines exceeding the upper limit of the vertical axis, i.e., 0.5\%, are not displayed for concentrating the valuable information on the others.}
	\label{Figure_Correlation-results-ETS}
\end{figure}

Specifically, in Model (\ref{Equation_Generation}), errors of all series at the bottom level are taken from standard normality, with $\sigma_i = 0.5$, and the correlation coefficient between any two of them varies from 0 to 0.9.
Figure~\ref{Figure_Correlation-results-ETS} reports the results for the reconciled forecasts under different correlations among HTS, where the base forecasts are obtained by ETS to mimic the misspecification of model.

As shown in the top panel of Figure~\ref{Figure_Correlation-results-ETS}, the performance curves of most competing methods increase with the level of correlation, indicating that stronger positive dependence among series tends to reduce the gains from reconciliation at the most disaggregated level relative to the benchmark. A similar pattern is observed for short-term forecasts, as illustrated in the left panel. In contrast, for mid- and long-term horizons, the effect of correlation is much less pronounced at aggregated levels and across the entire hierarchy, with the corresponding curves fluctuating around zero.
By contrast, the influence from the correlation is not significant for mid- and long-term forecasting in the aggregated levels or the whole hierarchy, as the related lines hover around the horizontal line at zero.

To further examine the impact of model misspecification, we also conduct simulations in which the base forecasts are generated using ARIMA models and the corresponding results are reported in Appendix~\ref{App_B3}.

The main conclusions from the correct ARIMA model for base forecasts, as shown in Figure~\ref{Figure_Correlation-results-ARIMA}, are consistent with those from the incorrect ETS model. A remarkable difference is the greater decrease of RMSE for the reconciled forecasts by ARIMA than that by ETS from the benchmark.
Nevertheless, the improvement of reconciliation procedure in this case in generally limited, which is caused by the fact that the base forecasts for a Gaussian HTS can be good enough, therefore, the improvement based on them would be reasonably small.

\subsection{Complexity of hierarchy} \label{Section_Complexity}

\begin{figure}[ht]
	\centering
	\includegraphics[width = 14cm]{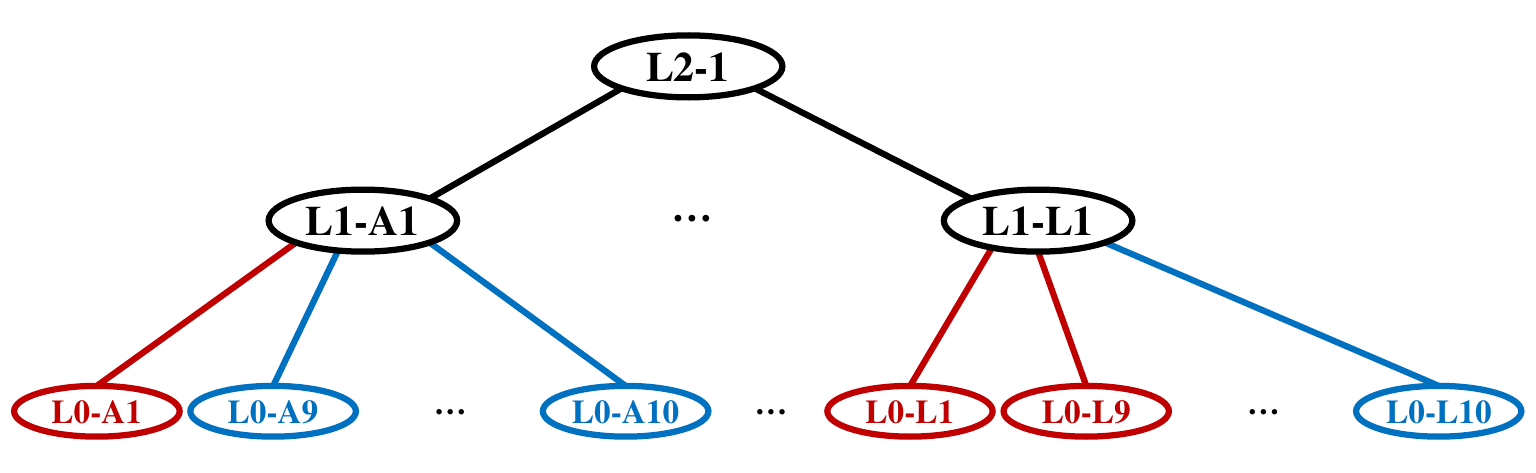} \\
	\caption{\rm Hierarchical tree for the numerical experiments in Section~\ref{Section_Complexity}. This hierarchy follows the similar aggregation rules of that in Figure~\ref{Figure_Proportion} that series in the middle level are aggregated in order by every 10 series at the bottom level. Some of the most disaggregated series (in red) are randomly selected to be non-Gaussian, and the others (in blue) are standard normal.}
	\label{Figure_Complexity}
\end{figure}

Finally, we discuss the complexity of hierarchy via a group of hierarchical structures with increasing scales of the bottom level from $n_b = 20$, 40 to 120, as illustrated in Figure~\ref{Figure_Complexity}.
In each hierarchy, we assume that 40\% of the most disaggregated series are out of order selected to be non-Gaussian, with equal probabilities, which further complicates the irregular series in HTS from Section~\ref{Section_Proportion}.

\begin{figure}[htbp]
	\begin{center}
		\subfigure[Bottom level: $h = 1$, $1 \sim 6$ and $1 \sim 12$.]{
			\begin{minipage}[t]{15.6cm}
				\includegraphics[width = 5cm]{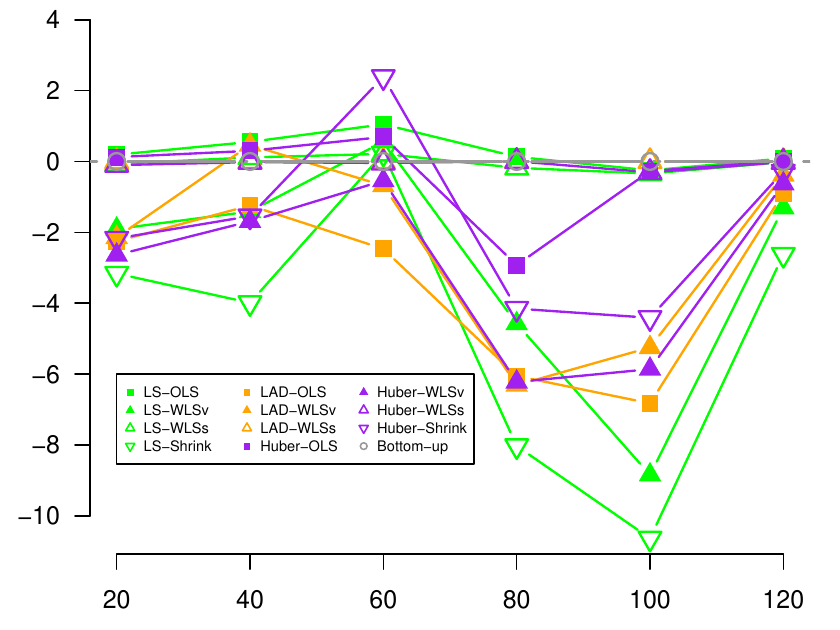}
				\
				\includegraphics[width = 5cm]{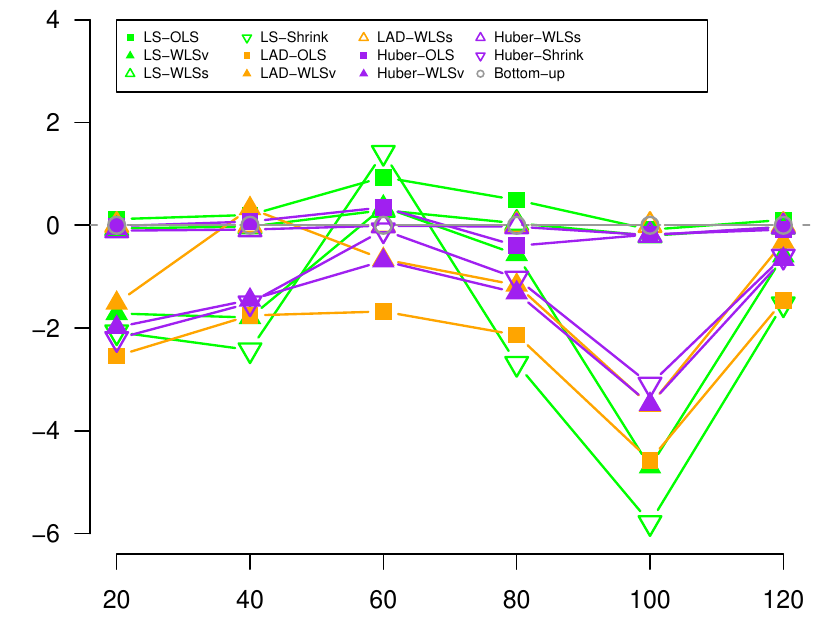}
				\
				\includegraphics[width = 5cm]{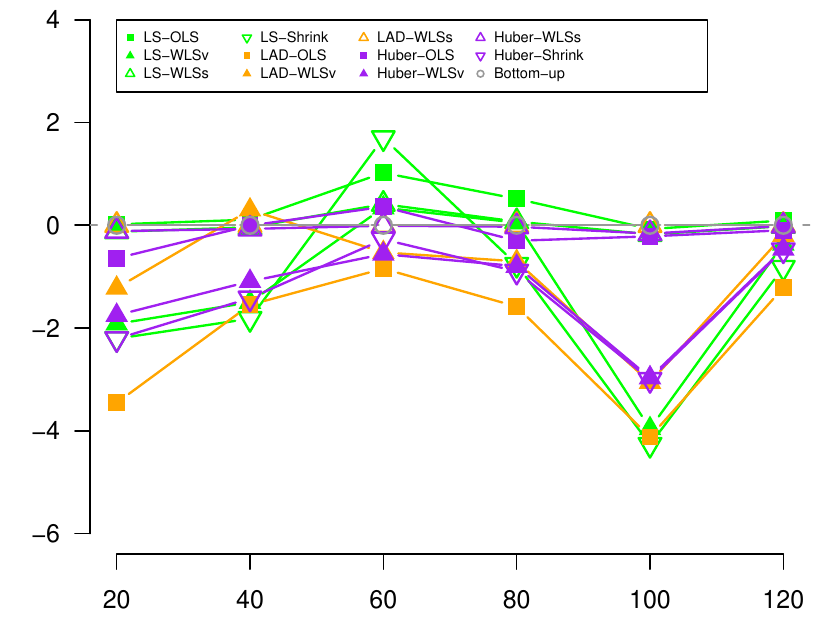}
			\end{minipage}
		}
	\end{center}
	\begin{center}
		\subfigure[Aggregated levels: $h = 1$, $1 \sim 6$ and $1 \sim 12$.]{
			\begin{minipage}[t]{15.6cm}
				\includegraphics[width = 5cm]{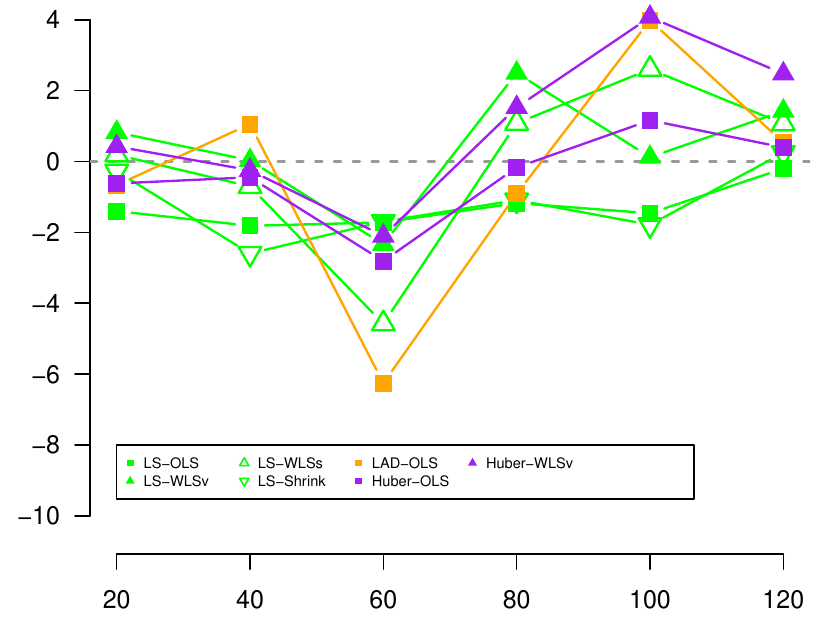}
				\
				\includegraphics[width = 5cm]{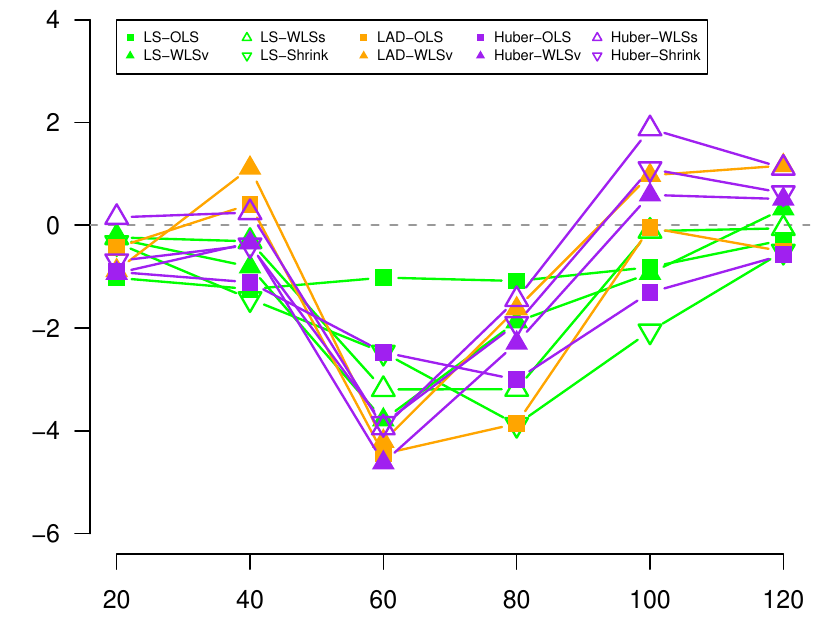}
				\
				\includegraphics[width = 5cm]{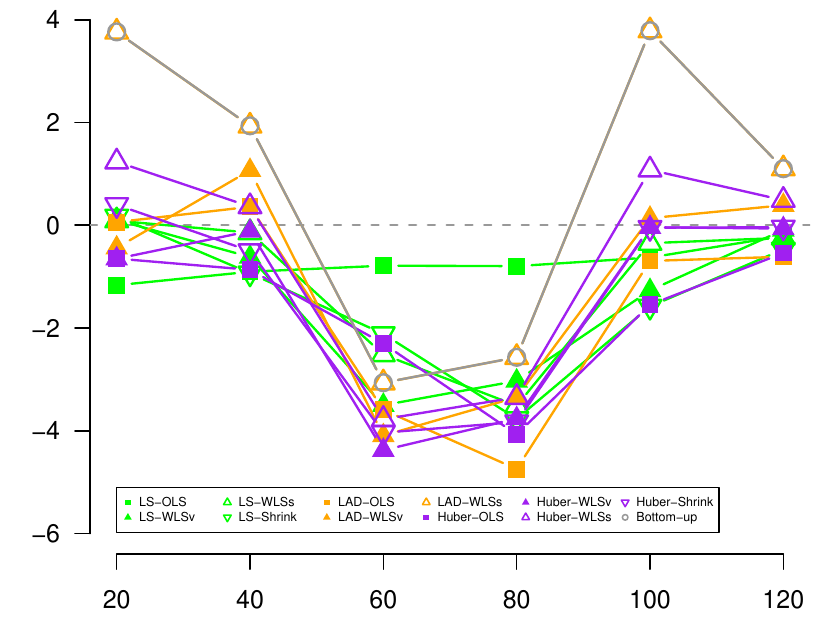}
			\end{minipage}
		}
	\end{center}
	\begin{center}
		\subfigure[Whole hierarchy: $h = 1$, $1 \sim 6$ and $1 \sim 12$.]{
			\begin{minipage}[t]{15.6cm}
				\includegraphics[width = 5cm]{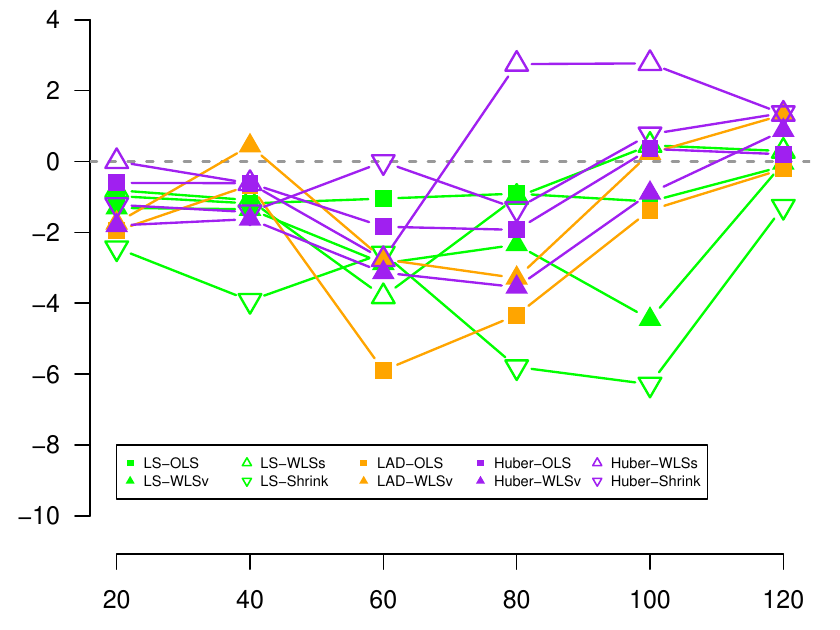}
				\
				\includegraphics[width = 5cm]{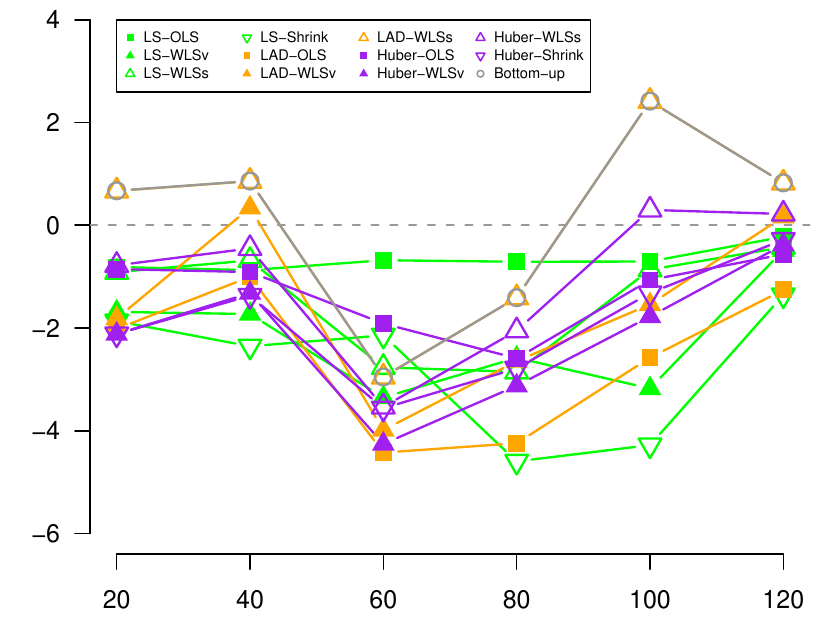}
				\
				\includegraphics[width = 5cm]{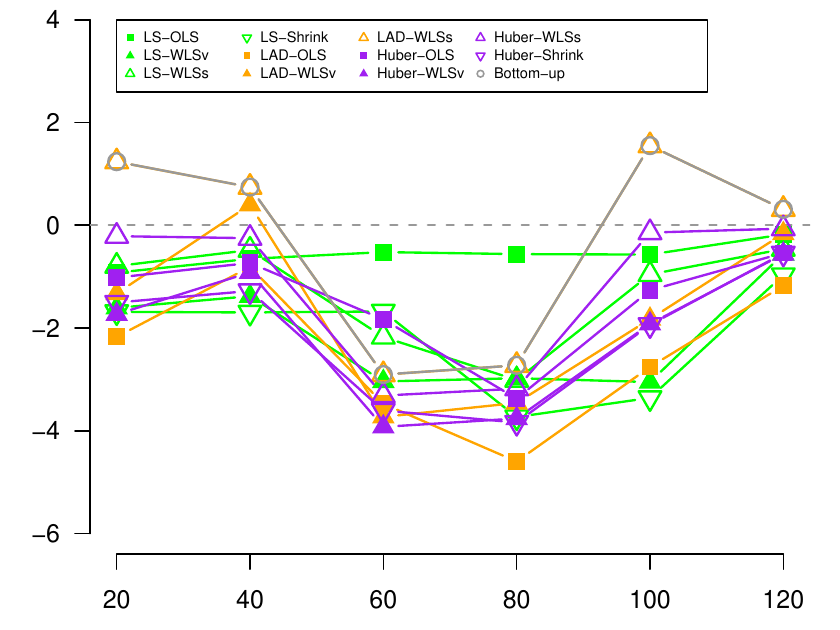}
			\end{minipage}
		}
	\end{center}
	\caption{\rm Line charts for the percentage increases by RoME from average RMSE of the base forecasts obtained by ARIMA against the complexity of hierarchy in Section~\ref{Section_Complexity}. The panels in the sub-figures, as well as the aggregated levels wherein, denote the same as Figure~\ref{Figure_Proportion-results}. The horizontal axis in every single figure denotes the number of the most disaggregated series in HTS. The lines exceeding the upper limit of the vertical axis, i.e., 4\%, are not displayed for concentrating the valuable information on the others.}
	\label{Figure_Complexity-results}
\end{figure}

Specifically, the determination of the distribution of related non-normal errors and the parameter settings of $\alpha_i$ and $\sigma_i$ in Model (\ref{Equation_Generation}) remain the same as Section~\ref{Section_Proportion}.
Figure~\ref{Figure_Complexity-results} draws the results for the reconciled results under different scales of hierarchies, where the reconciliations under Sample covariance design are not considered since the data are not enough for some large hierarchies.

As shown in Figure~\ref{Figure_Complexity-results}, most reconciliation models, including RoME and MinT, behave better than the benchmark (below 0), succeeding in reducing RMSE of the reconciled forecasts on average for the Aggregated levels and the whole hierarchy.
LAD-OLS (in orange square) takes the obvious leading place, with the maximum decrease of RMSE, and the family of MinT are generally reliable.
For the bottom level, the improvement of reconciliation procedure is limited (or sometime negative) when the scale of hierarchy is relatively small, say $n_b \leq 60$.
Given more structural and historical information on a larger HTS and its base forecasts, the reconciliation procedure would show more robustness to potential bad forecasts involved.

\section{Real-data study} \label{Section_Application}

In this section, we apply the proposed RoME reconciliation method to the well-known Australian Domestic Tourism dataset to demonstrate its feasibility and practical usefulness. For comparison, we also implement the MinT reconciliation approach and the baseline BU method.

\subsection{Australian Domestic Tourism Dataset: A multi-level geographical hierarchy} \label{sec:geo}

We consider the Australian Domestic Tourism dataset, which is widely used in HTS forecasting and has been extensively studied in the literature, e.g., \citet{Athanasopoulos2009Hierarchical, Karmy2019Hierarchical, Wickramasuriya2019Optimal,kourentzesCrosstemporalCoherentForecasts2019,Ashouri2021Fast}, among others. The dataset is organized according to a multi-level geographical hierarchy comprising 1 national series, 7 states, 27 zones, and 76 regions. However, as discussed by \citet{di2023cross,girolimetto2024cross}, six zones are each represented by a single region, resulting in a total of 105 unique time series under the geographical hierarchy. 
The forecasting variable is monthly ``\textit{visitor nights}'', defined as the total number of nights spent away from home, observed from January 1998 to December 2016. After removing the duplicated series, the resulting hierarchy becomes unbalanced. To account for this, we also conducted experiments using the \emph{FoReco} package, which supports point forecast reconciliation for unbalanced hierarchies \citep{FoReco2025}. The results are generally consistent with those obtained using MinT, except when the sample covariance matrix is employed. For clarity, we report only the MinT results in the main text, additional results are provided in Appendix~\ref{App_B5}.

To assess the robustness of the proposed method, we evaluate forecasting accuracy using a rolling-window experiment. Specifically, the initial training window consists of 96 monthly observations, and the window is expanded by one observation at each rolling iteration. 1- to 12-step-ahead base forecasts are generated using ARIMA and ETS models. RoME reconciliation is then applied under various combinations of loss functions and covariance specifications. For the Huber loss, the tuning parameter $k$ is selected based on the estimated standard deviation of the corresponding base forecast residuals.

Tables~\ref{Table_Tourism-ETS} and \ref{Table_Tourism-ARIMA} report the forecasting results for the geographical hierarchy, comparing RoME and MinT reconciled forecasts with the corresponding ETS- and ARIMA-based base forecasts. As shown in Table~\ref{Table_Tourism-ETS}, RoME exhibits clear advantages in mid- and long-term forecasting horizons. When ETS base forecasts are used, the LAD–OLS specification achieves the largest improvements in RMSE relative to the BU baseline for the 1–6, 6-step-ahead, and 1–12 horizons. For ARIMA-based forecasts, most alternative reconciliation methods slightly outperform the BU approach, leading to reduced RMSE at higher hierarchical levels, particularly at the national (``Australia") level.

\subsection{Australian Domestic Tourism Dataset: A cross-sectional hierarchy}\label{sec:cross}

In addition to the geographical structure, the dataset also exhibits a hierarchical organization based on the purpose of travel, which provides further insights into travel behavior across Australia. According to \citet{wickramasuriya2020optimal,di2023cross,girolimetto2024cross}, travel purposes are classified into four categories: holiday, visiting friends and relatives, business, and other. Incorporating travel purpose results in a cross-sectional hierarchy that combines geographical regions with travel categories. In this setting, the dataset comprises 304 bottom-level series and 525 total series. Further details of the hierarchical structures can be found in Table 7 of \citet{girolimetto2024cross}. 

The experimental setup is same as Section~\ref{sec:geo}. The forecasting results results for the cross-sectional hierarchical structure are reported in Tables~\ref{Table_Tourism-ETS-full} and \ref{Table_Tourism-ARIMA-full}. The RMSE values are summarized separately according to the geographical hierarchy and the travel-purpose hierarchy. Overall, the findings are consistent with those from the purely geographical setting, RoME delivers superior performance, especially for mid- and long-term forecasts. To avoid redundancy, detailed reconciliation results are deferred to~\ref{App_B5}.

In summary, the real data application confirms the practical value of the proposed RoME reconciliation framework for hierarchical time series forecasting. By incorporating robust M-estimation techniques, such as the LAD loss into the reconciliation step, RoME enhances forecast accuracy and robustness relative to existing approaches.

\begin{table}[htbp]
  \centering
  \caption{\rm Forecasting results of the proposed RoME reconciliation through ETS in the real-data study in Section~\ref{Section_Application}. The rows denote the same as Table~\ref{Table_NonGaussian-mixture}.  The columns ``Australia", ``States", ``Zones" and ``Regions" correspond to the aggregation levels. And the sub-columns indicate the results for the corresponding ahead-forecasting steps and out-of-sample forecasting periods. The best results are highlighted in bold.}
	\label{Table_Tourism-ETS}
	\setlength\tabcolsep{6pt}
	\renewcommand{\arraystretch}{0.8}
		\footnotesize
    \begin{tabular}{cccccccccc}
    \toprule
    \multirow{2}[4]{*}{$\rho$} & \multirow{2}[4]{*}{$\bm{W}_h$} & \multicolumn{4}{c}{Australia} & \multicolumn{4}{c}{States} \\
\cmidrule{3-10}          &       & $h=1$   & $h=6$   & $1 \sim 6$   & 1~12  & $h=1$   & $h=6$   & $1 \sim 6$   & $1 \sim 12$ \\
    \midrule
    \multirow{5}[1]{*}{LS} & OLS   & -0.45 & -0.40 & -0.43 & -0.39 & -0.49 & -0.44 & 0.54  & 0.22 \\
          & WLSv  & 1.54  & -0.48 & -0.97 & -1.26 & 0.58  & 1.86  & 2.10  & 0.45 \\
          & WLSs  & 0.28  & -1.07 & -1.24 & -1.32 & -0.20 & 0.83  & 1.35  & 0.13 \\
          & Sample & 3.99  & 4.01  & 1.37  & 1.85  & 2.83  & 7.66  & 4.96  & 4.53 \\
          & Shrink & \textbf{-6.34} & 9.25  & 0.85  & 7.07  & \textbf{-9.10} & 16.83 & 5.99  & 10.72 \\
          \midrule
    \multirow{5}[0]{*}{LAD} & OLS   & -1.59 & \textbf{-1.74} & \textbf{-2.10} & \textbf{-1.71} & -1.80 & \textbf{-1.91} & \textbf{-0.14} & \textbf{-0.70} \\
          & WLSv  & 3.08  & 1.67  & 0.17  & -0.67 & 2.69  & 4.10  & 3.68  & 1.58 \\
          & WLSs  & 3.07  & 1.65  & 0.16  & -0.67 & 2.69  & 4.09  & 3.68  & 1.58 \\
          & Sample & 7.13  & 20.56 & 7.08  & 8.77  & 7.05  & 21.04 & 10.29 & 11.45 \\
          & Shrink & 131.80 & 249.64 & 134.08 & 179.98 & 99.91 & 198.81 & 123.65 & 174.93 \\
          \midrule
    \multirow{5}[1]{*}{Huber} & OLS   & -0.45 & -0.40 & -0.43 & -0.39 & -0.49 & -0.44 & 0.54  & 0.22 \\
          & WLSv  & 1.88  & -0.10 & -0.74 & -1.13 & 0.98  & 2.38  & 2.46  & 0.72 \\
          & WLSs  & 0.71  & -0.65 & -0.98 & -1.28 & 0.15  & 1.34  & 1.68  & 0.32 \\
          & Sample & 9.02  & 34.88 & 12.58 & 13.54 & 6.78  & 31.85 & 15.27 & 16.29 \\
          & Shrink & 4.30  & 77.46 & 27.85 & 70.85 & -1.21 & 67.73 & 29.38 & 70.62 \\
    \midrule
    \multicolumn{2}{c}{Bottom-up} & 3.07  & 1.65  & 0.16  & -0.67 & 2.69  & 4.09  & 3.68  & 1.58 \\
    \midrule
    \multicolumn{2}{c}{Base} & 4554.78 & 4332.15 & 6112.76 & 6729.83 & 1050.14 & 1063.38 & 1288.16 & 1376.91 \\
    \midrule
    \midrule
    \multirow{2}[4]{*}{$\rho$} & \multirow{2}[4]{*}{$\bm{W}_h$} & \multicolumn{4}{c}{Zones} & \multicolumn{4}{c}{Regions} \\
\cmidrule{3-10}          &       & $h=1$   & $h=6$   & $1 \sim 6$   & 1~12  & $h=1$   & $h=6$   & $1 \sim 6$   & $1 \sim 12$ \\
    \midrule
    \multirow{5}[1]{*}{LS} & OLS   & -0.36 & -1.59 & -0.52 & -0.65 & -1.31 & -0.95 & -1.08 & -1.14 \\
          & WLSv  & -0.66 & -2.26 & -0.81 & -1.25 & -1.71 & -1.62 & -1.38 & -1.73 \\
          & WLSs  & -1.04 & -1.77 & -0.89 & -1.06 & -0.89 & -0.51 & -0.57 & -0.68 \\
          & Sample & 0.52  & 1.90  & 1.24  & 2.34  & -0.13 & 2.19  & 0.99  & 2.28 \\
          & Shrink & \textbf{-8.03} & 8.76  & 2.24  & 6.77  & \textbf{-11.36} & 0.07  & -3.01 & -0.14 \\
          \midrule
    \multirow{5}[0]{*}{LAD} & OLS   & -0.70 & \textbf{-3.44} & \textbf{-1.39} & \textbf{-1.55} & -5.04 & \textbf{-5.77} & \textbf{-4.62} & \textbf{-5.08} \\
          & WLSv  & 0.55  & -0.85 & 0.03  & -0.23 & 0.01  & 0.00  & 0.00  & 0.00 \\
          & WLSs  & 0.54  & -0.85 & 0.03  & -0.24 & 0.00  & 0.00  & 0.00  & 0.00 \\
          & Sample & 4.54  & 13.93 & 6.48  & 9.06  & 3.09  & 12.13 & 5.46  & 8.07 \\
          & Shrink & 79.10 & 152.13 & 100.88 & 148.78 & 56.66 & 115.88 & 77.98 & 122.85 \\
          \midrule
    \multirow{5}[1]{*}{Huber} & OLS   & -0.36 & -1.59 & -0.52 & -0.65 & -1.31 & -0.95 & -1.08 & -1.14 \\
          & WLSv  & -0.56 & -2.19 & -0.75 & -1.26 & -1.26 & -1.07 & -0.95 & -1.29 \\
          & WLSs  & -0.95 & -1.91 & -0.90 & -1.18 & -0.80 & -0.41 & -0.49 & -0.60 \\
          & Sample & 3.85  & 22.74 & 10.54 & 13.08 & 2.01  & 18.00 & 8.26  & 11.22 \\
          & Shrink & -2.03 & 46.30 & 20.94 & 56.38 & -7.29 & 30.77 & 12.05 & 43.19 \\
    \midrule
    \multicolumn{2}{c}{Bottom-up} & 0.54  & -0.85 & 0.03  & -0.23 & 0.00  & 0.00  & 0.00  & 0.00 \\
    \midrule
    \multicolumn{2}{c}{Base} & 374.61 & 399.32 & 451.98 & 472.04 & 183.07 & 199.95 & 215.82 & 223.92 \\
    \bottomrule
    \end{tabular}
\end{table}

\begin{table}[htbp]
 \centering
 \caption{\rm Forecasting results of the proposed RoME reconciliation through ARIMA in the real-data study in Section~\ref{Section_Application}. The rows, columns and sub-columns denote the same as Table~\ref{Table_Tourism-ETS}. The best results are highlighted in bold.}
	\label{Table_Tourism-ARIMA}
	\setlength\tabcolsep{6pt}
		\renewcommand{\arraystretch}{0.8}
		\footnotesize
    \begin{tabular}{cccccccccc}
    \toprule
    \multirow{2}[4]{*}{$\rho$} & \multirow{2}[4]{*}{$\bm{W}_h$} & \multicolumn{4}{c}{Australia} & \multicolumn{4}{c}{States} \\
\cmidrule{3-10}          &      & $h=1$   & $h=6$   & $1 \sim 6$   & $1 \sim 12$  & $h=1$  & $h=6$   & $1 \sim 6$  & $1 \sim 12$ \\
    \midrule
    \multirow{5}[1]{*}{LS} & OLS   & -3.47 & -0.74 & -1.39 & -0.64 & 1.39  & 0.33  & -0.08 & 0.39 \\
          & WLSv  & -13.89 & -2.07 & -5.04 & -2.78 & -8.90 & -0.89 & -4.07 & -1.88 \\
          & WLSs  & -12.24 & -2.15 & -4.64 & -2.32 & -7.00 & -0.73 & -3.40 & -1.33 \\
          & Sample & -14.36 & 3.56  & -8.24 & -7.15 & -12.86 & 0.27  & -8.41 & -6.34 \\
          & Shrink & \textbf{-18.64} & \textbf{-5.90} & \textbf{-13.42} & \textbf{-10.03} & \textbf{-18.87} & \textbf{-8.19} & \textbf{-14.59} & \textbf{-10.04} \\
          \midrule
    \multirow{5}[0]{*}{LAD} & OLS   & -11.67 & -1.70 & -3.97 & -1.84 & -5.01 & 0.09  & -2.12 & -0.56 \\
          & WLSv  & -11.83 & -0.86 & -3.92 & -2.38 & -8.00 & 0.49  & -3.02 & -1.41 \\
          & WLSs  & -11.81 & -0.85 & -3.91 & -2.38 & -8.00 & 0.49  & -3.02 & -1.41 \\
          & Sample & 64.44 & 73.25 & 49.48 & 29.01 & 55.27 & 52.60 & 42.05 & 27.28 \\
          & Shrink & 114.14 & 67.70 & 63.20 & 32.43 & 86.25 & 43.52 & 49.28 & 27.98 \\
          \midrule
    \multirow{5}[1]{*}{Huber} & OLS   & -3.63 & -0.74 & -1.45 & -0.66 & 1.16  & 0.33  & -0.13 & 0.38 \\
          & WLSv  & -13.98 & -2.03 & -4.98 & -2.76 & -9.07 & -0.84 & -4.02 & -1.86 \\
          & WLSs  & -13.40 & -2.13 & -4.62 & -2.29 & -7.74 & -0.71 & -3.45 & -1.33 \\
          & Sample & 1.13  & 19.27 & 6.12  & -0.64 & 0.36  & 11.77 & 3.62  & -0.32 \\
          & Shrink & 4.20  & 6.67  & -0.76 & -7.49 & -1.38 & -1.06 & -4.69 & -7.64 \\
    \midrule
    \multicolumn{2}{c}{Bottom-up} & -11.81 & -0.85 & -3.91 & -2.38 & -8.00 & 0.49  & -3.02 & -1.41 \\
    \midrule
    \multicolumn{2}{c}{Base} & 4787.88 & 4192.54 & 5955.78 & 6629.73 & 1055.61 & 985.16 & 1245.54 & 1327.75 \\
    \midrule
    \midrule
    \multirow{2}[4]{*}{$\rho$} & \multirow{2}[4]{*}{$\bm{W_h}$} & \multicolumn{4}{c}{Zones}     & \multicolumn{4}{c}{Regions} \\
\cmidrule{3-10}          &       & $h=1$   & $h=6$   & $1 \sim 6$   & $1 \sim 12$  & $h=1$  & $h=6$   & $1 \sim 6$  & $1 \sim 12$ \\
    \midrule
    \multirow{5}[1]{*}{LS} & OLS   & 3.28  & -0.25 & 0.56  & 0.53  & 4.88  & -0.02 & 1.42  & 0.75 \\
          & WLSv  & -2.39 & -1.42 & -1.36 & -0.71 & -0.37 & -0.85 & -0.79 & -0.55 \\
          & WLSs  & -1.05 & -1.05 & -0.90 & -0.33 & 0.31  & -0.49 & -0.34 & -0.21 \\
          & Sample & -7.61 & -1.84 & -6.83 & -5.74 & -4.69 & -1.07 & -4.47 & -4.34 \\
          & Shrink & \textbf{-12.32} & \textbf{-7.94} & \textbf{-11.84} & \textbf{-9.03} & \textbf{-9.17} & \textbf{-6.61} & \textbf{-9.68} & \textbf{-7.78} \\
          \midrule
    \multirow{5}[0]{*}{LAD} & OLS   & -1.07 & -0.48 & -0.47 & 0.04  & 4.33  & 0.15  & 1.60  & 0.99 \\
          & WLSv  & -1.66 & -0.43 & -0.40 & -0.08 & -0.01 & 0.01  & 0.00  & 0.00 \\
          & WLSs  & -1.65 & -0.43 & -0.40 & -0.08 & 0.00  & 0.00  & 0.00  & 0.00 \\
          & Sample & 55.56 & 41.56 & 38.92 & 25.75 & 46.74 & 35.46 & 34.92 & 23.24 \\
          & Shrink & 75.99 & 34.18 & 42.82 & 25.16 & 61.32 & 25.26 & 34.97 & 20.51 \\
          \midrule
    \multirow{5}[1]{*}{Huber} & OLS   & 3.14  & -0.25 & 0.53  & 0.52  & 4.74  & -0.02 & 1.39  & 0.74 \\
          & WLSv  & -2.51 & -1.41 & -1.36 & -0.70 & -0.43 & -0.82 & -0.75 & -0.53 \\
          & WLSs  & -1.43 & -1.05 & -0.94 & -0.34 & 0.15  & -0.48 & -0.34 & -0.21 \\
          & Sample & 3.00  & 6.89  & 2.81  & -0.68 & 3.09  & 5.83  & 3.34  & 0.06 \\
          & Shrink & 1.24  & -3.04 & -4.81 & -7.12 & 1.52  & -3.29 & -4.06 & -6.01 \\
    \midrule
    \multicolumn{2}{c}{Bottom-up} & -1.65 & -0.43 & -0.40 & -0.08 & 0.00  & 0.00  & 0.00  & 0.00 \\
    \midrule
    \multicolumn{2}{c}{Base} & 366.79 & 359.99 & 426.59 & 447.12 & 168.05 & 164.47 & 187.01 & 193.65 \\
    \bottomrule
    \end{tabular}
\end{table}

\section{Conclusion and discussion} \label{Section_Discussion}

This paper examined the robust reconciliation procedure with M-estimation, named RoME, for HTS forecasting. The reconciliation is initially formulated as the cost of transforming a group of base forecasts previously acquired from the historical information to fulfill the related aggregation constraints of the hierarchy, the procedure aims to find the optimal results with the minimum cost. From the perspective of constrained optimization, the proposed method introduces a general loss function with more robustness to measure the cost, such that the regression-based reconciliation procedure would be less sensitive to potential irregular series with bad forecasts. The related iterative computation, developed through the modified Newton-Raphson algorithm using LQA, possesses the property of global convergence. Furthermore, the proposed computational algorithm can be reformulated as the trace minimization problem with dynamic weights for each series in the hierarchy, which theoretically extends the existing LS-based reconciliation procedure in MinT to the robust framework.

In numerical studies and empirical application, the proposed robust forecast reconciliation is compared with MinT to provide a guideline on its optimal application domain.  Moreover, beyond geographical hierarchies, RoME is shown to perform effectively in cross-sectional hierarchical structures, highlighting its flexibility in accommodating complex hierarchical relationships. Although forecasting accuracy is improved, the proposed RoME framework incurs a substantial computational burden when applied to cross-sectional hierarchical time series. In particular, estimating high-dimensional error covariance matrices becomes computationally expensive. Incorporating penalized estimation methods \citep{taieb2017regularization} for high-dimensional error covariances may therefore help alleviate this issue.

A few extensions merit further exploration:

First, extending the proposed framework to hierarchical time series with non-stationary components is an important direction for future research, as the stationarity assumption is often violated in practice. In this paper, we focus on the theoretical optimization problem and iterative computation of RoME under stationarity, relaxing this assumption requires further investigation of both empirical performance and theoretical justification.

Second, although we consider three widely used and distinct loss functions in the proposed reconciliation framework, selecting an optimal loss function and tuning its associated parameters in M-estimation remains a challenging task and is likely application dependent. In addition, combining reconciled forecasts obtained from different loss functions offers a promising avenue for further improvement. In this paper, we explore simple combinations of reconciled forecasts based on LS and LAD losses. Simulation results reported in Appendix~\ref{App_B4} show modest but encouraging gains in forecasting accuracy. More advanced combination strategies, such as those using dynamic or data-driven weights, warrant systematic investigation in future work.

Third, beyond the point forecasts discussed above, interval and probabilistic forecasts that quantify the uncertainty of reconciled results constitute an important component of HTS forecasting. By providing probability distributions, such forecasts offer a richer characterization of predictive uncertainty, which can aid in assessing the likelihood of extreme events \citep{lee2025memory,rostami2025hierarchical}. The approximate covariance matrix of reconciled forecast errors at each iteration offers a potential avenue for deriving such uncertainty measures, although a rigorous theoretical justification remains a topic for future research. Existing studies have investigated probabilistic forecast reconciliation under both Gaussian frameworks \citep{panagiotelis2023probabilistic} and nonparametric frameworks \citep{girolimetto2024point}. Building on these approaches, integrating probabilistic reconciliation into the proposed RoME framework constitutes a promising avenue for future research.

Finally, extending the proposed RoME framework to accommodate more complex hierarchical structures constitutes an important direction for future research. Beyond purely sectional hierarchies, temporal and cross-temporal/sectional hierarchies are also commonly encountered in time series applications. Moreover, in practice, hierarchical structures do not always conform to a strict bottom-level aggregation scheme \citep{athanasopoulos2024forecast}. To address such settings, several studies have developed HTS forecasting approaches based on general linear constraints, including zero-sum constraints \citep{di2023cross, girolimetto2024cross}. These frameworks can be combined with machine learning methods to capture more complex relationships between different hierarchical levels, or integrated with clustering techniques to construct intermediate-level series that enhance structural flexibility \citep{zhang2025constructing}. In addition, in applications where the series are subject to non-negativity constraints, incorporating such constraints into the reconciliation process is essential \citep{wickramasuriya2020optimal}. All of these state-of-the-art developments offer promising opportunities for extending and enriching the RoME framework in future work.

\section*{Acknowledgment}
\vskip 0.3cm

The authors are grateful to the Editor, Associate Editor, and anonymous reviewers for their valuable suggestions, and thank Prof. Yangfei Kang and Dr. Bohan Zhang for their advice during the revision. This work was supported by the Key Program of the National Natural Science Foundation of China (Grant No. 72531002) and National Natural Science Foundation of China (Grant No. 72572010).

\section*{Conflicts of Interest}
The authors declare that they have no conflicts of interest.

\section*{Data Availability Statement}

The datasets used in this study are publicly available, which can obtained from the sources cited in the manuscript.

\section*{Ethics Statement}
N/A.

\FloatBarrier

\appendix

\renewcommand{\theequation}{\thesection.\arabic{equation}}
\renewcommand{\thefigure}{\thesection.\arabic{figure}}
\renewcommand{\thetable}{\thesection.\arabic{table}}
\setcounter{equation}{0}
\setcounter{figure}{0}
\setcounter{table}{0}

\section{Appendix: Proofs}\label{App_A}

\subsection{ Iterative solutions (\ref{Equation_y-RoME}) and (\ref{Equation_P-RoME})}\label{App_A1}

From (\ref{Equation_Condition}), the updated reconciled forecasts can be rewritten as
    \begin{equation} \nonumber
        \check{\bm{y}}_t^{\ast^{(\omega + 1)}} (h) = \hat{\bm{y}}_t^\ast (h) - \bm{D}_h^{(\omega)} \bm{W}_h^{1 / 2} \bm{U} \bm{\lambda},
    \end{equation}
    which implies that
    \begin{equation} \nonumber
        \bm{U}^{\top} \bm{W}_h^{1 / 2} \hat{\bm{y}}_t^\ast (h) - \bm{U}^{\top} \bm{W}_h^{1 / 2} \mathbf{D}_h^{(\omega)} \bm{W}_h^{1 / 2} \bm{U} \bm{\lambda} = \mathbf{0}_{m^\ast}
    \end{equation}
    since $\bm{U}^{\top} \bm{W}_h^{1 / 2} \check{\bm{y}}_t^{\ast^{(\omega)}} (h) = \mathbf{0}_{m^\ast}$.
Assume that $\bm{U}^{\top} \bm{W}_h^{1 / 2} \bm{D}_h^{(\omega)} \bm{W}_h^{1 / 2} \bm{U}$ is positive-definite, the Lagrangian multipliers can then be expressed as
    \begin{equation} \nonumber
        \bm{\lambda} = {\big (} \bm{U}^{\top} \bm{W}_h^{1 / 2} \bm{D}_h^{(\omega)} \bm{W}_h^{1 / 2} \bm{U} {\big )}^{-1} \bm{U}^{\top} \bm{W}_h^{1 / 2} \hat{\bm{y}}_t^\ast (h),
    \end{equation}
    which accounts for the solution to $\check{\bm{y}}_t^{\ast^{(\omega + 1)}} (h)$, i.e.,
    \begin{equation} \nonumber
        \check{\bm{y}}_t^{\ast^{(\omega + 1)}} (h) = {\big (} \mathbf{I}_m - \bm{D}_h^{(\omega)} \bm{W}_h^{1 / 2} \bm{U} {\big (} \bm{U}^{\top} \bm{W}_h^{1 / 2} \bm{D}_h^{(\omega)} \bm{W}_h^{1 / 2} \bm{U} {\big )}^{-1} \bm{U}^{\top} \bm{W}_h^{1 / 2} {\big )} \hat{\bm{y}}_t^\ast (h).
    \end{equation}
The standardization expressions, i.e., $\hat{\bm{y}}_t (h) = \bm{W}_h^{1 / 2} \hat{\bm{y}}_t^\ast (h)$ and $\check{\bm{y}}_t^{(\omega)} (h) = \bm{W}_h^{1 / 2} \check{\bm{y}}_t^{\ast^{(\omega)}} (h)$, finally lead to the iteration solutions.

\subsection{Iterative solution (\ref{Equation_P-RoME-equivalent})}\label{App_A2}

The proof of (\ref{Equation_P-RoME-equivalent}) is inspired by that of Theorem 1 in \cite{Wickramasuriya2019Optimal}.
Since no explicit expression for 
$\bm{G}_h$ is available with general loss function $\rho (| \cdot |)$, we begin with an obtained value of 
$\bm{G}_h$, say $\bm{G}_{\rm RoME}^{(\omega)} (h)$ at the iteration $\omega$.
Thus, it suffices to show that the minimization problem (\ref{Equation_Model-RoME}) can be approximately reformulated as
    \begin{equation} \label{Equation_Model-RoME-trace}
        \begin{array}{cl}
             \mathop{\text{minimize}}\limits_{\bm{G}_h \in \mathbb{R}^{n_b \times n}} & \text{tr} {\big (} \bm{S} \bm{G}_h \mathbb{W}_h^{(\omega)} \bm{G}_h^{\top} \bm{S}^{\top} {\big )} \\
            \text{s.t.} & \bm{G}_h \bm{S} = \mathbf{I}_n,
            
        \end{array}
    \end{equation}
where $\text{tr} (\cdot)$ denotes the trace of a matrix.

Actually, given the base forecast errors $\hat{\bm{\varepsilon}}_t (h)$, the reconciled forecast errors $\tilde{\bm{\varepsilon}}_t (h)$ can be equivalently expressed as $\bm{S} \bm{G}_h \hat{\bm{\varepsilon}}_t (h)$ from (\ref{Equation_Reconciliation}).
Suppose that $\bm{G}_{\rm RoME}^{(\omega)} (h)$ is close enough to the expected $\bm{G}_{\rm RoME} (h)$,
the related reconciled forecast error $\tilde{\varepsilon}_{t,i}^{(\omega)} (h)$ will approach the optimal value, i.e., $\tilde{\varepsilon}_{t,i} (h)$, synchronously for $i = 1, 2, \cdots, n$.
Following the quadratic approximation of (\ref{Equation_Approximation}), the general loss function at $\tilde{\varepsilon}_{t,i} (h)$ can be approximated as
    \begin{equation} \nonumber
        \rho {\big (} | \tilde{\varepsilon}_{t,i} (h) | {\big )} \approx \rho {\big (} | \tilde{\varepsilon}_{t,i}^{(\omega)} (h) | {\big )} + \frac{1}{2} \frac{\rho^\prime (| \tilde{\varepsilon}_{t,i}^{(\omega)} (h) |)} {| \tilde{\varepsilon}_{t,i}^{(\omega)} (h) |} {\big (} \tilde{\varepsilon}_{t,i}^2 (h) - (\tilde{\varepsilon}_{t,i}^{(\omega)} (h))^2 {\big )}.
    \end{equation}
Taking the expectation on $\rho (| \tilde{\varepsilon}_{t,i} (h) |)$ w.r.t. $\tilde{\varepsilon}_{t,i} (h)$ given the historical information $\mathcal{Y}_{\mathcal{I}_t}$, and omitting the terms irrelevant to $\tilde{\varepsilon}_{t,i} (h)$, we have that $\mathbb{E} [ \rho (| \tilde{\varepsilon}_{t,i} (h) |)]$ is proportional to
    \begin{equation} \nonumber
        \frac{\rho^\prime (| \tilde{\varepsilon}_{t,i}^{(\omega)} (h) |)} {| \tilde{\varepsilon}_{t,i}^{(\omega)} (h) |} \mathbb{E} {\big [} \tilde{\varepsilon}_{t,i}^2 (h) {\big |} \mathcal{Y}_{\mathcal{I}_t} {\big ]}.
    \end{equation}
Let $\bm{S}_i$ denote the $i$-th row of $\bm{S}$, then
    \begin{equation} \nonumber
        \mathbb{E} {\big [} \tilde{\varepsilon}_{t,i}^2 (h) {\big |} \mathcal{Y}_{\mathcal{I}_t} {\big ]}
        = \mathbb{E} {\big [} (\bm{S}_i \bm{G}_h \tilde{\bm{\varepsilon}}_t (h))^2 {\big |} \mathcal{Y}_{\mathcal{I}_t} {\big ]}
        = \mathbb{E} {\big [} \bm{S}_i \bm{G}_h \tilde{\bm{\varepsilon}}_t (h) \tilde{\bm{\varepsilon}}_t^{\top} (h) \bm{G}_h^{\top} \bm{S}_i^{\top} {\big |} \mathcal{Y}_{\mathcal{I}_y} {\big ]}
        = \bm{S}_i \bm{G}_h \bm{W}_h \bm{G}_h^{\top} \bm{S}_i^{\top}.
    \end{equation}
Thus, the objective function in (\ref{Equation_Optimization-RoME}) can be reformulated as
    \begin{equation} \nonumber
        \sum_{i = 1}^n \varpi_{hi}^{(\omega)} \mathbb{E} {\big [} (\tilde{\varepsilon}_{t,i} (h))^2 {\big |} \mathcal{Y}_{\mathcal{I}_t} {\big ]} = \sum_{i = 1}^n \bm{S}_i \bm{G}_h \mathbb{W}_h^{(\omega)} \bm{G}_h^{\top} \bm{S}_i^{\top},
    \end{equation}
    which coincides with the objective function in (\ref{Equation_Model-RoME-trace}).

The trace minimization problem (\ref{Equation_Model-RoME-trace}) falls into the paradigm of the typical optimization problem in MinT.
Following the conclusion made by \cite{Wickramasuriya2019Optimal}, we can directly obtain the iterative solution.

\section{Appendix: Simulation results}\label{App_B}

\subsection{Comparison on computation time}\label{App_B1}

\begin{figure}[htbp]
	\begin{center}
		\subfigure[$0.9 \cdot \mathcal{N} (0, 1) + 0.1 \cdot \mathcal{N} (0, 3^2)$.]{
			\resizebox{4.8cm}{3.6cm}{\includegraphics{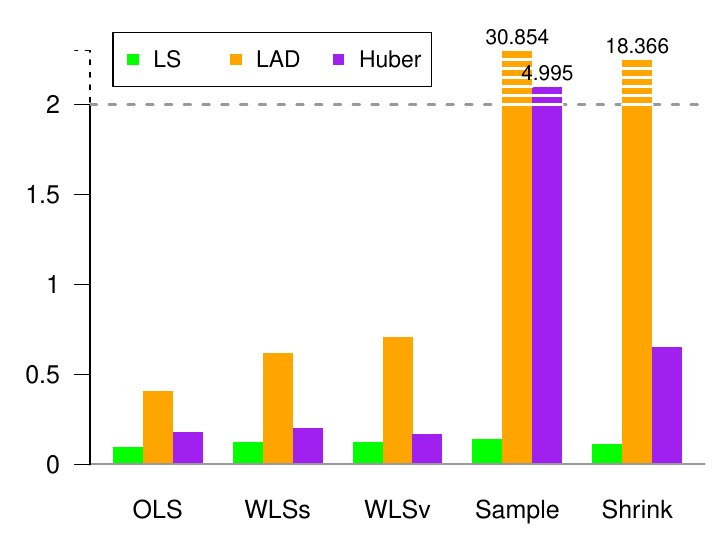}}
		}
		\subfigure[$t (3)$.]{
			\resizebox{4.8cm}{3.6cm}{\includegraphics{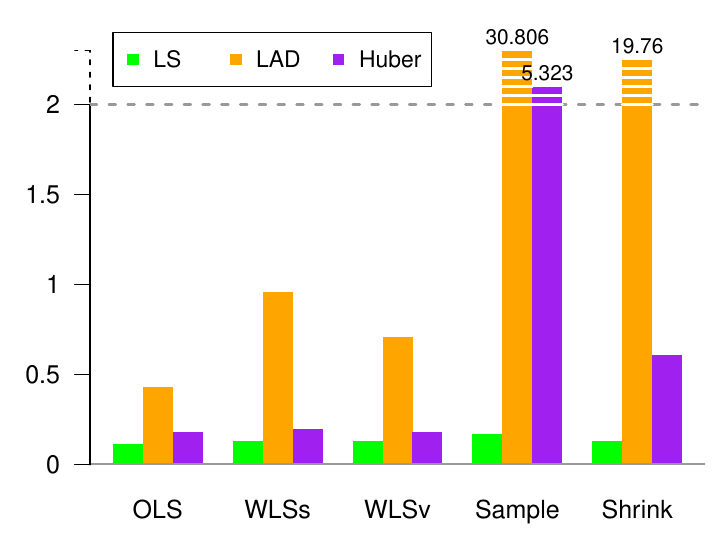}}
		}
		\subfigure[$\text{Cauchy} (0, 1)$.]{
			\resizebox{4.8cm}{3.6cm}{\includegraphics{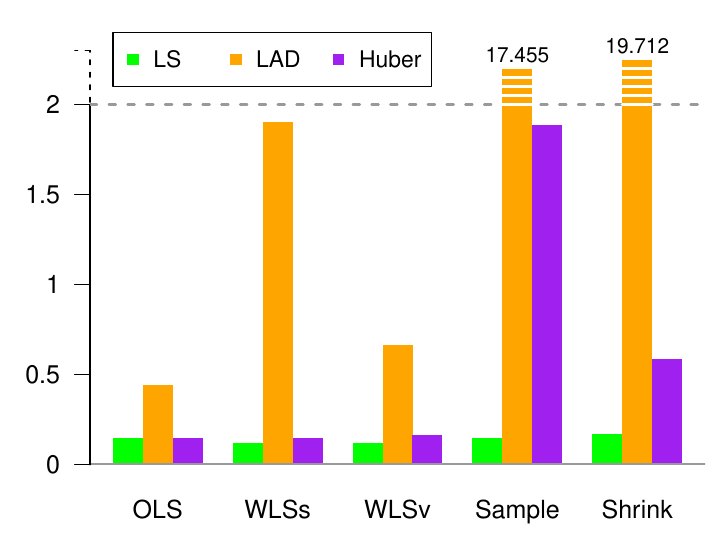}}
		}
	\end{center}
	\caption{\rm Bar charts for the average computation time (in millisecond) of the proposed robust reconciliation procedure under different types of non-Gaussian series in Section~\ref{Section_NonGaussian}. The ordinate values above the upper limit, i.e., 2, are displayed by specific numbers.}
	\label{Figure_NonGaussian-time}
\end{figure}

\begin{figure}[htbp]
	\begin{center}
		\resizebox{6cm}{4.5cm}{\includegraphics{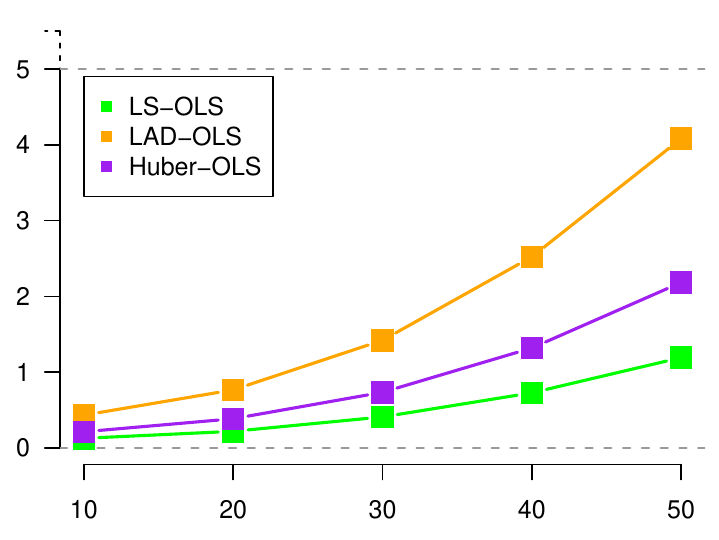}} \qquad
		\resizebox{6cm}{4.5cm}{\includegraphics{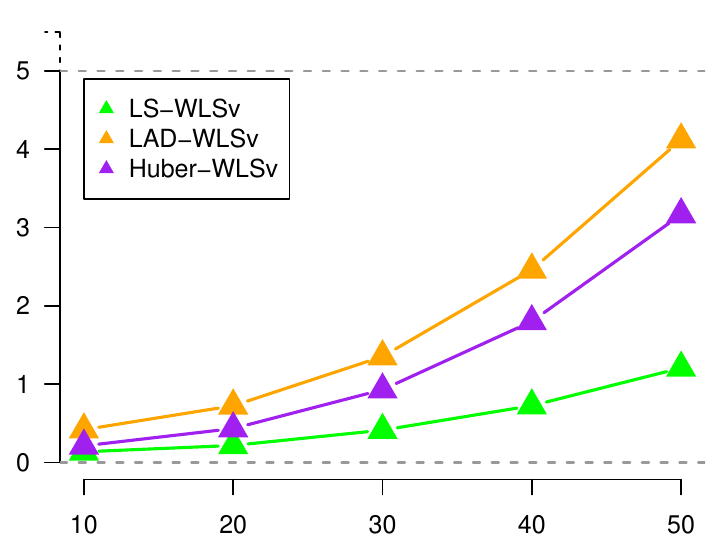}} \\
		\resizebox{6cm}{4.5cm}{\includegraphics{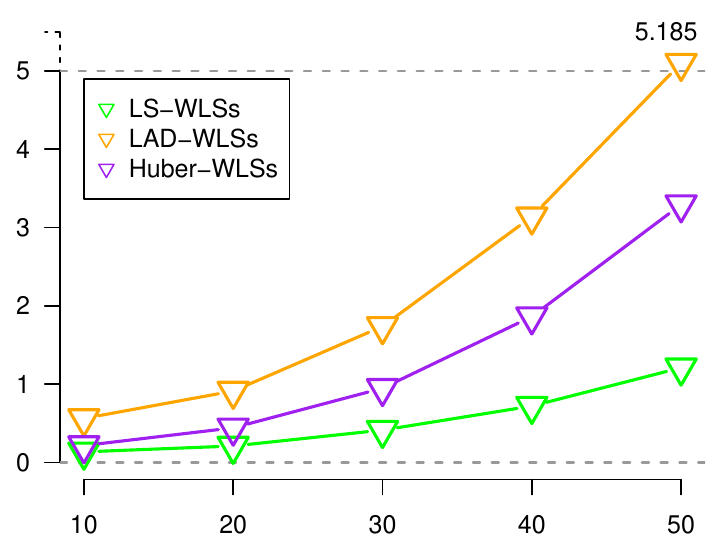}} \qquad
		\resizebox{6cm}{4.5cm}{\includegraphics{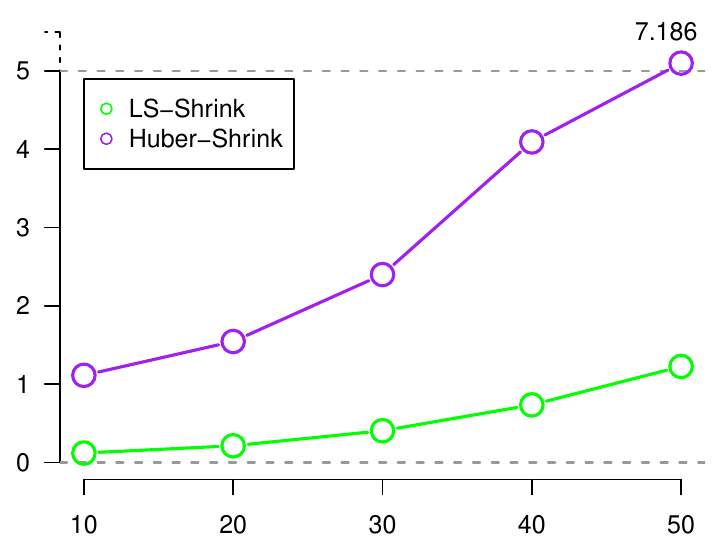}}
	\end{center}
	\caption{\rm Line charts for the average computation time (in millisecond) of the proposed robust reconciliation procedure against different scales of hierarchy in Section~\ref{Section_EfficiencyLoss}. The horizontal axis indicates the number of the most disaggregated series. The ordinate values above the upper limit, i.e., 5, are displayed by specific number, and the results for LAD-Shrink are not shown since they greatly exceed the upper limit.}
	\label{Figure_EfficiencyLoss-time}
\end{figure}

We compare the computation cost of the proposed robust reconciliation procedure for two issues.
That is, the type of non-Gaussian series, and scale of hierarchy, which correspond to the numerical experiments in Sections~\ref{Section_NonGaussian} and \ref{Section_EfficiencyLoss}, respectively.
Figures~\ref{Figure_NonGaussian-time} and \ref{Figure_EfficiencyLoss-time} report the computation time of the alternative RoME models, including MinT, for two issues, where different ahead-forecasting steps are considered equally.

For the type of non-Gaussian series, as shown in Figure~\ref{Figure_NonGaussian-time}, the computation time of the robust reconciliation procedure does not show remarkable differences in three types of non-normal distributions.
Under the simple covariance designs such as OLS, WLSv and WLSs, the computation time for RoME with Huber's loss function is close to that for MinT; by contrast, the robust procedure with LAD loss function would take relatively more time.
Since the computation time for RoME with LAD loss function is less than 2 milliseconds on average under those simple designs, the extra computation cost is acceptable, compared with the increase in forecasting accuracy.

For the scale of hierarchy, as shown in Figure~\ref{Figure_EfficiencyLoss-time}, the computation time for three types of loss functions grows synchronously under the simple covariance designs; specifically, RoME with LAD loss function takes 3 or 4 times than MinT, and that with Huber's loss function reduces to be 2 or 3 times.
Since using LAD and Huber's loss functions under simple covariance designs may obtain nice reconciled results, there is not much need of introducing a complicate covariance design when they are employed.

\subsection{Implementation solutions to LAD loss function}\label{App_B2}

We evaluate the performances of two implementation solutions developed in Section~\ref{Section_Issue} through the numerical experiments in Section~\ref{Section_NonGaussian}, and Table~\ref{Table_Implementation-LAD} compares the forecasting accuracy and computation cost of them.

As shown in Table~\ref{Table_Implementation-LAD}, the reconciled results for RoME with LAD loss function by two implementation solutions are almost the same in forecasting accuracy and computation time in most cases.
However, when OLS covariance design is employed, i.e., for LAD-OLS of most interest, the computation time for (a) Perturbation would be significantly larger than that for (b) Huber approximation.
From the perspectives of the overall performance, this paper adopts the Huber approximation in implementation, and set $k = 10^{-4} \cdot \hat{\sigma}$, where $\hat{\sigma}$ is the estimated standard deviation of base forecast residuals.

\begin{table}[hp]
	\caption {\rm Comparison between two implementation solutions to RoME with LAD loss function for the whole hierarchy. The sub-columns ``$h = 1$", ``$1 \sim 6$" and ``$1 \sim 12$" denote the same as Table~\ref{Table_NonGaussian-mixture}, and ``Time" denotes the average computation time (in brackets in millisecond) of the related procedures.}
	\label{Table_Implementation-LAD}
	\setlength\tabcolsep{7pt}
	\renewcommand{\arraystretch}{1.2}
	\begin{center}
		\footnotesize
		\begin{tabular}{cccccccccccc}
			\toprule
			\multirow{2}{*}{$\rho$} & \multirow{2}{*}{$\bm{W}_h$} & & \multicolumn{4}{c}{(a) Perturbation} & & \multicolumn{4}{c}{(b) Huber approximation} \\
			\cline{4-7} \cline{9-12}
			& & & $h = 1$ & $1 \sim 6$ & $1 \sim 12$ & Time & & $h = 1$ & $1 \sim 6$ & $1 \sim 12$ & Time \\
			\midrule
			\multicolumn{12}{c}{Mixture normal distribution: $0.9 \cdot \mathcal{N} (0, 1) + 0.1 \cdot \mathcal{N} (0, 3^2)$} \\

			\multirow{5}[1]{*}{LAD} & OLS & & -0.748 & -1.693 & -2.406 & (0.379) & & -0.748 & -1.692 & -2.407 & (0.410) \\
			& WLSv & & -0.136 & -0.527 & -0.794 & (0.639) & & -0.136 & -0.527 & -0.794 & (0.646) \\
			& WLSs & & -0.136 & -0.527 & -0.794 & (0.683) & & -0.136 & -0.527 & -0.794 & (0.675) \\
			& Sample & & 38.007 & 60.603 & 71.807 & (29.012) & & 39.310 & 60.244 & 71.473 & (30.468) \\
			& Shrink & & 11.838 & 20.741 & 24.340 & (17.192) & & 11.668 & 20.859 & 24.447 & (18.119) \\
            \midrule
			\multicolumn{12}{c}{Standard \emph{t} distribution: $t (3)$} \\
			\multirow{5}[1]{*}{LAD} & OLS & & -0.825 & -1.798 & -2.370 & (0.358) & & -0.825 & -1.796 & -2.367 & (0.431) \\
			& WLSv & & -0.507 & -0.346 & -0.417 & (0.898) & & -0.507 & -0.347 & -0.417 & (0.934) \\
			& WLSs & & -0.545 & -0.295 & -0.303 & (0.695) & & -0.546 & -0.300 & -0.305 & (0.655) \\
			& Sample & & 37.633 & 67.726 & 78.038 & (28.471) & & 38.247 & 67.975 & 77.907 & (29.963) \\
			& Shrink & & 9.361 & 18.783 & 22.323 & (18.302) & & 9.323 & 18.769 & 22.299 & (19.209) \\
            \midrule
			\multicolumn{12}{c}{Standard Cauchy distribution$^a$: $\text{Cauchy} (0, 1)$} \\
			\multirow{3}[1]{*}{LAD} & OLS & & -2.620 & -1.963 & -1.401 & (0.400) & & -2.621 & -1.942 & -1.392 & (0.445) \\
			& WLSv & & -2.008 & -1.091 & -0.743 & (1.832) & & -2.010 & -1.086 & -0.753 & (1.863) \\
			& WLSs & & 0.079 & -0.619 & -0.618 & (0.759) & & 0.078 & -0.619 & -0.618 & (0.660) \\
			\bottomrule
			\multicolumn{12}{l}{$^a$ Results under Sample and Shrink covariance designs are not reported since they are extremely large.}
		\end{tabular}
	\end{center}
\end{table}

Conducting the Huber approximation avoids introducing the extra perturbation to $\mathbf{D}_h^{(\omega)}$, i.e., $\varrho = 10^{-8}$, which would be another advantage of this solution.

\subsection{Comparison for the misspecification of model}\label{App_B3}

\begin{figure}[htbp]
	\begin{center}
		\subfigure[Bottom level: $h = 1$, $1 \sim 6$ and $1 \sim 12$.]{
			\begin{minipage}[t]{15.6cm}
				\includegraphics[width = 5cm]{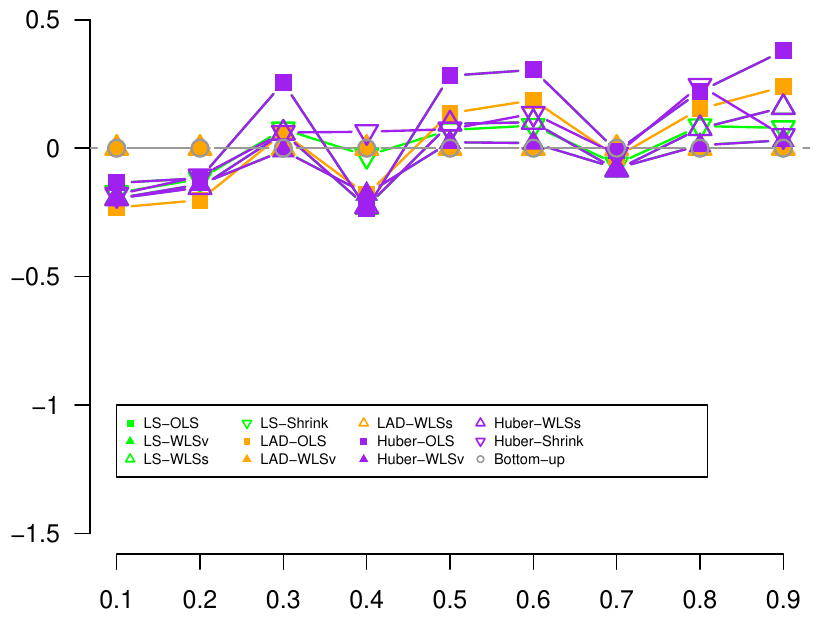}
				\
				\includegraphics[width = 5cm]{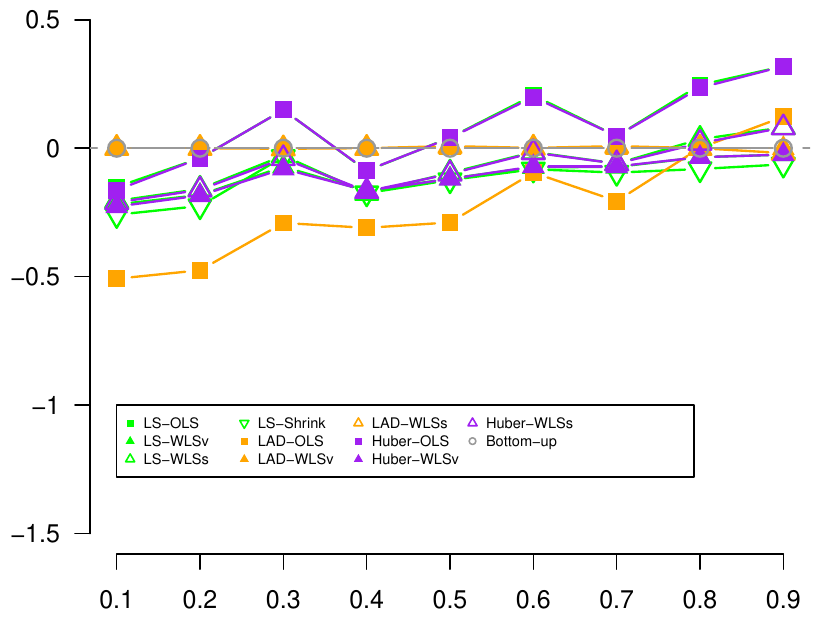}
				\
				\includegraphics[width = 5cm]{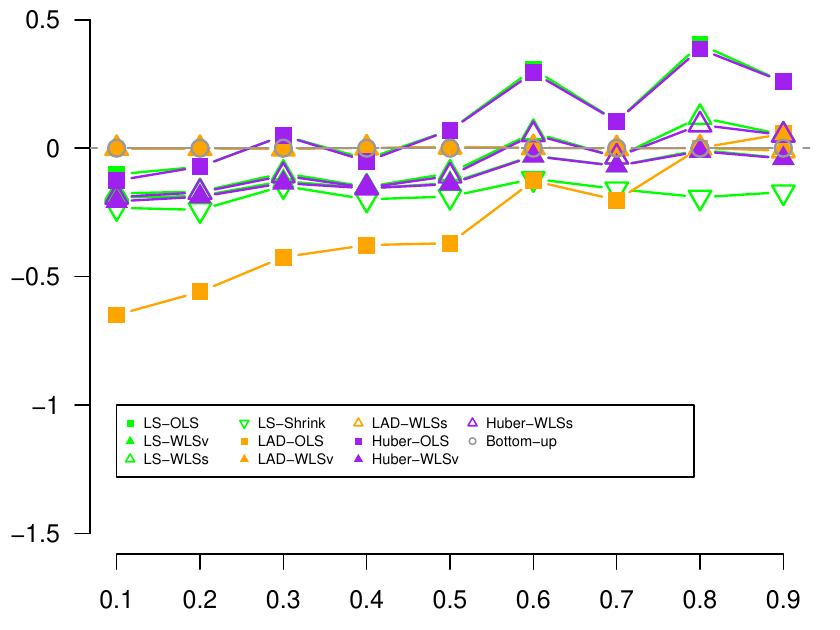}
			\end{minipage}
		}
	\end{center}
	
	\begin{center}
		\subfigure[Aggregated levels: $h = 1$, $1 \sim 6$ and $1 \sim 12$.]{
			\begin{minipage}[t]{15.6cm}
				\includegraphics[width = 5cm]{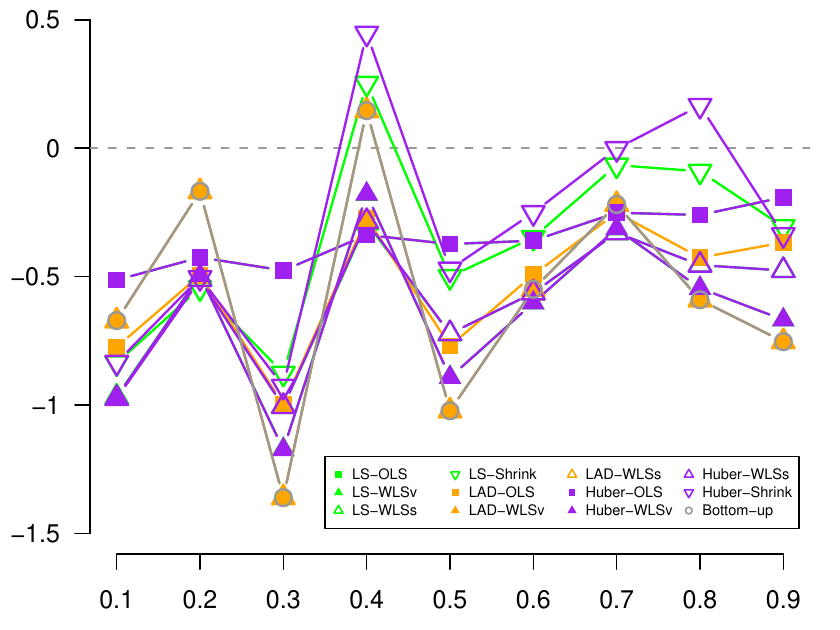}
				\
				\includegraphics[width = 5cm]{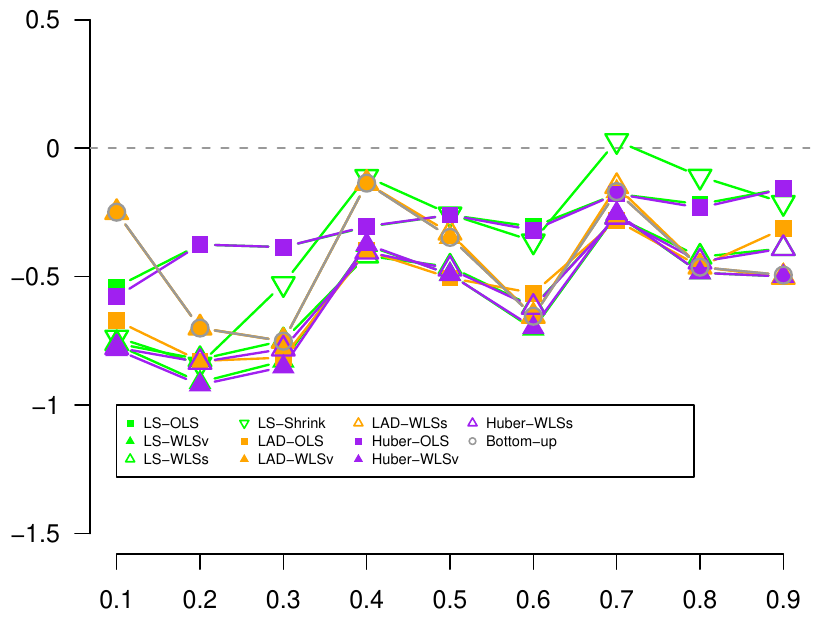}
				\
				\includegraphics[width = 5cm]{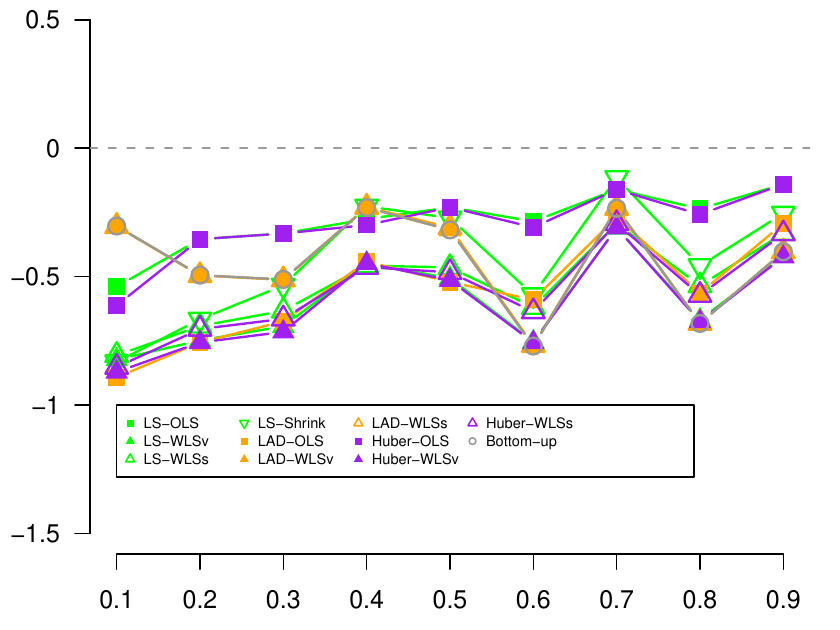}
			\end{minipage}
		}
	\end{center}
	
	\begin{center}
		\subfigure[Whole hierarchy: $h = 1$, $1 \sim 6$ and $1 \sim 12$.]{
			\begin{minipage}[t]{15.6cm}
				\includegraphics[width = 5cm]{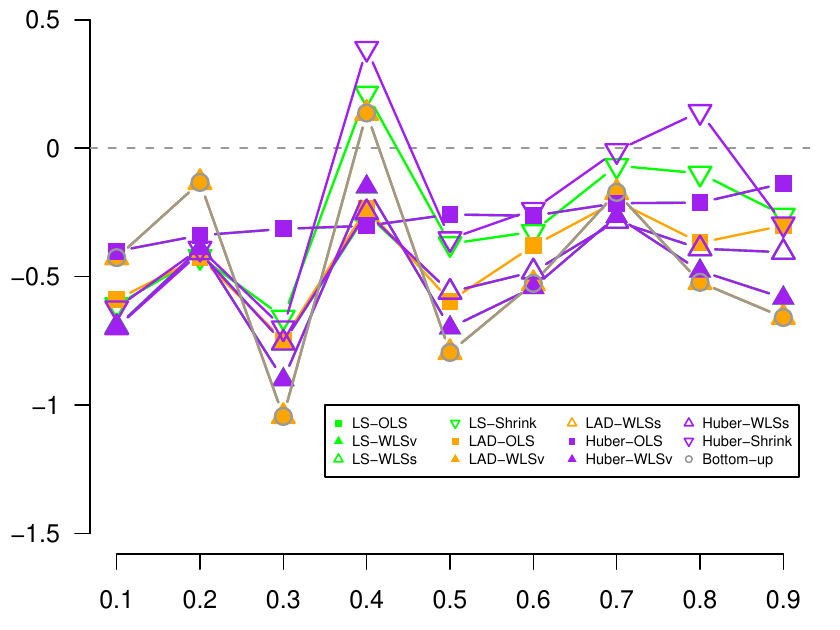}
				\
				\includegraphics[width = 5cm]{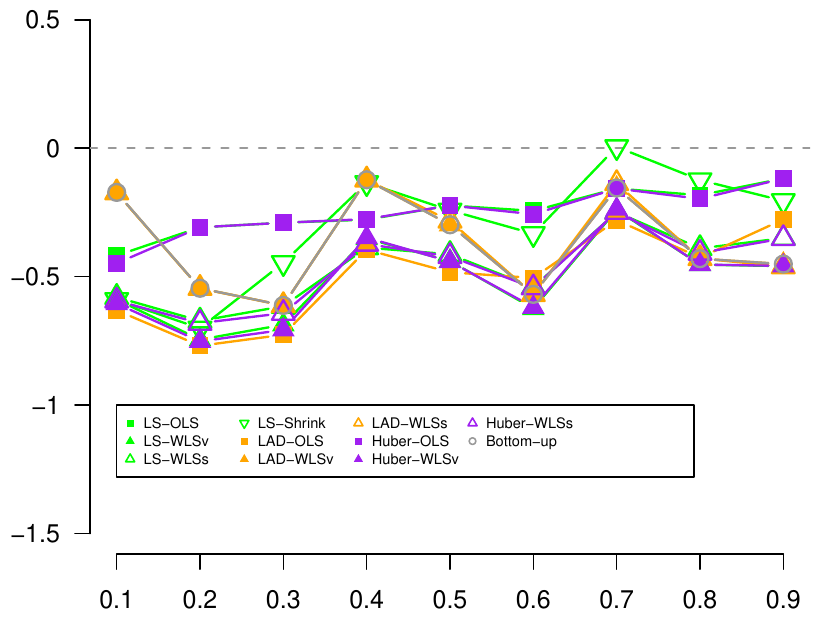}
				\
				\includegraphics[width = 5cm]{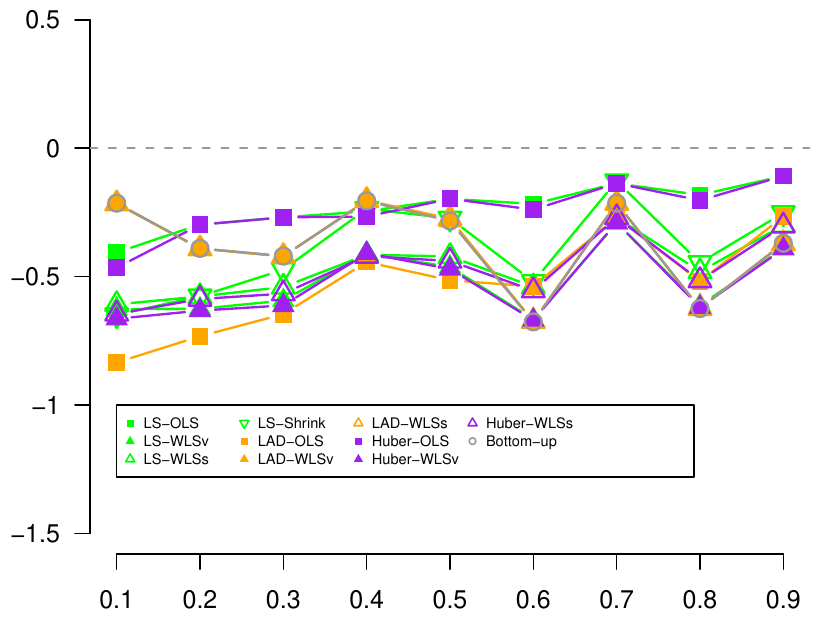}
			\end{minipage}
		}
	\end{center}
	\caption{\rm Line charts for the percentage increases by RoME from average RMSE of the base forecasts obtained by ARIMA against the correlation among HTS in Section~\ref{Section_Correlation}. The panels in the sub-figures, as well as the aggregated levels wherein, and the horizontal axis in every single figure denotes the same as Figure~\ref{Figure_Correlation-results-ETS}. The lines exceeding the upper limit of the vertical axis, i.e., 0.5\%, are not displayed for concentrating the valuable information on the others.}
	\label{Figure_Correlation-results-ARIMA}
\end{figure}

Figure~\ref{Figure_Correlation-results-ARIMA} report the results for reconciled forecasts under different correlations among HTS in Section~\ref{Section_Correlation}, where the base forecasts are obtained by the auto-ARIMA algorithm.

\subsection{Combination forecasts}\label{App_B4}

Let $\pi_{\rm LS} (h)$ and $\pi_{\rm LAD} (h)$ be the combination weights for the reconciled forecasts obtained by using LS and LAD loss functions, respectively, which satisfies $\pi_{\rm LS} (h) + \pi_{\rm LAD} (h) = 1$, we consider three patterns of these weights.
Specifically, for an $H$-term out-of-sample forecasting, they are defined for $h = 1, 2, \cdots, H$ as
    \begin{enumerate}[{\ (a)}]
        \setlength{\itemsep}{0pt}
        \setlength{\parsep}{0pt}
        \setlength{\parskip}{0pt}
        \item Average: $\pi_{\rm LS} (h) = \pi_{\rm LAD} (h) = 0.5$;
        \item One-way: $\pi_{\rm LS} (h) = 1 / (h + 1)$, and $\pi_{\rm LAD} (h) = h / (h + 1)$;
        \item Two-way: $\pi_{\rm LS} (h) = (H + 1 - h) / (H + 1)$, and $\pi_{\rm LAD} (h) = h / (H + 1)$.
    \end{enumerate}
The aforementioned patterns of combination weights are motivated by the simulation result that MinT with LS loss function is more suitable for short-term forecasting, while RoME with LAD loss function behaves better for the mid- and long-term ones.
Tables \ref{Table_Combination-mixture}--\ref{Table_Combination-cauchy} report the results for the numerical experiments in Section~\ref{Section_NonGaussian} under those patterns, where the reconciled forecasts for the related single procedures with LS and LAD loss functions are shown for comparison.

As shown in Tables \ref{Table_Combination-mixture}--\ref{Table_Combination-cauchy}, the combined results in some cases may be better than those by a single procedure, especially for the ``One-way" pattern that realizes the best average increase in 12 out of 27 groups (3 subsets of HTS $\times$ 3 ahead-forecasting periods $\times$ 3 non-normal distributions) of the alternative models.

\begin{table}[htbp]
	\caption {\rm Out-of-sample forecasting performances of RoME with the base forecasts obtained by ARIMA for three patterns of combination weights in Appendix~\ref{App_B2} under the mixture normal distribution in Section~ \ref{Section_NonGaussian}. The rows, columns and sub-columns denote the same as Table~\ref{Table_NonGaussian-mixture}. The best results are highlighted in bold.}
	\label{Table_Combination-mixture}
	\setlength\tabcolsep{4pt}
	\renewcommand{\arraystretch}{1}
	\begin{center}
		\footnotesize
		\begin{tabular}{cccccccccccccc}
			\toprule
			\multirow{2}{*}{Pattern} & \multirow{2}{*}{$\bm{W}_h$} & & \multicolumn{3}{c}{Changeable} & & \multicolumn{3}{c}{Stable} & & \multicolumn{3}{c}{All} \\
			\cline{4-6} \cline{8-10} \cline{12-14}
			& & & $h = 1$ & $1 \sim 6$ & $1 \sim 12$ & & $h = 1$ & $1 \sim 6$ & $1 \sim 12$ & & $h = 1$ & $1 \sim 6$ & $1 \sim 12$ \\
			\midrule
			\multirow{5}[1]{*}{Average} & OLS & & -0.927 & -1.550 & -1.748 & & -0.499 & -0.976 & -1.397 & & \textbf{-0.923} & -1.562 & -1.993 \\
			& WLSv & & -0.205 & -0.901 & -0.903 & & -0.237 & -0.861 & -1.049 & & -0.497 & -0.954 & -1.298 \\
			& WLSs & & -0.165 & -0.850 & -0.843 & & -0.306 & -0.897 & -1.100 & & -0.561 & -1.004 & -1.345 \\
			& Sample & & 17.472 & 16.750 & 18.075 & & 13.398 & 17.374 & 19.757 & & 17.746 & 23.069 & 26.224 \\
			& Shrink & & 4.448 & 3.852 & 3.729 & & 3.119 & 3.737 & 4.471 & & 4.693 & 6.528 & 6.839 \\
            \midrule
			\multirow{5}[1]{*}{One-way} & OLS & & -0.927 & \textbf{-1.598} & \textbf{-1.866} & & -0.499 & -1.234 & -1.796 & & \textbf{-0.923} & \textbf{-1.669} & -2.232 \\
			& WLSv & & -0.205 & -0.283 & -0.005 & & -0.237 & -0.643 & -0.675 & & -0.497 & -0.465 & -0.583 \\
			& WLSs & & -0.165 & -0.261 & 0.011 & & -0.306 & -0.681 & -0.709 & & -0.561 & -0.502 & -0.612 \\
			& Sample & & 17.472 & 35.188 & 42.938 & & 13.398 & 37.477 & 48.684 & & 17.746 & 47.676 & 61.805 \\
			& Shrink & & 4.448 & 10.870 & 13.181 & & 3.119 & 10.701 & 15.147 & & 4.693 & 16.223 & 21.039 \\
            \midrule
			\multirow{5}[1]{*}{Two-way} & OLS & & -0.893 & -1.462 & -1.805 & & -0.310 & -0.798 & -1.529 & & -0.812 & -1.472 & -2.087 \\
			& WLSv & & -0.906 & -1.121 & -0.583 & & -0.445 & -0.952 & -0.916 & & -0.830 & -1.154 & -1.063 \\
			& WLSs & & -0.847 & -1.055 & -0.539 & & -0.480 & -0.957 & -0.950 & & -0.884 & -1.190 & -1.099 \\
			& Sample & & 1.929 & 9.500 & 26.875 & & 2.006 & 9.288 & 29.868 & & 1.693 & 12.745 & 38.784 \\
			& Shrink & & -0.804 & 1.432 & 6.930 & & -0.224 & 1.270 & 8.279 & & -0.587 & 2.856 & 11.720 \\
            \midrule
			\multirow{5}[1]{*}{LS} & OLS & & -0.863 & -1.126 & -1.078 & & -0.259 & -0.246 & -0.396 & & -0.776 & -1.110 & -1.269 \\
			& WLSv & & \textbf{-0.987} & -1.423 & -1.490 & & -0.470 & -1.027 & -1.275 & & -0.869 & -1.412 & -1.764 \\
			& WLSs & & -0.927 & -1.333 & -1.381 & & -0.490 & -0.906 & -1.143 & & -0.914 & -1.377 & -1.693 \\
			& Sample & & 0.930 & 0.567 & 0.933 & & 1.544 & 0.375 & 0.038 & & 0.820 & 0.208 & -0.114 \\
			& Shrink & & -1.093 & -1.463 & -1.568 & & -0.336 & -1.231 & -1.643 & & -0.852 & -1.565 & -2.060 \\
            \midrule
			\multirow{3}[1]{*}{LAD} & OLS & & -0.677 & -1.541 & -1.828 & & \textbf{-0.517} & \textbf{-1.320} & \textbf{-1.862} & & -0.860 & -1.653 & \textbf{-2.244} \\
			& WLSv & & 1.148 & 0.257 & 0.386 & & 0.169 & -0.424 & -0.499 & & 0.154 & -0.022 & -0.254 \\
			& WLSs & & 1.148 & 0.258 & 0.386 & & 0.169 & -0.425 & -0.499 & & 0.154 & -0.022 & -0.254 \\
            \midrule
			\multirow{3}[1]{*}{Huber} & OLS & & -0.863 & -1.143 & -1.157 & & -0.259 & -0.269 & -0.489 & & -0.776 & -1.129 & -1.354 \\
			& WLSv & & -0.987 & -1.411 & -1.522 & & -0.470 & -1.068 & -1.352 & & -0.869 & -1.388 & -1.815 \\
			& WLSs & & -0.927 & -1.323 & -1.409 & & -0.490 & -0.963 & -1.265 & & -0.914 & -1.361 & -1.769 \\
			\midrule
			\multicolumn{2}{c}{Bottom-up} & & 1.150 & 0.258 & 0.386 & & 0.170 & -0.424 & -0.499 & & 0.155 & -0.021 & -0.253 \\
			\midrule
			\multicolumn{2}{c}{Base} & & 0.714 & 1.156 & 1.300 & & 0.678 & 1.027 & 1.146 & & 0.880 & 1.344 & 1.492 \\
			\bottomrule
		\end{tabular}
	\end{center}
\end{table}

\begin{table}[htbp]
	\caption {\rm Out-of-sample forecasting performances of RoME with the base forecasts obtained by ARIMA for three patterns of combination weights in Appendix~\ref{App_B2} under the standard $t$ distribution in Section~ \ref{Section_NonGaussian}. The rows, columns and sub-columns denote the same as Table~\ref{Table_NonGaussian-mixture}. The best results are highlighted in bold.}
	\label{Table_Combination-t}
	\setlength\tabcolsep{4pt}
	\renewcommand{\arraystretch}{0.9}
	\begin{center}
		\footnotesize
		\begin{tabular}{cccccccccccccc}
			\toprule
			\multirow{2}{*}{Pattern} & \multirow{2}{*}{$\bm{W}_h$} & & \multicolumn{3}{c}{Changeable} & & \multicolumn{3}{c}{Stable} & & \multicolumn{3}{c}{All} \\
			\cline{4-6} \cline{8-10} \cline{12-14}
			& & & $h = 1$ & $1 \sim 6$ & $1 \sim 12$ & & $h = 1$ & $1 \sim 6$ & $1 \sim 12$ & & $h = 1$ & $1 \sim 6$ & $1 \sim 12$ \\
			\midrule
			\multirow{5}[1]{*}{Average} & OLS & & -1.027 & -1.595 & -1.736 & & -0.487 & -1.134 & -1.584 & & -0.954 & -1.634 & -2.050 \\
			& WLSv & & -1.260 & -0.792 & -0.604 & & -0.587 & -0.676 & -0.929 & & -0.734 & -0.864 & -1.096 \\
			& WLSs & & \textbf{-1.299} & -0.712 & -0.496 & & -0.617 & -0.782 & -1.044 & & -0.789 & -0.899 & -1.093 \\
			& Sample & & 19.576 & 22.271 & 22.313 & & 13.022 & 20.756 & 22.828 & & 17.219 & 27.649 & 30.560 \\
			& Shrink & & 1.513 & 3.067 & 3.043 & & 2.317 & 4.247 & 4.705 & & 3.348 & 5.588 & 6.284 \\
            \midrule
			\multirow{5}[1]{*}{One-way} & OLS & & -1.027 & -1.775 & -2.007 & & -0.487 & -1.288 & -1.885 & & \textbf{-0.954} & \textbf{-1.737} & \textbf{-2.216} \\
			& WLSv & & -1.260 & -0.213 & 0.293 & & -0.587 & -0.374 & -0.508 & & -0.734 & -0.336 & -0.307 \\
			& WLSs & & \textbf{-1.299} & -0.105 & 0.435 & & -0.617 & -0.435 & -0.560 & & -0.789 & -0.315 & -0.214 \\
			& Sample & & 19.576 & 44.752 & 51.987 & & 13.022 & 42.033 & 52.921 & & 17.219 & 55.542 & 69.284 \\
			& Shrink & & 1.513 & 9.955 & 12.329 & & 2.317 & 10.963 & 14.602 & & 3.348 & 14.735 & 19.522 \\
            \midrule
			\multirow{5}[1]{*}{Two-way} & OLS & & -0.567 & -1.426 & -1.807 & & -0.277 & -0.984 & -1.705 & & -0.811 & -1.527 & -2.110 \\
			& WLSv & & -1.189 & -1.009 & -0.234 & & -0.665 & -0.772 & -0.802 & & -0.939 & -1.057 & -0.811 \\
			& WLSs & & -1.289 & -0.943 & -0.113 & & -0.628 & -0.857 & -0.894 & & -0.948 & -1.088 & -0.772 \\
			& Sample & & 2.871 & 13.142 & 32.881 & & 1.834 & 12.367 & 33.405 & & 1.748 & 16.228 & 44.373 \\
			& Shrink & & -1.203 & 0.860 & 6.425 & & -0.435 & 1.950 & 8.183 & & -0.714 & 2.418 & 11.014 \\
            \midrule
			\multirow{5}[1]{*}{LS} & OLS & & -0.455 & -0.916 & -0.880 & & -0.225 & -0.525 & -0.661 & & -0.771 & -1.162 & -1.365 \\
			& WLSv & & -1.131 & -1.216 & -1.116 & & \textbf{-0.668} & -0.966 & -1.241 & & \textbf{-0.954} & -1.333 & -1.619 \\
			& WLSs & & -1.246 & -1.220 & -1.129 & & -0.610 & -0.970 & -1.219 & & -0.950 & -1.358 & -1.622 \\
			& Sample & & 2.395 & 1.433 & 1.216 & & 1.323 & 1.080 & 0.889 & & 0.985 & 0.903 & 0.786 \\
			& Shrink & & -0.932 & -1.338 & -1.402 & & -0.552 & -1.043 & -1.393 & & -0.856 & -1.408 & -1.779 \\
            \midrule
			\multirow{3}[1]{*}{LAD} & OLS & & -1.201 & \textbf{-1.804} & \textbf{-2.036} & & -0.559 & \textbf{-1.304} & \textbf{-1.902} & & -0.948 & -1.725 & -2.207 \\
			& WLSv & & -0.767 & 0.321 & 0.679 & & -0.362 & -0.141 & -0.324 & & -0.219 & 0.114 & 0.039 \\
			& WLSs & & -0.775 & 0.432 & 0.820 & & -0.362 & -0.141 & -0.324 & & -0.268 & 0.182 & 0.174 \\
            \midrule
			\multirow{3}[1]{*}{Huber} & OLS & & -0.455 & -0.950 & -0.933 & & -0.225 & -0.551 & -0.748 & & -0.771 & -1.190 & -1.440 \\
			& WLSv & & -1.133 & -1.320 & -1.238 & & \textbf{-0.668} & -0.973 & -1.242 & & \textbf{-0.954} & -1.366 & -1.635 \\
			& WLSs & & -1.246 & -1.282 & -1.188 & & -0.610 & -1.013 & -1.269 & & -0.950 & -1.396 & -1.634 \\
			\midrule
			\multicolumn{2}{c}{Bottom-up} & & -0.774 & 0.433 & 0.820 & & -0.362 & -0.140 & -0.324 & & -0.267 & 0.182 & 0.175 \\
			\midrule
			\multicolumn{2}{c}{Base} & & 0.821 & 1.400 & 1.611 & & 0.670 & 1.015 & 1.131 & & 0.929 & 1.451 & 1.632 \\
			\bottomrule
		\end{tabular}
	\end{center}
\end{table}

\begin{table}[htbp]
	\caption {\rm Out-of-sample forecasting performances of RoME with the base forecasts obtained by ARIMA for three patterns of combination weights in Appendix~\ref{App_B2} under the standard Cauchy distribution in Section~\ref{Section_NonGaussian}. The rows, columns and sub-columns denote the same as Table~\ref{Table_NonGaussian-mixture}. The best results are highlighted in bold.}
	\label{Table_Combination-cauchy}
	\setlength\tabcolsep{4pt}
	\renewcommand{\arraystretch}{1.2}
	\begin{center}
		\footnotesize
		\begin{tabular}{cccccccccccccc}
			\toprule
			\multirow{2}{*}{Pattern} & \multirow{2}{*}{$\bm{W}_h$} & & \multicolumn{3}{c}{Changeable} & & \multicolumn{3}{c}{Stable} & & \multicolumn{3}{c}{All} \\
			\cline{4-6} \cline{8-10} \cline{12-14}
			& & & $h = 1$ & $1 \sim 6$ & $1 \sim 12$ & & $h = 1$ & $1 \sim 6$ & $1 \sim 12$ & & $h = 1$ & $1 \sim 6$ & $1 \sim 12$ \\
			\midrule
			\multirow{3}[1]{*}{Average} & OLS & & \textbf{-4.928} & -3.832 & -3.765 & & 2.787 & 6.070 & 7.930 & & -4.449 & -3.942 & -3.808 \\
			& WLSv & & -0.085 & -0.779 & -0.872 & & -0.507 & -0.548 & -0.736 & & -1.824 & -2.232 & -2.192 \\
			& WLSs & & -3.688 & -2.844 & -2.551 & & 1.667 & 2.736 & 3.863 & & -2.142 & -2.479 & -2.350 \\
            \midrule
			\multirow{3}[1]{*}{One-way} & OLS & & \textbf{-4.928} & \textbf{-4.225} & \textbf{-4.206} & & 2.787 & 1.226 & 0.479 & & -4.449 & \textbf{-4.332} & \textbf{-4.295} \\
			& WLSv & & -0.085 & 0.188 & 0.265 & & -0.507 & -0.320 & -0.434 & & -1.824 & -1.332 & -1.206 \\
			& WLSs & & -3.688 & -1.714 & -1.264 & & 1.667 & 0.482 & 0.174 & & -2.142 & -0.969 & -0.587 \\
            \midrule
			\multirow{3}[1]{*}{Two-way} & OLS & & -3.791 & -3.408 & -3.825 & & 7.831 & 9.369 & 6.077 & & -3.297 & -3.530 & -3.914 \\
			& WLSv & & -1.242 & -1.014 & -0.436 & & -0.624 & -0.630 & -0.667 & & -2.681 & -2.482 & -1.848 \\
			& WLSs & & -4.850 & -3.085 & -2.030 & & 5.164 & 4.602 & 2.970 & & -3.648 & -2.870 & -1.662 \\
            \midrule
			\multirow{3}[1]{*}{LS} & OLS & & -3.463 & -2.113 & -1.956 & & 8.854 & 16.767 & 21.093 & & -3.001 & -2.360 & -2.113 \\
			& WLSv & & -1.326 & -1.231 & -1.377 & & -0.636 & -0.772 & -0.958 & & -2.738 & -2.745 & -2.615 \\
			& WLSs & & -4.861 & -3.275 & -2.945 & & 5.903 & 8.831 & 11.505 & & -3.726 & -3.283 & -3.056 \\
            \midrule
			\multirow{3}[1]{*}{LAD} & OLS & & -4.627 & -4.127 & -4.126 & & \textbf{-0.651} & -0.708 & \textbf{-1.009} & & \textbf{-4.538} & -4.302 & -4.260 \\
			& WLSv & & 2.909 & 1.135 & 0.840 & & -0.252 & -0.137 & -0.293 & & 0.367 & -0.482 & -0.705 \\
			& WLSs & & 0.073 & -0.652 & -0.640 & & -0.269 & -0.151 & -0.296 & & 1.865 & 0.422 & 0.252 \\
            \midrule
			\multirow{3}[1]{*}{Huber} & OLS & & -3.456 & -2.228 & -2.152 & & 8.661 & 12.051 & 13.849 & & -2.997 & -2.511 & -2.346 \\
			& WLSv & & -1.367 & -1.412 & -1.588 & & -0.633 & \textbf{-0.774} & -0.969 & & -2.810 & -2.812 & -2.719 \\
			& WLSs & & -4.858 & -3.342 & -2.982 & & 5.268 & 5.558 & 6.183 & & -3.702 & -3.260 & -3.020 \\
			\midrule
			\multicolumn{2}{c}{Bottom-up} & & 0.142 & -0.637 & -0.636 & & -0.226 & -0.108 & -0.277 & & 1.911 & 0.438 & 0.257 \\
			\midrule
			\multicolumn{2}{c}{Base} & & 9.564 & 28.426 & 46.187 & & 0.667 & 0.998 & 1.123 & & 6.635 & 19.393 & 31.439 \\
			\bottomrule
		\end{tabular}
	\end{center}
\end{table}

These simulation results show the potential of forecast combination for RoME.
By taking into account the strengths of RoME with LS (i.e., MinT) and LAD loss functions in short-, mid- and long-term forecasting, we may improve the forecasting accuracy by combining the reconciled forecasts obtained by single procedures.

\subsection{Results for real-data study}\label{App_B5}

Tables~\ref{ETS-FeReco} and~\ref{ARIMA-FeReco} represent results of MinT forecasts (denoted as LS) obtained using the \emph{hts} package and FeReCo forecasts produced by the \emph{FeReco} package.
Tables~\ref{Table_Tourism-ETS-full} and~\ref{Table_Tourism-ARIMA-full} present the results of the reconciled forecasts obtained using RoME and MinT, with ETS and ARIMA base forecasts, respectively, for the Australian Domestic Tourism dataset under the cross-sectional hierarchical structure.

\begin{table}[htbp]
  \centering
  \caption {\rm Comparison of MinT (LS) forecasts from the \emph{hts} package and FoReCo forecasts from the \emph{RoReco} package, with ETS base forecasts.}
	\label{ETS-FeReco}
	\setlength\tabcolsep{4pt}
	\renewcommand{\arraystretch}{1.2}
    \footnotesize
	\renewcommand{\arraystretch}{1}
    \begin{tabular}{cccccccccc}
    \toprule
    \multirow{2}[4]{*}{$\rho$} & \multirow{2}[4]{*}{$\bm{W}_h$} & \multicolumn{4}{c}{Australia} & \multicolumn{4}{c}{States} \\
\cmidrule{3-10}          &       & $h$=1   & $h$=6   & $1\sim 6$   & $1\sim 12$  & $h$=1   & $h$=6   & $1\sim 6$   & $1\sim 12$ \\
    \midrule
    \multirow{5}[2]{*}{FoReco} & OLS   & -0.45 & -0.40 & -0.43 & -0.39 & -0.49 & -0.44 & 0.54  & 0.22 \\
          & WLSv  & 1.54  & -0.48 & -0.97 & -1.26 & 0.58  & 1.86  & 2.10  & 0.45 \\
          & WLSs  & 0.28  & -1.07 & -1.24 & -1.32 & -0.20 & 0.83  & 1.35  & 0.13 \\
          & Sample & 117.84 & 211.68 & 116.05 & 152.53 & 92.10 & 174.65 & 108.20 & 150.01 \\
          & Shrink & -6.34 & 9.25  & 0.85  & 7.07  & -9.10 & 16.83 & 5.99  & 10.72 \\
    \midrule
    \multirow{5}[2]{*}{LS} & OLS   & -0.45 & -0.40 & -0.43 & -0.39 & -0.49 & -0.44 & 0.54  & 0.22 \\
          & WLSv  & 1.54  & -0.48 & -0.97 & -1.26 & 0.58  & 1.86  & 2.10  & 0.45 \\
          & WLSs  & 0.28  & -1.07 & -1.24 & -1.32 & -0.20 & 0.83  & 1.35  & 0.13 \\
          & Sample & 3.99  & 4.01  & 1.37  & 1.85  & 2.83  & 7.66  & 4.96  & 4.53 \\
          & Shrink & -6.34 & 9.25  & 0.85  & 7.07  & -9.10 & 16.83 & 5.99  & 10.72 \\
    \midrule
    \multirow{2}[4]{*}{$\rho$} & \multirow{2}[4]{*}{$\bm{W}_h$} & \multicolumn{4}{c}{Zones}     & \multicolumn{4}{c}{Regions} \\
\cmidrule{3-10}          &       & $h$=1   & $h$=6   & $1\sim 6$   & $1\sim 12$  & $h$=1   & $h$=6   & $1\sim 6$   & $1\sim 12$  \\
    \midrule
    \multirow{5}[2]{*}{FoReco} & OLS   & -0.36 & -1.59 & -0.52 & -0.65 & -1.31 & -0.95 & -1.08 & -1.14 \\
          & WLSv  & -0.66 & -2.26 & -0.81 & -1.25 & -1.71 & -1.62 & -1.38 & -1.73 \\
          & WLSs  & -1.04 & -1.77 & -0.89 & -1.06 & -0.89 & -0.51 & -0.57 & -0.68 \\
          & Sample & 81.44 & 144.03 & 95.42 & 137.67 & 59.51 & 110.10 & 73.70 & 112.35 \\
          & Shrink & -8.03 & 8.76  & 2.24  & 6.77  & -11.36 & 0.07  & -3.01 & -0.14 \\
    \midrule
    \multirow{5}[2]{*}{LS} & OLS   & -0.36 & -1.59 & -0.52 & -0.65 & -1.31 & -0.95 & -1.08 & -1.14 \\
          & WLSv  & -0.66 & -2.26 & -0.81 & -1.25 & -1.71 & -1.62 & -1.38 & -1.73 \\
          & WLSs  & -1.04 & -1.77 & -0.89 & -1.06 & -0.89 & -0.51 & -0.57 & -0.68 \\
          & Sample & 0.52  & 1.90  & 1.24  & 2.34  & -0.13 & 2.19  & 0.99  & 2.28 \\
          & Shrink & -8.03 & 8.76  & 2.24  & 6.77  & -11.36 & 0.07  & -3.01 & -0.14 \\
    \bottomrule
    \end{tabular}
\end{table}

\begin{table}[htbp]
  \centering
  \caption {\rm Comparison of MinT (LS) forecasts from the \emph{hts} package and FoReCo forecasts from the \emph{FoReco} package, with ARIMA base forecasts.}
	\label{ARIMA-FeReco}
	\setlength\tabcolsep{4pt}
	\renewcommand{\arraystretch}{1.2}
    \footnotesize
	\renewcommand{\arraystretch}{1}
    \begin{tabular}{cccccccccc}
    \toprule
    \multirow{2}[4]{*}{$\rho$} & \multirow{2}[4]{*}{$\bm{W}_h$} & \multicolumn{4}{c}{Australia} & \multicolumn{4}{c}{States} \\
\cmidrule{3-10}          &       & $h$=1   & $h$=6   & $1\sim 6$   & $1\sim 12$  & $h$=1   & $h$=6   & $1\sim 6$   & $1\sim 12$  \\
    \midrule
    \multirow{5}[1]{*}{FoReco} & OLS   & -3.47 & -0.74 & -1.39 & -0.64 & 1.39  & 0.33  & -0.08 & 0.39 \\
          & WLSv  & -13.89 & -2.07 & -5.04 & -2.78 & -8.90 & -0.89 & -4.07 & -1.88 \\
          & WLSs  & -12.24 & -2.15 & -4.64 & -2.32 & -7.00 & -0.73 & -3.40 & -1.33 \\
          & Sample & -14.01 & 3.59  & -7.72 & -8.55 & -14.19 & -1.66 & -9.93 & -8.40 \\
          & Shrink & -18.64 & -5.90 & -13.42 & -10.03 & -18.87 & -8.19 & -14.59 & -10.04 \\
          \midrule
    \multirow{5}[1]{*}{LS} & OLS   & -3.47 & -0.74 & -1.39 & -0.64 & 1.39  & 0.33  & -0.08 & 0.39 \\
          & WLSv  & -13.89 & -2.07 & -5.04 & -2.78 & -8.90 & -0.89 & -4.07 & -1.88 \\
          & WLSs  & -12.24 & -2.15 & -4.64 & -2.32 & -7.00 & -0.73 & -3.40 & -1.33 \\
          & Sample & -14.36 & 3.56  & -8.24 & -7.15 & -12.86 & 0.27  & -8.41 & -6.34 \\
          & Shrink & -18.64 & -5.90 & -13.42 & -10.03 & -18.87 & -8.19 & -14.59 & -10.04 \\
    \midrule
    \multirow{2}[4]{*}{$\rho$} & \multirow{2}[4]{*}{$\bm{W}_h$} & \multicolumn{4}{c}{Zones}     & \multicolumn{4}{c}{Regions} \\
\cmidrule{3-10}          &       & $h$=1   & $h$=6   & $1\sim 6$   & $1\sim 12$  & $h$=1   & $h$=6   & $1\sim 6$   & $1\sim 12$  \\
    \midrule
    \multirow{5}[1]{*}{FoReco} & OLS   & 3.28  & -0.25 & 0.56  & 0.53  & 4.88  & -0.02 & 1.42  & 0.75 \\
          & WLSv  & -2.39 & -1.42 & -1.36 & -0.71 & -0.37 & -0.85 & -0.79 & -0.55 \\
          & WLSs  & -1.05 & -1.05 & -0.90 & -0.33 & 0.31  & -0.49 & -0.34 & -0.21 \\
          & Sample & -7.59 & -2.74 & -7.78 & -7.22 & -4.50 & -2.10 & -5.55 & -5.77 \\
          & Shrink & -12.32 & -7.94 & -11.84 & -9.03 & -9.17 & -6.61 & -9.68 & -7.78 \\
          \midrule
    \multirow{5}[1]{*}{LS} & OLS   & 3.28  & -0.25 & 0.56  & 0.53  & 4.88  & -0.02 & 1.42  & 0.75 \\
          & WLSv  & -2.39 & -1.42 & -1.36 & -0.71 & -0.37 & -0.85 & -0.79 & -0.55 \\
          & WLSs  & -1.05 & -1.05 & -0.90 & -0.33 & 0.31  & -0.49 & -0.34 & -0.21 \\
          & Sample & -7.61 & -1.84 & -6.83 & -5.74 & -4.69 & -1.07 & -4.47 & -4.34 \\
          & Shrink & -12.32 & -7.94 & -11.84 & -9.03 & -9.17 & -6.61 & -9.68 & -7.78 \\
    \bottomrule
    \end{tabular}

\end{table}

\begin{table}[htbp]
    \centering
	\caption {\rm Forecasting results of the proposed RoME reconciliation through ETS in the real-data study in Section~\ref{sec:cross}. The best results are highlighted in bold.}
	\label{Table_Tourism-ETS-full}
	\setlength\tabcolsep{3pt}
	\renewcommand{\arraystretch}{1.25}
    \footnotesize
	\scalebox{0.6}{
    \begin{tabular}{cccccccccccccccccccc}
    \toprule
    \multirow{2}[4]{*}{$\rho$} & \multirow{2}[4]{*}{$\bm{W}_h$} & \multicolumn{4}{c}{Australia} & \multicolumn{4}{c}{Australia by purpose of travel} & \multirow{2}[4]{*}{$\rho$}& \multirow{2}[4]{*}{$\bm{W}_h$}& \multicolumn{4}{c}{States}    & \multicolumn{4}{c}{States by purpose of travel} \\
\cmidrule{3-10}\cmidrule{13-20}          &       & $h=1$   & $h=6$   & $1 \sim 6$   & $1 \sim 12$  & $h=1$   & $h=6$   & $1 \sim 6$   & $1 \sim 12$  &       &       & $h=1$   & $h=6$   & $1 \sim 6$   & $1 \sim 12$  & $h=1$   & $h=6$   & $1 \sim 6$   & $1 \sim 12$2 \\
    \midrule
    \multirow{5}[1]{*}{LS} & OLS   & -0.69 & -0.74 & -0.51 & -0.49 & 0.78  & 0.77  & 0.51  & 0.46  & \multirow{5}[1]{*}{LS} & OLS   & -1.38 & -2.27 & -0.71 & -1.00 & -0.72 & -1.43 & -0.46 & -0.69 \\
          & WLSv  & -0.40 & -0.71 & -0.79 & -0.82 & 2.04  & 2.11  & 0.73  & 0.85  &       & WLSv  & -1.73 & -1.54 & -0.37 & -1.38 & -1.36 & -0.60 & -0.33 & -0.87 \\
          & WLSs  & -0.35 & -1.02 & -0.58 & -0.82 & 1.38  & 1.59  & 0.50  & 0.63  &       & WLSs  & -1.58 & -1.83 & -0.30 & -1.30 & -1.34 & -1.20 & -0.63 & -1.04 \\
          & Sample & 193.82 & 867.39 & 395.23 & 669.73 & 188.52 & 840.87 & 407.47 & 698.52 &       & Sample & 196.14 & 825.97 & 427.47 & 743.85 & 197.69 & 815.14 & 434.91 & 759.71 \\
          & Shrink & \textbf{-15.15} & 12.80 & 1.79  & 11.34 & \textbf{-13.13} & 9.46  & 1.88  & 11.05 &       & Shrink & \textbf{-16.99} & 19.94 & 4.32  & 12.59 & \textbf{-15.03} & 14.73 & 3.11  & 10.58 \\
          \midrule
    \multirow{5}[0]{*}{LAD} & OLS   & -1.63 & \textbf{-1.96}& -1.28 & \textbf{-1.08} & 0.89  & \textbf{0.81} & 0.31  & \textbf{0.34} & \multirow{5}[0]{*}{LAD} & OLS   & -1.78 & \textbf{-4.21} & \textbf{-1.43} & \textbf{-1.80} & -1.43 & \textbf{-3.39} & \textbf{-1.45} & \textbf{-1.75} \\
          & WLSv  & -1.35 & 0.31  & -1.63 & -0.80 & 1.70  & 3.23  & 0.10  & 1.15  &       & WLSv  & -2.47 & -1.00 & -0.97 & -1.36 & -1.68 & 0.10  & -0.82 & -0.62 \\
          & WLSs  & -1.69 & 0.67  & \textbf{-1.67} & -0.78 & 1.31  & 3.39  & \textbf{0.05} & 1.14  &       & WLSs  & -2.62 & -0.90 & -1.07 & -1.34 & \textbf{-1.85} & 0.31  & -0.84 & -0.55 \\
          & Sample & 11.46 & 17.99 & 16.22 & 21.01 & 12.70 & 18.29 & 16.02 & 22.48 &       & Sample & 8.63  & 13.69 & 15.41 & 20.24 & 8.89  & 14.15 & 14.87 & 21.56 \\
          & Shrink & 155.24 & 387.00 & 196.34 & 267.28 & 143.18 & 343.59 & 188.62 & 261.35 &       & Shrink & 119.47 & 300.03 & 173.86 & 246.61 & 109.95 & 269.79 & 160.79 & 230.98 \\
          \midrule
    \multirow{5}[1]{*}{Huber} & OLS   & -0.70 & -0.77 & -0.52 & -0.50 & 0.78  & 0.77  & 0.51  & 0.45  & \multirow{5}[1]{*}{Huber} & OLS   & -1.41 & -2.38 & -0.75 & -1.05 & -0.74 & -1.48 & -0.48 & -0.73 \\
          & WLSv  & -0.88 & -1.01 & -1.09 & -0.93 & 1.62  & 1.74  & 0.42  & 0.76  &       & WLSv  & -2.21 & -2.05 & -0.75 & -1.55 & -1.78 & -1.12 & -0.70 & -1.01 \\
          & WLSs  & -0.57 & -1.12 & -0.73 & -0.93 & 1.31  & 1.62  & 0.46  & 0.60  &       & WLSs  & -1.86 & -2.14 & -0.55 & -1.46 & -1.46 & -1.32 & -0.72 & -1.11 \\
          & Sample & 136.87 & 424.40 & 232.48 & 368.60 & 132.49 & 403.60 & 235.75 & 383.13 &       & Sample & 125.41 & 398.99 & 239.39 & 403.13 & 124.08 & 383.99 & 238.04 & 405.63 \\
          & Shrink & -1.17 & 105.45 & 40.39 & 90.92 & -0.13 & 89.65 & 37.49 & 88.35 &       & Shrink & -6.43 & 85.98 & 36.69 & 85.44 & -5.63 & 73.93 & 32.09 & 79.08 \\
    \midrule
    \multicolumn{2}{c}{Bottom-up} & -1.71 & 0.61  & -1.70 & -0.80 & 1.29  & 3.34  & 0.03  & 1.13  & \multicolumn{2}{c}{Bottom-up}    & -2.63 & -0.91 & -1.07 & -1.34 & -1.86 & 0.30  & -0.84 & -0.55 \\
    \midrule
    \multicolumn{2}{c}{Base} & 4539.91 & 4298.87 & 6008.76 & 6600.51 & 1962.75 & 1887.26 & 2473.27 & 2687.77 & \multicolumn{2}{c}{Base}   & 1039.83 & 1039.49 & 1274.49 & 1357.36 & 448.98 & 449.45 & 537.00 & 570.31 \\
    \midrule
    \midrule
    \multirow{2}[4]{*}{$\rho$} & \multirow{2}[4]{*}{$\bm{W}_h$} & \multicolumn{4}{c}{Zones}     & \multicolumn{4}{c}{Zones by purpose of travel} & \multirow{2}[4]{*}{$\rho$} & \multirow{2}[4]{*}{$\bm{W}_h$} & \multicolumn{4}{c}{Regions}   & \multicolumn{4}{c}{Regions by purpose of travel} \\
\cmidrule{3-10}\cmidrule{13-20}          &       & $h=1$   & $h=6$   & $1 \sim 6$   & $1 \sim 12$  & $h=1$   & $h=6$   & $1 \sim 6$   & $1 \sim 12$  &       &       & $h=1$   & $h=6$   & $1 \sim 6$   & $1 \sim 12$  & $h=1$   & $h=6$   & $1 \sim 6$   & $1 \sim 12$2 \\
    \midrule
    \multirow{5}[1]{*}{LS} & OLS   & -1.61 & 0.69  & -0.16 & -0.25 & -1.11 & -1.38 & -1.05 & -0.84 & \multirow{5}[1]{*}{LS} & OLS   & -4.14 & -5.22 & -4.47 & -4.45 & 2.08  & 1.68  & 2.04  & 1.22 \\
          & WLSv  & -1.84 & -1.01 & -0.37 & -1.50 & -2.04 & -2.10 & -1.67 & -1.92 &       & WLSv  & -7.80 & -10.00 & -8.18 & -8.35 & 0.42  & 0.52  & 0.98  & 0.15 \\
          & WLSs  & -1.97 & -0.37 & -0.32 & -1.21 & -1.88 & -2.10 & -1.61 & -1.76 &       & WLSs  & -6.32 & -7.89 & -6.48 & -6.74 & 0.98  & 0.62  & 1.05  & 0.29 \\
          & Sample & 224.52 & 886.78 & 485.28 & 857.07 & 230.71 & 889.36 & 507.24 & 910.68 &       & Sample & 206.33 & 750.59 & 442.84 & 807.89 & 244.21 & 883.98 & 529.77 & 940.87 \\
          & Shrink & \textbf{-13.87} & 18.23 & 4.07  & 11.27 & \textbf{-12.36} & 11.87 & 1.51  & 8.36  &       & Shrink & \textbf{-17.82} & -0.55 & -6.40 & -0.56 & \textbf{-9.05} & 9.47  & 2.73  & 7.16 \\
          \midrule
    \multirow{5}[0]{*}{LAD} & OLS   & -1.52 & \textbf{-2.17} & \textbf{-0.97} & \textbf{-1.54} & -1.94 & \textbf{-3.56} & \textbf{-2.40} & \textbf{-2.38} & \multirow{5}[0]{*}{LAD} & OLS   & -8.42 & \textbf{-13.03} & \textbf{-10.04} & \textbf{-9.94} & -0.46 & \textbf{-3.14} & \textbf{-1.12} & \textbf{-2.26} \\
          & WLSv  & -2.08 & -0.74 & -0.53 & -1.37 & -1.98 & -1.32 & -1.68 & -1.52 &       & WLSv  & -8.59 & -10.49 & -9.22 & -8.54 & 0.06  & 0.09  & 0.07  & -0.05 \\
          & WLSs  & -2.09 & -0.84 & -0.61 & -1.32 & -2.15 & -1.44 & -1.79 & -1.48 &       & WLSs  & -8.49 & -10.43 & -9.22 & -8.44 & -0.01 & 0.00  & 0.00  & -0.01 \\
          & Sample & 10.10 & 12.56 & 16.53 & 22.08 & 10.43 & 12.61 & 15.80 & 23.62 &       & Sample & 2.76  & 3.75  & 7.45  & 14.43 & 13.26 & 15.60 & 18.72 & 25.56 \\
          & Shrink & 100.51 & 262.09 & 157.72 & 231.30 & 87.05 & 223.52 & 138.73 & 209.94 &       & Shrink & 71.97 & 176.07 & 111.89 & 176.14 & 77.80 & 186.95 & 121.31 & 184.05 \\
          \midrule
    \multirow{5}[1]{*}{Huber} & OLS   & -1.57 & 0.47  & -0.20 & -0.39 & -1.12 & -1.52 & -1.10 & -0.95 & \multirow{5}[1]{*}{Huber} & OLS   & -4.72 & -6.17 & -5.14 & -5.20 & 1.82  & 1.21  & 1.71  & 0.87 \\
          & WLSv  & -1.83 & -1.06 & -0.38 & -1.52 & -2.05 & -2.14 & -1.69 & -1.96 &       & WLSv  & -8.50 & -10.88 & -8.88 & -8.91 & -0.17 & -0.30 & 0.36  & -0.29 \\
          & WLSs  & -1.86 & -0.55 & -0.33 & -1.29 & -1.86 & -2.20 & -1.64 & -1.83 &       & WLSs  & -7.29 & -9.05 & -7.42 & -7.59 & 0.56  & 0.10  & 0.63  & -0.04 \\
          & Sample & 139.13 & 420.97 & 266.50 & 462.37 & 137.06 & 407.95 & 266.51 & 475.01 &       & Sample & 116.27 & 339.98 & 227.43 & 414.76 & 136.41 & 393.43 & 268.10 & 473.94 \\
          & Shrink & -5.85 & 72.20 & 31.49 & 75.90 & -5.70 & 57.70 & 25.06 & 67.11 &       & Shrink & -12.81 & 38.62 & 14.08 & 52.52 & -4.18 & 48.34 & 22.70 & 60.49 \\
    \midrule
    \multicolumn{2}{c}{Bottom-up} & -2.07 & -0.83 & -0.59 & -1.31 & -2.15 & -1.43 & -1.79 & -1.47 & \multicolumn{2}{c}{Bottom-up} & -8.47 & -10.43 & -9.21 & -8.43 & 0.00  & 0.00  & 0.00  & 0.00 \\
    \midrule
    \multicolumn{2}{c}{Base} & 362.46 & 360.99 & 422.73 & 442.55 & 160.84 & 162.74 & 184.11 & 190.35 & \multicolumn{2}{c}{Base} & 181.73 & 197.91 & 215.33 & 221.66 & 76.26 & 80.27 & 87.10 & 90.93 \\
    \bottomrule
    \end{tabular}}
\end{table}

\begin{table}[htbp]
  \centering
	\caption {\rm Forecasting results of the proposed RoME reconciliation through ARIMA in the real-data study in Section~\ref{sec:cross}.  The best results are highlighted in bold.}
	\label{Table_Tourism-ARIMA-full}
	\setlength\tabcolsep{3pt}
	\renewcommand{\arraystretch}{1.25}
    \footnotesize
	\scalebox{0.6}{
    \begin{tabular}{cccccccccccccccccccc}
    \toprule
    \multirow{2}[4]{*}{$\rho$} & \multirow{2}[4]{*}{$\bm{W}_h$} & \multicolumn{4}{c}{Australia} & \multicolumn{4}{c}{Australia by purpose of travel} & \multirow{2}[4]{*}{$\rho$}& \multirow{2}[4]{*}{$\bm{W}_h$}& \multicolumn{4}{c}{States}    & \multicolumn{4}{c}{States by purpose of travel} \\
\cmidrule{3-10}\cmidrule{13-20}          &       & $h=1$   & $h=6$   & $1 \sim 6$   & $1 \sim 12$  & $h=1$   & $h=6$   & $1 \sim 6$   & $1 \sim 12$  &       &       & $h=1$   & $h=6$   & $1 \sim 6$   & $1 \sim 12$  & $h=1$   & $h=6$   & $1 \sim 6$   & $1 \sim 12$2 \\
    \midrule
    \multirow{5}[1]{*}{LS} & OLS   & -7.48 & -1.04 & -3.08 & -1.44 & 1.50  & -0.04 & 0.56  & 0.33  & \multirow{5}[1]{*}{LS} & OLS   & -1.78 & -0.38 & -1.96 & -0.65 & 3.41  & 0.00  & 1.30  & 1.23 \\
          & WLSv  & -14.85 & -1.95 & -5.77 & -2.99 & -8.60 & -1.17 & -2.00 & -1.01 &       & WLSv  & -10.58 & -2.03 & -5.79 & -2.70 & -6.19 & -2.04 & -2.36 & -0.50 \\
          & WLSs  & -14.73 & -2.28 & -6.26 & -3.01 & -7.93 & -1.41 & -2.51 & -1.17 &       & WLSs  & -9.87 & -1.88 & -5.66 & -2.49 & -4.90 & -1.94 & -2.25 & -0.42 \\
          & Sample & 187.90 & 167.28 & 124.66 & 115.44 & 195.75 & 158.09 & 132.99 & 121.58 &       & Sample & 207.92 & 169.42 & 134.53 & 131.13 & 211.11 & 161.07 & 144.57 & 138.55 \\
          & Shrink & \textbf{-27.72} & \textbf{-8.41} & \textbf{-17.48} & \textbf{-13.39} & \textbf{-26.70} & \textbf{-10.50} & \textbf{-16.69} & \textbf{-12.21} &       & Shrink & \textbf{-27.26} & \textbf{-11.06} & \textbf{-19.24} & \textbf{-13.82} & \textbf{-24.02} & \textbf{-11.06} & \textbf{-16.29} & \textbf{-11.46} \\
          \midrule
    \multirow{5}[0]{*}{LAD} & OLS   & -14.70 & -1.75 & -5.69 & -2.50 & -7.65 & -1.02 & -1.62 & -0.59 & \multirow{5}[0]{*}{LAD} & OLS   & -9.98 & -1.16 & -5.04 & -1.94 & -4.32 & -1.07 & -1.21 & 0.32 \\
          & WLSv  & -12.14 & -0.10 & -3.92 & -2.18 & -5.82 & -0.57 & -0.60 & -0.37 &       & WLSv  & -8.56 & -1.18 & -4.51 & -2.04 & -4.67 & -1.66 & -1.41 & -0.03 \\
          & WLSs  & -11.98 & 0.27  & -3.55 & -2.02 & -5.45 & -0.35 & \textbf{-0.07} & -0.16 &       & WLSs  & -8.74 & -1.07 & -4.40 & -1.99 & \textbf{-4.70} & -1.61 & -1.25 & 0.03 \\
          & Sample & -11.89 & 0.30  & -3.48 & -2.01 & -5.39 & -0.33 & -0.02 & -0.16 &       & Sample & -8.67 & -1.03 & -4.36 & -1.97 & -4.67 & -1.58 & -1.22 & 0.04 \\
          & Shrink & 31.46 & 27.11 & 15.70 & -0.06 & 31.34 & 18.78 & 15.61 & 0.78  &       & Shrink & 23.81 & 13.40 & 9.52  & -0.84 & 22.36 & 10.55 & 10.16 & 0.61 \\
          \midrule
    \multirow{5}[1]{*}{Huber} & OLS   & -10.37 & -1.02 & -3.82 & -1.66 & -1.23 & -0.12 & 0.12  & 0.24  & \multirow{5}[1]{*}{Huber} & OLS   & -4.05 & -0.38 & -2.58 & -0.85 & 1.76  & -0.03 & 0.91  & 1.12 \\
          & WLSv  & -14.69 & -1.85 & -5.59 & -2.93 & -8.29 & -1.18 & -1.81 & -0.95 &       & WLSv  & -10.43 & -1.95 & -5.64 & -2.63 & -5.94 & -2.01 & -2.21 & -0.44 \\
          & WLSs  & -15.27 & -2.19 & -6.13 & -2.95 & -8.22 & -1.42 & -2.27 & -1.05 &       & WLSs  & -10.38 & -1.82 & -5.69 & -2.48 & -5.09 & -1.96 & -2.13 & -0.34 \\
          & Sample & -11.89 & 0.30  & -3.48 & -2.01 & -5.39 & -0.33 & -0.02 & -0.16 &       & Sample & -8.67 & -1.03 & -4.36 & -1.97 & -4.67 & -1.58 & -1.22 & 0.04 \\
          & Shrink & -12.31 & 1.08  & -7.49 & -12.92 & -12.33 & -4.41 & -7.94 & -12.39 &       & Shrink & -14.58 & -6.00 & -11.49 & -13.34 & -13.13 & -7.17 & -10.05 & \textbf{-11.57} \\
    \midrule
    \multicolumn{2}{c}{Bottom-up} & -11.89 & 0.30  & -3.48 & -2.01 & -5.39 & -0.33 & -0.02 & -0.16 & \multicolumn{2}{c}{Bottom-up} & -8.67 & -1.03 & -4.36 & -1.97 & -4.67 & -1.58 & -1.22 & 0.04 \\
    \midrule
    \multicolumn{2}{c}{Base} & 4809.72 & 4160.65 & 5895.83 & 6527.62 & 1985.49 & 1883.16 & 2384.58 & 2641.53 & \multicolumn{2}{c}{Base} & 1040.96 & 957.75 & 1229.66 & 1306.29 & 436.83 & 422.50 & 504.26 & 537.46 \\
    \midrule
    \multirow{2}[4]{*}{$\rho$} & \multirow{2}[4]{*}{$\bm{W}_h$} & \multicolumn{4}{c}{Zones}     & \multicolumn{4}{c}{Zones by purpose of travel} & \multirow{2}[4]{*}{$\rho$} & \multirow{2}[4]{*}{$\bm{W}_h$} & \multicolumn{4}{c}{Regions}   & \multicolumn{4}{c}{Regions by purpose of travel} \\
\cmidrule{3-10}\cmidrule{13-20}           &       & $h=1$   & $h=6$   & $1 \sim 6$   & $1 \sim 12$  & $h=1$   & $h=6$   & $1 \sim 6$   & $1 \sim 12$  &       &       & $h=1$   & $h=6$   & $1 \sim 6$   & $1 \sim 12$  & $h=1$   & $h=6$   & $1 \sim 6$   & $1 \sim 12$2 \\
    \midrule
    \multirow{5}[1]{*}{LS} & OLS   & 0.76  & -0.91 & -0.62 & -0.19 & 1.36  & 0.57  & 0.65  & 0.44  & \multirow{5}[1]{*}{LS} & OLS   & 2.85  & -0.31 & 0.65  & 0.36  & 3.84  & 0.95  & 1.70  & 0.95 \\
          & WLSv  & -4.70 & -2.57 & -3.08 & -1.69 & -3.91 & -0.78 & -1.63 & -0.91 &       & WLSv  & -3.11 & -1.83 & -2.24 & -1.35 & -0.63 & -0.14 & -0.58 & -0.37 \\
          & WLSs  & -4.04 & -2.22 & -2.68 & -1.37 & -2.90 & -0.61 & -1.24 & -0.66 &       & WLSs  & -2.50 & -1.49 & -1.78 & -1.01 & -0.02 & -0.01 & -0.16 & -0.09 \\
          & Sample & 225.43 & 175.98 & 153.63 & 150.00 & 230.03 & 177.85 & 165.61 & 160.88 &       & Sample & 220.49 & 163.12 & 155.00 & 151.61 & 237.74 & 171.24 & 169.70 & 164.00 \\
          & Shrink & \textbf{-20.68} & \textbf{-9.51} & \textbf{-15.89} & \textbf{-12.26} & \textbf{-19.18} & \textbf{-7.94} & \textbf{-14.21} & \textbf{-10.97} &       & Shrink & \textbf{-16.35} & \textbf{-8.62} & \textbf{-13.15} & \textbf{-10.42} & \textbf{-13.06} & \textbf{-6.55} & \textbf{-10.77} & \textbf{-8.64} \\
          \midrule
    \multirow{5}[0]{*}{LAD} & OLS   & -3.35 & -1.57 & -2.02 & -0.80 & -2.18 & 0.08  & -0.46 & -0.03 & \multirow{5}[0]{*}{LAD} & OLS   & -2.07 & -0.86 & -1.19 & -0.44 & 0.43  & 0.68  & 0.48  & 0.49 \\
          & WLSv  & -3.69 & -2.32 & -2.38 & -1.20 & -3.48 & -0.58 & -1.17 & -0.59 &       & WLSv  & -1.99 & -1.58 & -1.55 & -0.89 & 0.09  & 0.05  & -0.07 & -0.05 \\
          & WLSs  & -3.71 & -2.44 & -2.40 & -1.28 & -3.38 & -0.76 & -1.15 & -0.63 &       & WLSs  & -2.16 & -1.57 & -1.52 & -0.87 & -0.02 & -0.03 & -0.01 & -0.01 \\
          & Sample & -3.66 & -2.42 & -2.37 & -1.26 & -3.35 & -0.73 & -1.13 & -0.62 &       & Sample & -2.11 & -1.54 & -1.49 & -0.86 & 0.00  & 0.00  & 0.00  & 0.00 \\
          & Shrink & 23.17 & 11.41 & 10.24 & 0.87  & 16.78 & 9.60  & 8.44  & 0.89  &       & Shrink & 18.89 & 8.06  & 8.48  & 1.12  & 17.57 & 8.98  & 8.53  & 2.28 \\
          \midrule
    \multirow{5}[1]{*}{Huber} & OLS   & -0.59 & -0.91 & -0.91 & -0.30 & 0.59  & 0.56  & 0.49  & 0.39  & \multirow{5}[1]{*}{Huber} & OLS   & 1.55  & -0.31 & 0.31  & 0.23  & 3.12  & 0.94  & 1.51  & 0.88 \\
          & WLSv  & -4.64 & -2.56 & -3.02 & -1.65 & -3.85 & -0.79 & -1.56 & -0.87 &       & WLSv  & -3.02 & -1.80 & -2.15 & -1.31 & -0.49 & -0.11 & -0.48 & -0.33 \\
          & WLSs  & -4.25 & -2.21 & -2.71 & -1.37 & -2.99 & -0.62 & -1.22 & -0.64 &       & WLSs  & -2.59 & -1.47 & -1.79 & -1.00 & 0.00  & -0.01 & -0.13 & -0.07 \\
          & Sample & -3.66 & -2.42 & -2.37 & -1.26 & -3.35 & -0.73 & -1.13 & -0.62 &       & Sample & -2.11 & -1.54 & -1.49 & -0.86 & 0.00  & 0.00  & 0.00  & 0.00 \\
          & Shrink & -11.31 & -6.33 & -10.33 & -11.86 & -12.38 & -5.76 & -10.28 & -10.99 &       & Shrink & -9.63 & -6.66 & -9.38 & -10.40 & -7.60 & -5.01 & -7.94 & -8.76 \\
    \midrule
    \multicolumn{2}{c}{Bottom-up} & -3.66 & -2.42 & -2.37 & -1.26 & -3.35 & -0.73 & -1.13 & -0.62 & \multicolumn{2}{c}{Bottom-up} & -2.11 & -1.54 & -1.49 & -0.86 & 0.00  & 0.00  & 0.00  & 0.00 \\
    \midrule
    \multicolumn{2}{c}{Base} & 356.80 & 346.71 & 410.61 & 428.60 & 158.16 & 152.21 & 174.60 & 181.15 & \multicolumn{2}{c}{Base} & 165.46 & 161.31 & 183.71 & 190.03 & 74.43 & 72.83 & 81.22 & 83.76 \\
    \bottomrule
    \end{tabular}}
\end{table}

\FloatBarrier 
\bibliography{Reference_RoME}
\bibliographystyle{apalike}

\end{document}